\documentclass[preprint,authoryear]{elsarticle}

\usepackage[a4paper, total={18.35cm, 24cm}]{geometry}

\usepackage{amssymb}
\usepackage{amsmath}
\usepackage{booktabs,multirow,makecell,caption}
\usepackage{longtable}
\usepackage{pdflscape}
\usepackage[caption=false, font=footnotesize]{subfig}
\usepackage[export]{adjustbox}
\usepackage{enumitem}
\usepackage{microtype}
\usepackage{csquotes}
\usepackage{floatrow}
\usepackage{xurl}
\usepackage{chngcntr}
\usepackage{tabularx} 
\newlength\targetwidth
\newcolumntype{C}{>{\centering\arraybackslash}X}
\newcolumntype{P}{>{\raggedright}p{\targetwidth}}

\newlist{compactenum}{enumerate}{4}
\setlist[compactenum,1]{nolistsep}

\makeatletter
\patchcmd{\csq@bquote@i}{{#6}}{{\emph{#6}}}{}{}
\makeatother
\usepackage{hyperref}
\hypersetup{colorlinks,urlcolor=black}
\makeatletter
\providecommand{\doi}[1]{%
  \begingroup
    \let\bibinfo\@secondoftwo
    \urlstyle{rm}%
    \href{http://dx.doi.org/#1}{%
      \discretionary{}{}{}%
      \nolinkurl{#1}%
    }%
  \endgroup
}
\makeatother

\usepackage{color,soul}
\usepackage{changepage}
\usepackage[acronym]{glossaries-extra}
\glssetcategoryattribute{acronym}{displayglossary}{true}
\setabbreviationstyle[acronym]{long-short}
\makeglossaries

\journal{}
            
\begin{document}
\newacronym{gb}{GB}{Great Britain}
\newacronym{beis}{BEIS}{Department for Business, Energy \& Industrial Strategy}
\newacronym[
  longplural={Privacy-Preserving Techniques},
  plural={PPTs}
]{ppt}{PPT}{Privacy-Preserving Technique}

\newacronym{uk}{UK}{United Kingdom}
\newacronym{ofgem}{OFGEM}{Office of Gas and Electricity Markets}
\newacronym{ons}{ONS}{Office for National Statistics}
\newacronym{ico}{ICO}{Information Commissioner’s Office}
\newacronym{dcc}{DCC}{Data Communications Company}

\newacronym{smip}{SMIP}{Smart Meter Implementation Program}
\newacronym{gdpr}{GDPR}{General Data Protection Regulation}
\newacronym{dapf}{DAPF}{Data Access and Protection Framework}
\newacronym{smets}{SMETS}{Smart Metering Equipment Technical Specification}
\newacronym{mhhs}{MHHS}{Market-Wide Half-Hourly Settlement}

\newacronym[plural={IHDs},longplural={In-Home Displays}]{ihd}{IHD}{In-Home Display}
\newacronym{han}{HAN}{Home Area Network}
\newacronym{wan}{WAN}{Wide Area Network}
\newacronym{cad}{CAD}{Consumer Access Device}
\newacronym{csp}{CSP}{Communication Service Provider}

\newacronym{dno}{DNO}{Distribution Network Operator}
\newacronym{lse}{LSE}{Load-Serving Entity}
\newacronym{sm}{SM}{Smart Meter}
\newacronym{hh}{HH}{Half-Hourly}
\newacronym{gsp}{GSP}{Grid Supply Point}
\newacronym{sp}{SP}{Settlement Period}
\newacronym{dlc}{DLC}{Daily Load Coefficient}
\newacronym{nhhs}{NHHS}{Non-Half-Hourly Settlement}
\newacronym{hhs}{HHS}{Half-Hourly Settlement}
\newacronym{pv}{PV}{Photovoltaics}
\newacronym{ev}{EV}{Electric Vehicles}
\newacronym[plural={TVTs}, longplural={Time-Varying Tariffs}]{tvt}{TVT}{Time-Varying Tariff}
\newacronym{tou}{ToU}{Time-of-Use}

\newacronym{nilm}{NILM}{Non-Intrusive Load Monitoring}
\newacronym{psi-ca}{PSI-CA}{Private Set Intersection-Cardinality}
\newacronym{rum}{RUM}{Random Utility Maximisation}
\newacronym{dsa}{DSA}{Data Sharing Attitude}
\newacronym{dp}{DP}{Differential Privacy}
\newacronym{ddp}{DDP}{Discounted Differential Privacy}
\newacronym{fl}{FL}{Federated Learning}
\newacronym{p2p}{P2P}{Peer-to-Peer}
\newacronym{mpc}{MPC}{Multi-Party Computation}

\newacronym{ann}{ANN}{Artificial Neural Networks}
\newacronym{milp}{MILP}{Mixed-Integer Linear Program}
\newacronym{miqp}{MIQP}{Mixed-Integer Quadratic Program}
\newacronym{dro}{DRO}{Distributionally-Robust Optimisation}
\newacronym{cvar}{CVaR}{Conditional Value-at-Risk}

\newacronym{kld}{KLD}{Kullback-Leibler Divergence}
\newacronym{jsd}{JSD}{Jensen-Shannon Divergence}
\newacronym{tvd}{TVD}{Total Variation Distance}
\newacronym{ks}{KS}{Kolmogorov-Smirnov Metric}
\newacronym{pdf}{PDF}{Probability Density Function}
\newacronym{cdf}{CDF}{Cumulative Distribution Function}
\newacronym{emd}{EMD}{Earth Mover's Distance}
\newacronym{wass}{WD}{Wasserstein Distance}

\newacronym{iid}{I.I.D.}{Independent and Identically Distributed}
\newacronym{iia}{IIA}{Independence from Irrelevant Alternatives}

\newacronym{wape}{WAPE}{Weighted Absolute Percentage Error}
\newacronym{mape}{MAPE}{Mean Absolute Percentage Error}
\newacronym{mse}{MSE}{Mean Squared Error}
\newacronym{rmse}{RMSE}{Root Mean Squared Error}
\newacronym{mae}{MAE}{Mean Absolute Error}
\newacronym{mpl}{MPL}{Mean Pinball Loss}
\newacronym{lasso}{LASSO}{Least Absolute Shrinkage and Selection Operator}

\newacronym[plural=SEGs, longplural=Socio-Economic Groups]{seg}{SEG}{Socio-Economic Group}
\newacronym{wtp}{WTP}{Willingness-to-Pay}
\newacronym{wta}{WTA}{Willingness-to-Accept}
\newacronym{wts}{WTS}{Willingness-to-Share}
\newacronym{iwts}{IWTS}{Initial Willingness-to-Share}
\newacronym{wtpa}{WTP/A}{Willingness-to-Pay/Accept}
\newacronym{smd}{SMD}{Smart Meter Demand}
\newacronym{mnl}{MNL}{Multinomial Logit Model}
\newacronym[plural={MXLs}, longplural={Mixed Logit Models}]{mxl}{MXL}{Mixed Logit Model}
\newacronym{lc}{LC}{Latent Class Model}
\newacronym{rct}{RCT}{Randomised Control Trial}
\newacronym{dce}{DCE}{Discrete Choice Experiment}
\newacronym{misocp}{MISOCP}{Mixed Integer Second Order Conic Program}

\newacronym{ic}{IC}{Incentive Compatibility}
\newacronym{ir}{IR}{Individual Rationality}
\newacronym{bf}{BF}{Budget Feasibility}
\newacronym{ttp}{TTP}{Trusted-Third Party}

\begin{frontmatter}

\title{Privacy, Informed Consent and the Demand for Anonymisation of Smart Meter Data}

\author[label1]{Saurab Chhachhi\corref{cor1}}
\author[label1]{Fei Teng} 
\affiliation[label1]{organization={Department of Electrical and Electronic Engineering, Imperial College London},
        addressline={Exhibition Road}, 
            city={London},
            postcode={SW7 2AZ}, 
            state={},
            country={UK}}
\cortext[cor1]{Corresponding author: saurab.chhachhi11@imperial.ac.uk}

\begin{abstract}
Access to smart meter data offers system-wide benefits but raises significant privacy concerns due to the personal information it contains. Privacy-preserving techniques could facilitate wider access, though they introduce privacy–utility trade-offs. Understanding consumer valuations for anonymisation can help identify appropriate trade-offs. However, existing studies do not focus on anonymisation specifically or account for information asymmetries regarding privacy risks, raising questions about the validity of informed consent under current regulations.

We use a mixed-methods approach to estimate non-monetary (willingness-to-share and smart metering demand) and monetary (willingness-to-pay/accept) preferences for anonymisation, based on a representative sample of 965 GB bill payers. An embedded randomised control trial examines the effect of providing information about privacy implications.

On average, consumers are willing to pay for anonymisation, are more willing to share data when anonymised and less willing to share non-anonymised data once anonymisation is presented as an option. However, a significant minority remains unwilling to adopt smart meters, despite anonymisation. We find strong evidence of information asymmetries that suppress demand for anonymisation and identify substantial variation across demographic and electricity supply characteristics. Qualitative responses corroborate the quantitative findings, underscoring the need for stronger privacy defaults, user-centric design, and consent mechanisms that enable truly informed decisions.
\end{abstract}


\begin{highlights}
\item Mean willingness-to-accept to anonymise half-hourly data is 12\% of electricity bills.
\item Providing an anonymisation option reduces willingness-to-share non-anonymised data.
\item Information asymmetries depress demand for anonymisation and hinder informed consent.
\item Lack of default anonymisation elicits moral outrage and lowers willingness-to-pay.
\item Demand for anonymisation varies significantly across socio-demographics.
\end{highlights}


\end{frontmatter}

    \section{Introduction}\label{sec:intro}
    Smart meters are central to building a more dynamic, cost-reflective, and decarbonised electricity system. They enable high-resolution data logging, support the integration of smart appliances and automated load control, and create opportunities for innovative business models and pricing strategies \citep{FARUQUI20106222}. However, these benefits depend on both widespread meter adoption \citep{Hledik2018} and consumers' \gls{wts} granular data \citep{Teng2022}. In \gls{gb}, the rollout has fallen short of targets: as of March 2025, only 68\% of households had smart meters installed \citep{BEIS2025}. Even fewer are sharing data at the resolution needed to unlock full system benefits \citep{Citizen2019}.
    
    A major barrier to adoption is concern over data privacy and potential misuse \citep{Sovacool2017, Wilson2017}. In the Netherlands, privacy-related legal action halted the mandatory rollout \citep{Cuijpers2013}. Smart meter data can reveal sensitive personal information ranging  from occupancy and daily routines to financial and socio-demographic information \citep{Stankovic2016, Satre-Meloy2018, Wang2019, Beckel2014}. These risks increase with higher spatio-temporal granularity, linked datasets and advancing analytical techniques \citep{Veliz2018, Teng2022}.
    
    Informed consent is central to both the \gls{gdpr} and the \gls{dapf} \citep{BEIS2018}, which underpin \gls{gb}'s current data sharing and processing regulations for smart meter data. Yet, meaningful consent is difficult when consumers lack the information or expertise to assess privacy risks. The technical complexity of smart meter data and the pace of innovation in analytics create substantial information asymmetries between consumers and data users \citep{Waerdt2020}, undermining trust and limiting informed decision-making.  
    
    Unlike in countries such as the U.S., \gls{gb} consumers cannot anonymise  their data using \glspl{ppt} before sharing it \citep{SustainabilityFirstandCSE2018}.  \glspl{ppt} have been proposed to widen access to smart meter data while mitigating privacy risks \citep{Teng2022, Jawurek2012}. Techniques such as differential privacy are already being adopted by private firms like Apple \citep{Apple2017} and public institutions such as the U.S. Census Bureau \citep{Hawes2020a}. However, these methods introduce trade-offs between data utility (e.g. reduced accuracy or granularity) and privacy guarantees \citep{Acquisti2016}. 
    
    Assessing these trade-offs requires an understanding not only of the benefits smart meter data provide to stakeholders (e.g. improved procurement decisions for suppliers \citep{Chhachhi2024JED}) but also of how consumers value privacy protections, including anonymisation. While existing studies quantify consumers' \gls{wts} data \citep[e.g.][]{Citizen2019} or \gls{wtpa} to avoid sharing \citep[e.g.][]{Loessl2023}, they often overlook consumers’ specific demand for anonymisation or particular \glspl{ppt}, which is essential to inform appropriate regulatory and technical design \citep{Teng2022}.
    
    These questions have become increasingly salient in light of \gls{mhhs}, a policy reform by \gls{ofgem} aimed at increasing the use of granular consumption data \citep{OFGEM2021a}. Alongside this, there have been growing calls to expand access to smart meter data for system operation and public interest use cases \citep{SustainabilityFirstandCSE2021, EnergySystemsCatapult2023}. While \gls{gb}’s current Privacy by Design framework emphasises strong privacy defaults and user-centric protections, the \gls{mhhs} proposal to adopt an opt-out model without \glspl{ppt} would represent a shift away from that principle \citep{ICO2016, ICO2023, CitizensAdvice2018a}.  These added privacy safeguards were specifically ruled out due to a lack of evidence about the costs and benefits of such mechanisms \citep{OFGEM2019}. 
    
    This study addresses these knowledge gaps by quantifying \gls{gb} consumers' demand for anonymisation of smart meter data shared with energy suppliers for operational purposes. We employ a mixed-methods approach combining monetary valuations (\gls{wtpa}), non-monetary measures (\gls{wts} and \gls{smd}), and qualitative analysis with a nationally representative sample of 965 respondents. Crucially, we investigate information asymmetries through an embedded \gls{rct} examining the effects of informing consumers about privacy implications. We analyse four dimensions: demand for anonymisation, effects of information asymmetries, policy framing, and demographic variation. Our findings provide crucial evidence for policymakers evaluating privacy-preserving options in smart meter data governance. 

\section{Literature Review} \label{sec:litrev}
Despite a growing body of work on \glspl{ppt} for smart meter data \citep{Teng2022}, few studies have explicitly measured consumers' demand for anonymisation. Table \ref{tab:lite_rev} summarises existing survey work showing that most \gls{gb} studies focus on \gls{wts} or \gls{wtp} to avoid sharing, and are often industry-commissioned for regulatory consultations (e.g. \gls{mhhs}).

\begin{table*}
\centering
    \caption{Relevant Survey Studies}
    \label{tab:lite_rev}
    \resizebox{\textwidth}{!}{
    \setlength\targetwidth{1.2\textwidth}
    \begin{tabularx}{\targetwidth}{@{}lccccccCcc@{}}
    \toprule
        Source & Year & Loc. & Size & Rep. & Type & Mode & Measure & \multirow{2}{*}{\makecell{Demand for \\Anonymisation}} & \multirow{2}{*}{\makecell{Privacy\\ Implications}} \\\\
        \midrule
            \citet{GERPOTT2013483} & 2011 & GER & 453 & Y & A & O & SMD (WTP)  & N & N  \\
            \citet{Jakobi2019}     & 2014 & GER & 205 & N & A & P, FG & WTS  & N & Y  \\
            \citet{Horne2015}      & 2015 & USA &708 & Y & A & O & SMD  &N & Y  \\
            \citet{Richter2018}    & 2015 & GB & 1,892 & Y & A & O & WTP  &N & N  \\
            \citet{Dickman2017}    & 2017 & GB & 120 & N & I & FG & WTS  &Y & N  \\
            \citet{Skatova2019}    & 2017 & GB & 265 & N & A & O & WTP  & N & N  \\
            \citet{Knight2018a}    & 2018 & GB & 1,467 & Y & I & F2F & WTS, Change in WTS  &Y & N  \\
            \citet{Gosnell2023}    & 2019 & GB & 2,430 & Y & A & O & SMD (WTP)  &N & N  \\
            \citet{SSEN2020}       & 2019 & GB & 1,000 & N & I & O & WTS & Y & N  \\
            \citet{Citizen2019}    & 2019 & GB & 3,221 & Y & I & O & WTS, SMD  &N & Y* \\
            \cite{Grunewald2020c}  & 2019 & UK & 701 & Y & A & O & WTS & N & Y*  \\
            \citet{Loessl2023}     & 2021 & GER & 1,063 & Y & A & O & WTA  &N  & N  \\
            \citet{Pelka2024}      & 2022 & GER & 962 & Y & A & O & WTS & N & N  \\
            This study             & 2021 & GB & 965 & Y & A & O & WTS, Change in WTS, WTP/A, SMD & Y & Y  \\
    \bottomrule
    \multicolumn{10}{@{}P@{}}{Notes: Type - Industry(I) or Academia(A); Mode - Face-to-face (F2F), Focus Group (FG), Online (O), Post (P). *Both \citet{Citizen2019} and \citet{Grunewald2020c} present potential privacy risks of sharing data but do not explicitly state that these are realisable.}
    \end{tabularx}
    }
\end{table*}

\subsection{Privacy Valuations and Demand for Anonymisation}
Existing literature predominantly assesses consumers' \gls{wts} smart meter data through explicit trade-offs for tangible benefits. For instance, an \gls{ofgem}-commissioned survey indicated that 55\% of respondents were willing to share half-hourly data with energy suppliers in exchange for direct benefits such as ongoing discounts \citep{Knight2018a}. Similarly, focus groups by \citet{Dickman2017} reported that the majority were comfortable sharing individual-level, non-aggregated data, though a notable minority (12\%) expressed discomfort. In addition, \citet{Citizen2019} found that 63\% of respondents were willing to share monthly smart meter data, dropping to 43\% for higher resolution data (e.g. half-hourly or real-time).  \citet{Pelka2024}  further highlighted that consumers prefer demand-response services involving data sharing if the benefits, like reduced costs and greater appliance control, outweighed privacy considerations. However, \gls{wts} consistently declines when data is used for marketing or is accessed by less trusted entities, such as third parties or intermediaries. Similarly,  \citet{Grunewald2020c} found that trust and whether the data will be used for its intended purpose plays a significant role in determining \gls{wts}. 

Explicit studies quantifying \gls{wtpa} to protect privacy reveal significant variation. \citet{Loessl2023} found that German consumers required an average compensation of €2.54/month to allow their energy supplier to analyse their data but demanded significantly more (€21.33/month) when third parties were involved, with valuations strongly linked to general privacy concerns. In \gls{gb}, earlier studies reported \gls{wtp} for third-party data sharing ranging from £1.00 to £7.27/month, underscoring substantial heterogeneity and sensitivity to specific conditions and the options presented \citep{Richter2018,Skatova2019}. Notably, existing GB studies typically assume supplier access as granted and thus lack detailed insights into consumers' valuation of supplier-specific privacy concerns.

Data privacy concerns significantly impact overall \gls{smd}, given its voluntary deployment. \citet{Gosnell2023} found privacy fears were a prominent reason for smart meter rejection, with privacy-sensitive consumers demanding an average compensation of £192 to accept installation. This aligns with earlier findings from Germany, where increased trust in a supplier's data protection raised consumer \gls{wtp} for smart meters \citep{GERPOTT2013483}.

Very few surveys directly investigate \glspl{ppt} like data aggregation and anonymisation though existing studies clearly indicate consumer interest. For example, \citet{Knight2018a} reported that anonymisation encouraged 41\% of respondents to share data, while a \gls{dno}-commissioned study found 79\% considered anonymisation important \citep{SSEN2020}. \citet{Dickman2017} similarly confirmed a majority's comfort with sharing aggregated, anonymised data. However, explicit monetary valuations (\gls{wtpa}) specifically for anonymisation remain, to the best of our knowledge, absent from the existing literature. 

Moreover, the framing of privacy choices (such as the availability of anonymisation, options to restrict supplier data access, and data resolution), the data user, and the proposed use significantly influence perceived privacy importance. Thus, evaluating both non-monetary (\gls{wts}) measures and explicit monetary valuations (\gls{wtpa}) is crucial for accurately capturing consumer preferences regarding anonymisation. Additionally, given the voluntary nature of the \gls{gb} smart meter rollout, understanding how privacy preferences impact overall smart meter demand (\gls{smd}) is equally important. 

\subsection{Informed Consent and Information Asymmetries}
Consumer \gls{wts} smart meter data is heavily influenced by perceived sensitivity, yet studies consistently reveal a mismatch between actual and perceived risks. While smart meter data can reveal intimate details (e.g. financial habits, medical conditions, occupancy patterns; \citet{Teng2022}), it is often viewed as non-sensitive \citep{SSEN2020}. For instance, consumers rank it below financial, location, or medical records \citep{Knight2018a, Skatova2019}, and \gls{wtp} to avoid sharing it (£1.00/month) is half that of broader personal data (£2.11/month) \citep{Richter2018}. This suggests bounded rationality \citep{Simon1990} with consumers lacking awareness of embedded risks resulting in information asymmetries that distort consent \citep{Waerdt2020}.

When privacy implications are explicitly explained, behaviour shifts:  \citet{Horne2015} observed a 20\% drop in demand for smart meters, and \citet{Jakobi2019}  saw 86\% of respondents revise their favoured utility subscriptions after learning the privacy risks. Similarly, \citet{Grunewald2020c} observed an 11 percentage-point rise in unease about data sharing during their survey, underscoring how even small amounts of contextual information alter preferences.  Conversely, transparency about data use can reduce concerns \citep{Loessl2023,Dickman2017}, highlighting a tension: passive disclosure dampens privacy concerns, while active education heightens them.

Critically, an estimated 37\% of \gls{gb} smart meter owners are unaware of their data-sharing options \citep{Citizen2019}, and energy-sector apathy further undermines informed consent \citep{Sovacool2017}. This  raises fundamental questions about whether prevailing policies, and much of the literature assessing them, fail to adequately address information asymmetries, instead relying on (and potentially exploiting) consumer ignorance rather than ensuring truly informed decision-making.

\subsection{Framing Effects}
The current data-sharing framework assumes consumers can rationally weigh privacy risks against the benefits of sharing smart meter data. However, this premise is undermined by framing effects and asymmetric valuations. Studies often contextualise privacy valuations within dynamic tariffs \citep{Richter2018, Loessl2023} or demand response schemes \citep{Pelka2024}, despite them not being contingent on high-resolution data sharing \citep{McKenna2015,Teng2022}. Consumers may accept data sharing for perceived benefits like financial savings \citep{Dickman2017} or market efficiency \citep{Knight2018a}, yet many core smart meter advantages (e.g., automated billing, energy feedback) need not entail disclosing identifiable data \citep{Teng2022}. 

Privacy valuations are highly sensitive to framing. A U.S. survey on general data privacy concerns revealed stark disparities between \gls{wtp} and \gls{wta} \citep{Winegar2019}, reflecting an endowment effect \citep{Kahneman1991}. This asymmetry may stem from moral outrage: consumers perceive privacy as a right to be protected by default, not a commodity to be purchased. Such framing has critical policy implications, for example, whether data sharing adopts opt-in (privacy-preserving by default) or opt-out (exploiting inertia;  \citet{Kahneman1991}) models. 

Consumer preferences are further shaped by available options and the contextual information provided. \citep{Palinski2021},  found that merely educating users about \gls{gdpr} rights increased privacy valuations for ride-hailing data, suggesting information framing alters perceived fairness.  Existing studies may underestimate privacy demand due to presentation bias. For instance, \citep{Knight2018a} measured \gls{wts} non-anonymised data without first informing respondents of anonymisation options, a design choice that likely suppressed true privacy valuations. The broader literature confirms ordering effects and default settings significantly influence decisions, raising questions about whether current policy frameworks and literature genuinely reflect consumer preferences \citep{Acquisti2013}. Critically, these preferences emphasise control: over 90\% of respondents in \citet{Citizen2019} considered having data-sharing choices essential for smart meter adoption. 

\subsection{Heterogeneity in Privacy Preferences}

Existing research demonstrates substantial variation in privacy preferences concerning smart meter data across demographic and socio-economic lines. Studies consistently show that women, older adults, and individuals from lower \gls{seg} express stronger privacy concerns and are less convinced by anonymisation options \citep{Knight2018a}. Age appears particularly significant, with older respondents demonstrating both lower trust in data sharing arrangements and greater discomfort with high-resolution data collection. Interestingly, this discomfort is amplified among those without smart meters, with only 21\% willing to share half-hourly consumption data \citep{Citizen2019}.

The perception of data sensitivity also varies across demographics. For example, younger respondents tend to view smart meter data as more sensitive than older generations \citep{SSEN2020}. Information asymmetries compound these differences, disproportionately affecting vulnerable groups; lower-\gls{seg} individuals are significantly less likely to understand the full implications of smart meter data collection, such as the ability to determine household occupancy patterns \citep{Citizen2019}.

Economic factors and technology literacy further complicate this landscape.  Clustering analysis in \citet{Richter2018}  found that women, technology savy, and higher-income individuals exhibit systematically higher privacy valuations correlate. However, these patterns are not universal, as a German study found no significant socio-demographic differences, suggesting important national context effects in privacy preferences \citep{Loessl2023}.

This complex interplay of demographic characteristics, technological literacy, trust levels, and economic circumstances creates a heterogeneous privacy preference landscape. The variation differs across different measurement approaches. Such findings underscore the importance of considering multiple dimensions of difference when designing privacy frameworks and communication strategies for smart meter programs. The lack of consistent patterns across studies highlights the contextual nature of privacy concerns and the need for flexible policy approaches that can accommodate diverse population needs.

The extant literature reveals three critical gaps in understanding consumer privacy valuations for smart meter data. First, while numerous studies quantify general \gls{wts} or \gls{wtpa} for data access, none explicitly measure demand for anonymisation despite evidence that consumers value it. Second, pervasive information asymmetries undermine informed consent: consumers systematically underestimate smart meter data sensitivity, and studies rarely account for how privacy education alters valuations. Third, existing valuations may be methodologically biased either by omitting anonymisation options or by framing choices in ways that privilege institutional over consumer preferences.

This study addresses these gaps by rigorously quantifying \gls{gb} consumers’ demand for anonymisation when sharing smart meter data with energy suppliers, employing a mixed-methods design that:
\begin{enumerate}
    \item Measures both monetary (\gls{wtpa}) and non-monetary (\gls{wts}, \gls{smd}) valuations of anonymisation;
    \item Tests the impact of information asymmetries via an embedded \gls{rct};
    \item Focuses on operational data uses (e.g. forecasting, settlement) to directly inform \gls{mhhs} policy debates.
\end{enumerate}
By integrating qualitative analysis of open-ended responses, we further illuminate the reasoning behind privacy preferences, a dimension absent from prior quantitative surveys. Our findings provide actionable evidence for designing \glspl{ppt} that balances utility and consumer protection.
 
The remainder of the paper is organised as follows. Section \ref{sec:method} summarises the methodology including the survey design, modelling framework and sample. Section \ref{sec:results} details consumers' \gls{wts} and \gls{wtpa}, their heterogeneity, and the results of the information treatment. Finally, the policy implications of informed consent and framing effects are discussed, and conclusions are drawn in Section \ref{sec:conc}.

\section{Methodology}\label{sec:method}
The study aims to quantify the demand for anonymisation of smart meter data when shared with energy suppliers in \gls{gb}. We assess policy implications by examining \gls{wts}, \gls{wtp} (opt-out), \gls{wta} (opt-in), and overall \gls{smd}. We investigate how providing information on privacy risks influences responses and explore heterogeneity across socio-demographic and other pertinent characteristics. 

\subsection{Survey Overview}
The survey consisted of three parts\footnote{The full questionnaire listed in Table \ref{tab:survey_qs}.}; (1) socio-demographic screening for eligibility, (2) electricity supply characteristics and general data-sharing attitudes; (3) a \gls{dce} with an embedded \gls{rct} assessing preferences for electricity supply contract.

The sample was nationally representative of \gls{gb} energy bill-paying adults, with quotas based on the broader \gls{gb} population (gender, age, ethnicity, \gls{seg}, region; as in \citet{Richter2018}), plus a soft quota for smart meter ownership \footnote{Quotas details can be found Table \ref{tab:full.samp}. Minimum survey completion time was 4 minutes for quality assurance.}. Collected background information included income, tenure, monthly bills, tariffs, fuel supply, smart meter data-sharing choices, and engagement with \gls{ihd}\footnote{Details in Table \ref{tab:survey_qs}, questions 10, 23-27.}. Finally, we measure respondents' general \glspl{dsa} based on their existing data sharing practices\footnote{Question 11 in Table \ref{tab:survey_qs}. We employ a generic statement and avoid \textit{privacy} completely, to avoid priming respondents prior to the \gls{dce} \citep{CACCIATORE2012673}.}:
\begin{compactenum}[label=(\alph*)] 
    \item Basic Sharing (BA): provide minimum data required to access service.
    \item Marketing \& Research (MR): allow data to be used for marketing, research, forecasting etc.
    \item Third-Party (TP): allowing data to be passed to third parties.
\end{compactenum}

\subsection{Experimental Design}\label{sec:survey_exp}
Since anonymisation is not currently offered in \gls{gb}, we employ a \gls{dce} to evaluate respondent preferences and \gls{wtpa} for anonymised half-hourly data sharing. An embedded \gls{rct} educates the treatment group on privacy risks associated with smart meter data. Similar methods have been employed to test the impact of informational interventions in the context of data privacy more broadly. Of particular note are: \citet{Palinski2021}, which investigated how privacy trade-offs for ride-hailing services are affected by knowledge of \gls{gdpr} rights, \citet{Glasgow2021}, which investigated whether the inclusion of data sharing as a choice attribute within the \gls{dce} as opposed to a general condition affected survey response bias, and finally, \citet{Loessl2023} which is most closely linked to our study but relies on respondents' own perceptions of what their \textit{private} smart meter data contains to estimate privacy valuations.

\subsubsection{Willingness-to-Pay/Accept} \label{sec:exp_wtpa}
The inferable personal information from smart meter data is dependent on: the temporal resolution, spatial resolution (aggregation), and whether it is anonymised\footnote{Following feedback from initial pilot of 46 respondents, aggregation was excluded due to difficulties in discerning differences between aggregation and anonymisation. This is an interesting finding in itself as aggregation and anonymisation do not offer the same notion of privacy preservation \citep{Teng2022}.}. We develop an unlabelled \gls{dce}, accounting for the choices available under the \gls{dapf}. As summarised in Table \ref{tab:sv-attrs}, each option consisted of three attributes: (1) the change in bill\footnote{Respondents who did not provide their bill were assigned the national average bill of £57/month. Calculated based on \gls{ofgem}'s average electricity consumption, (2,900 kWh, for a medium household in 2020 (https://www.ofgem.gov.uk/information-consumers/energy-advice-households/average-gas-and-electricity-use-explained) and the corresponding average electricity rates, 23.5 p/kWh (https://www.ofgem.gov.uk/energy-data-and-research/data-portal/retail-market-indicators/).}, (2) the frequency of data sharing, and (3) whether data is anonymised. The experimental design consisted of 12 blocks of 8 choice tasks per respondent with two unlabelled alternatives in each choice task\footnote{Restrictions were placed to ensure the exclusion of dominated alternatives. These are summarised in Table \ref{tab:sv_restrict}. A blocked fractional-factorial design was generated using the SAS \texttt{\%Choiceff} macro \citep{Kuhfeld2010} with priors based on the pilot study and selected based on D-efficiency criterion.}.

\begin{table}
    \centering
    \caption{Choice Attributes and Levels} \label{tab:sv-attrs}
    \begin{tabular}{p{0.12\linewidth}p{0.22\linewidth}p{0.48\linewidth}}
        \toprule
        Attribute & Levels & Description \\
        \midrule
        Bill & -20\% to + 20\% in 5\% intervals & Expected bill change in £ linked to actual monthly electricity bill.\\
        Anon & Yes, No & Anonymised data cannot be linked to a particular person and therefore cannot be used to build profiles or identify individuals.\\
        Freq & Real-Time, Half-Hourly, Daily & The resolution of the smart meter data shared.\\
        \bottomrule
    \end{tabular}   
\end{table}

\begin{figure}
    \centering
    \includegraphics[width = \linewidth]{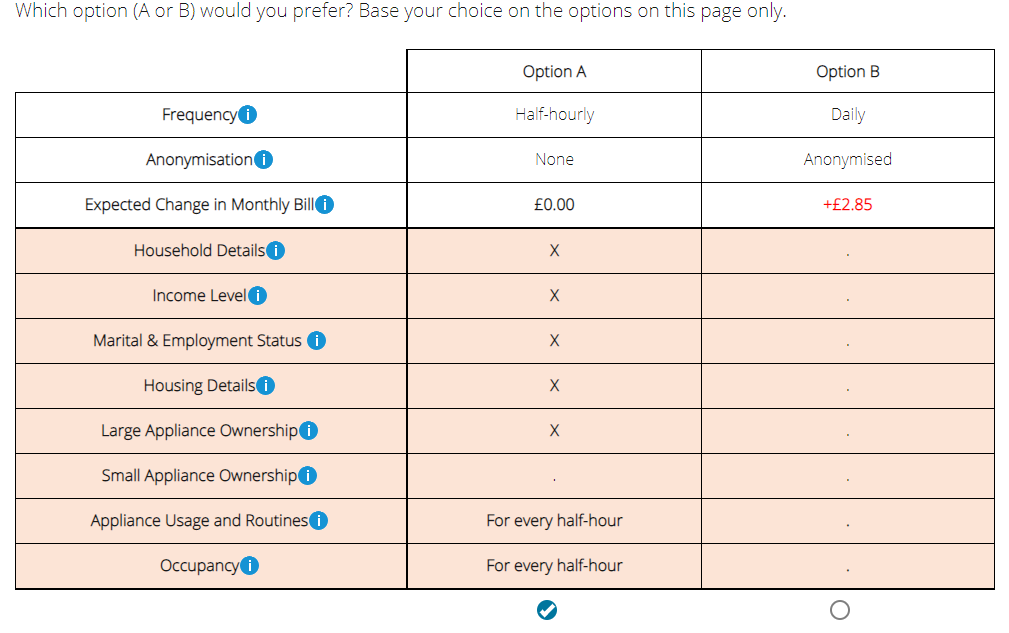}
    \caption{Example Choice Task}
    \label{fig:sv-card}
\end{figure}

The control group received the benefits of smart metering mimicking promotional material disseminated by suppliers and government \citep{SMGB2021}, followed by an explanation of the attributes\footnote{See illustrative screens in Table \ref{tab:choice_screen}.}. Importantly, these descriptions did not include what personal information may be embedded within smart meter data. Instead, they only mention that suppliers would have access to their energy consumption data at the specified resolution. The treatment group were given educational material about the implications of personal information being shared under each data sharing option\footnote{See screens 7 to 10 in Table \ref{tab:choice_screen}.}. They were shown the different type of personal information embedded within smart meter data, a labelled chart pointing out the energy usage patterns of different appliances, and a table showing the dependence of inferable personal information on data resolution. A combination of energy related information and non-energy related information was selected to highlight the broad range of personal information embedded within smart meter data\footnote{Included appliance usage, occupancy, income level and marital status. Their dependence on the data sharing options are summarised in Table \ref{tab:sv_restrict} with mapping based on \citet{Teng2022}. Further details can be found on screen 8 in Table \ref{tab:choice_screen}.}. Figure \ref{fig:sv-card} shows an example choice task with the control group shown only the first three rows (shown in white) while the treatment group were also shown the privacy implications of each option (shown in the beige rows).

To measure respondents' comprehension of the educational material they were asked to answer three true/false statements, and provide structured and open-ended feedback\footnote{Summarised in Table \ref{tab:samp.manip} and Table \ref{tab:samp.feed}.}. Strictly speaking, both groups receive a treatment through the information provided on benefits of smart metering. In terms of privacy risks the control groups' perceptions rely on their existing knowledge (or lack thereof). This mimics the setting in which most existing survey studies have been carried out \citep[e.g.][]{Knight2018a,Richter2018,Loessl2023} and the information landscape consumers currently face when installing a smart meter.

\subsubsection{Willingness-to-Share}
Respondent's \gls{iwts} half-hourly smart meter data was captured on a 5-point Likert scale, prior to the \gls{dce} introduction\footnote{Specifically:
\textit{
How willing would you be to share your half-hourly electricity consumption data with your energy supplier?
}
[Very willing, Quite willing, Indifferent, Not very willing, Not at all willing]
}.
Following the introduction to the \gls{dce} and educational material, respondents gave their \gls{wts} \textbf{non-anonymised} half-hourly data on a 3-point scale\footnote{Specifically, \textit{Considering the information you have just read, would you be more or less likely to share your \textbf{half-hourly} electricity consumption data if it was \textbf{not anonymised} before being shared?}[More likely, It makes no difference, Less likely]}. This was repeated for sharing \textbf{anonymised} half-hourly smart meter data. The responses to the change in \gls{wts} measure both the effect of offering the option to anonymise, in the case of the control group, and the effect of the informational treatment. 

\subsubsection{Demand for Smart Metering}
In order to ensure we elicit data sharing preferences from those who do not want a smart meter, we exclude the option to opt of smart metering in the \gls{dce}. Instead, after the choice tasks respondents were asked whether they would get a smart meter with one of the options in the choice tasks or not have one\footnote{See question 15 in Table \ref{tab:survey_qs}.}. All blocks of choice tasks included at least two anonymised options without a fee. Thus, the response provides an indication of the demand for a smart meter, \gls{smd}, given the possibility to anonymise.

\subsection{Modelling Framework}
The study examines four measures: (1) \gls{wtpa} for anonymisation and data-sharing frequencies, (2) \gls{iwts} for half-hourly data, (3) changes in \gls{wts} under different anonymisation conditions, and (4) the \gls{smd}. Hypotheses were developed based on existing literature on smart meter data privacy and valuations, assuming demand for anonymisation will follow the same trends as data privacy. These are summarised in Table \ref{tab:hypotheses}. For each of the four measures and corresponding hypotheses a series of modelling approaches were employed to ensure the robustness of our results\footnote{A summary of the robustness tests can be found in Table \ref{tab:robust}.}. All analysis was performed in \texttt{R}.

\begin{table*}
    \centering
    \caption{Hypotheses}
    \label{tab:hypotheses}
    \setlength\targetwidth{\textwidth}
    \begin{tabularx}{\targetwidth}{@{}CCp{0.7\textwidth}@{}}
        \toprule
        No. & Measure & Description \\
        \midrule
        \textbf{H1} & \gls{wtpa} for Frequency & Higher-frequency data allows more personal information to be inferred; therefore, the average \gls{wtpa} will be higher for sharing lower frequency data.\\
    
        \textbf{H2} & \gls{wtpa} for Anonymisation & The average \gls{wtpa} for anonymisation will be lower at lower data sharing frequencies, as this reduces privacy risks.\\
        
        \textbf{H3} & \gls{wtpa} for Anonymisation & Respondents will exhibit an endowment effect, resulting in the \gls{wta} being greater than the \gls{wtp}.\\
        
        \textbf{H4} & Change in \gls{wts} & A greater proportion of respondents will be less likely to share data when it is not anonymised, compared to when it is anonymised.\\
        
        \textbf{H5} & All & Measures will vary by socio-demographic characteristics, reflecting differing privacy concerns. Higher privacy concerns (i.e. higher \gls{wtpa}, lower \gls{iwts}, lower \gls{smd}) are expected among older individuals, women, those in lower \gls{seg}, non-smart meter owners, and individuals who share less data, based on existing literature. In addition, those on \gls{tvt} tariffs and who frequently engage with their \gls{ihd} may show greater awareness of data sensitivity and therefore have higher concerns.\\
        
        \textbf{H6} & All exc. \gls{iwts} & The information treatment will increase awareness of the personal nature of smart meter data, heightening privacy concerns leading to higher \gls{wtpa}, reduced \gls{wts}, and lower \gls{smd}.\\
        
        \textbf{H7} & All exc. \gls{iwts} & The treatment effect will be moderated by pre-existing privacy attitudes, with smaller effects observed among those more comfortable sharing data.\\
        
        \textbf{H8} & Change in \gls{wts} & The treatment effect on \gls{wts} will be less pronounced when data is anonymised, due to reduced perceived privacy risk.\\
        \bottomrule
        \multicolumn{3}{@{}P@{}}{Note: Post-hoc analysis also considered \gls{iwts} as a proxy for baseline privacy attitudes for \textbf{H7}.}
    \end{tabularx}
\end{table*}

\subsubsection{Willingness-to-Pay/Accept}
To estimate respondents’ \gls{wtpa}, we employ the \gls{rum} framework \citep{mcfadden1972conditional}, assuming a linear utility function comprising observable utility (from \gls{dce} attributes) and an unobserved component. Under \gls{rum}, respondents choose the alternative that maximises utility in each \gls{dce} task. To capture potential endowment effects, we differentiate bill changes as either an effective fee (bill increase) or discount (bill decrease) \citep{Lanz2010}, expressed as percentage changes rather than monetary amounts. Interaction terms between data sharing frequency and anonymisation are included.

The latent utility for each individual, $i$, for each alternative, $j$, in each choice task, $k$, is \footnote{The first manipulation (Table \ref{tab:samp.manip}) check tested whether respondents correctly interpreted the bill change attribute. To account for mis-interpretation (i.e. respondents considering a fee to be a discount or vice versa) an interaction term is included for monetary variables, in line with methods to deal with attribute non-attendance \citep{HESS2010781}. We also include results for models without this correction in Table \ref{tab:poe.freq.robust}.}:
\begin{align}
    \begin{split}
    U_{i,j,k}=&\ \alpha_{i,1} Fee_{j,k} + \alpha_{i,2} Disc_{j,k} + \beta_{i,1} HH_{j,k}\\
    & + \beta_{i,2} Daily_{j,k}  + \beta_{i,3} Anon_{j,k}\\
    & + \beta_{i,4} Anon_{j,k}\times {HH}_{j,k}\\
    & + \beta_{i,5} Anon_{j,k} \times Daily_{j,k}  + \epsilon_{i,j,k}
    \end{split}
\end{align}
where, $\alpha_{i,\star}$ are the parameters of the monetary attributes\footnote{$Fee$ - percentage point increase in bill, $Disc$ - percentage point decrease in bill.}, $\beta_{i,\star}$ for the non-monetary parameters\footnote{$Anon$ - 1 indicates data is anonymised, $HH$ - 1 indicates half-hourly sharing, $Daily$ - 1 indicates daily data sharing. The reference case is non-anonymised real-time data sharing.}, and $\epsilon_{i,j,k}$ is a type 1 extreme value error term. 

We estimate parameters via a \gls{mxl}, allowing for preference heterogeneity and relaxing the \gls{iia} assumption \citep{Train2009}. Non-monetary parameters follow a normal distribution, while monetary ones follow a symmetric, zero-bounded triangular distribution\footnote{Alternative distributions were assessed; log-normal and log-uniform yielded implausible means; normal has undefined moments; fixed parameters lacked realism. The chosen distribution balances plausibility and model fit \citep{Train2005,Daly2012,Hess2017CorrelationModels}.}.  

Mean \gls{wtpa} is calculated as the expected value of the ratio of non-monetary to monetary parameters. For example, the \gls{wtp} for switching from non-anonymised real-time data to anonymised half-hourly data, for individual $i$ would be:
\begin{align}
    WTP = \mathbb{E}\left[\frac{\beta_{i,1} + \beta_{i,3} + \beta_{i,4}}{\alpha_{i,1}}\right]
\end{align}
where, the expectation is computed over the joint distribution of the estimated parameters using the parametric bootstrapping procedure in \citet{Krinsky1986}. We estimate separate models for each experimental group, both overall and stratified by whether respondents typically share data with third parties\footnote{Pooled models with scale parameters are also estimated. See Table \ref{tab:mixl.split}.}. Differences in mean \gls{wtpa} are tested using the complete combinatorial test \citep{Poe2005ComputationalDistributions}, following \citet{Loessl2023}. All \glspl{mxl} were estimated using the \texttt{apollo} package \citep{Hess2019}.

To explore socio-demographic heterogeneity, we estimate a \gls{mxl} with mean-shifting interactions for anonymisation and frequency\footnote{Summarised in Table \ref{tab:mixl.het}.}. Group-specific means (e.g. women) are derived via bootstrapping and weighted marginal means \citep{MAYER2024103545}, using the sample's representativeness. Segmented models for each demographic group are also estimated for robustness \citep{Badole2023}. Results are reported in Table \ref{tab:seg.het}.

\subsubsection{Willingness-to-Share \& Demand for Smart Meters}
We analyse \gls{wts} and \gls{smd} using ordinal, binary, and \glspl{mnl}, depending on the outcome type, with estimates presented as marginal mean predicted probabilities \citep{MAYER2024103545}\footnote{Calculated using the \texttt{emmeans} package.}. This is supplemented by descriptive statistics and non-parametric tests\footnote{Mann-Whitney U and Wilcoxon paired tests for two-group comparisons; Kruskal-Wallis followed by Dunn tests for multiple groups, with Holm-adjusted and unadjusted p-values reported.}.

As the \gls{iwts} is captured on a 5-point ordinal scale from ‘Not at all willing’ to ‘Very willing’, we estimate a partial proportional odds model with socio-demographic covariates\footnote{Estimated using the \texttt{ordinal} package \citep{christensen2023}.}. This specification was chosen after testing and rejecting the proportional odds assumption for a number of covariates\footnote{Assessed via likelihood ratio tests and comparison with a \gls{mnl} (estimated using \texttt{mblogit} \citep{Elf2024}) and binary logistic regressions (estimated using \texttt{glm}). See Table \ref{tab:wtsg.lr}.}.

The \gls{smd} is a binary outcome, where 1 indicates a preference not to have a smart meter. We estimate a logistic regression using the same covariates, plus the \gls{iwts} and experimental group\footnote{Interaction models between respondents' \gls{dsa}/\gls{iwts} and the treatment were also tested but showed no significant effects. Full models in Table \ref{tab:wtsg.lr}.}.

To assess changes in \gls{wts} between anonymised and non-anonymised framings, we treat responses as repeated measures and estimate a random-effects \gls{mnl}, with outcomes ‘More likely’, ‘No difference’, and ‘Less likely’. Covariates include socio-demographics, \gls{dsa}, \gls{iwts}, treatment group (TR), and framing (Anon), with random intercepts for respondents \citep{Hensher_Rose_Greene_2015} \footnote{Ordinal models with random-effects were initially considered with proportional as well as partial proportional odds \citep{agresti2010analysis}. However, the proportional odds assumption did not hold for most covariates with the random-effects \gls{mnl} providing significantly higher explanatory power. This suggests the ordinal structure was not meaningful due to the repeated measurement specification. Separate (for anonymised and non-anonymised data) fixed-effects partial proportional models result in similar or more parsimonious models than their \gls{mnl} counterparts. See Tables \ref{tab:wtsn.ord} and \ref{tab:wtsa.ord}.}. The utility for each individual, $i$, for each alternative, $j$ is\footnote{Separate models for anonymised and non-anonymised data sharing estimated opposite signs for the effect of smart meter ownership. We therefore included an interaction term between the question framing and smart meter ownership in the combined model.}:
\begin{align}
\begin{split}
    &U_{i,j} =  \beta_{1,j}Anon + \beta_{2,j} TR + \beta_{3,i} IWTS\\
    &\qquad + \beta_{4,j} Anon \times TR + \beta_{4,j} Anon \times IWTS\\
    &\qquad + \beta_{5,j} Anon \times DSA + \beta_{6,j} TR \times IWTS\\
    &\qquad+ \beta_{7,j} TR \times DSA  + \sum_{m \in \mathcal{M}}\beta_{m,j}x_m + \epsilon_{i,j} + u_i
\end{split}
\end{align}
where, $\mathcal{M}$ is the set of socio-demographics covariates, $\epsilon_{i,j}$ is a type 1 extreme value error term and $u_i \sim N(0,\sigma^2)$ is the individual-level intercept\footnote{The model is estimated using the \texttt{mclogit} package \citep{Elf2024}, with results in Table \ref{tab:wtsna.ord}.}.

\subsection{Survey Sample}\label{sec:survey_sample}

The survey was administered online via Accent Market Research and panel partner Sevanta ComRes\footnote{Respondents were drawn from a target population using a matching algorithm on general population studies. Participants were invited through an online platform and compensated proportionally (approx. £0.50 for expected completion time).}. Fieldwork took place from 25 March to 12 April 2021, during which 2,810 respondents began the survey. Of these, 598 were screened out based on eligibility, 94 for completing the survey in under four minutes\footnote{Average completion times of 9.54 and 10.37 minutes for the control and treatment groups, respectively. See Figure \ref{fig:samp_time} for full distribution.}, and 99 excluded to meet nationally representative quotas. The final sample received by us comprised 965 respondents; 477 in the control group and 488 in the treatment group\footnote{The study protocol was reviewed and approved by the Imperial College London Research Governance and Integrity Team (RGIT) under SETREC number 21IC6603.}.

Missing data accounted for 2.16\% of relevant variables\footnote{13 respondents lacked age, gender or \gls{seg} data; 13 were unsure about smart meter ownership; 109 were unsure of tariff type.}. Where appropriate, unknown electricity characteristics were assumed to reflect default conditions (e.g. standard tariff or no smart meter). Missing socio-demographic values were imputed using multiple imputation via ordinal logistic regression on income\footnote{Imputed using \texttt{mice} package \citep{mice2011}. Full pre-imputation sample characteristics are shown in Tables \ref{tab:full.samp} and \ref{tab:samp.elec}.}. Electricity bill data, central to the \gls{dce}, were provided by 66.0\% of respondents with remaining respondents assigned the average monthly bill of £57\footnote{Average reported monthly bill was £65.84, with minimal differences between groups (£67.50 control; £64.30 treatment), as shown in Figure \ref{fig:bill}.}.

Both groups are nationally representative across key socio-demographic quotas (age, gender, ethnicity, \gls{seg}, region), with no significant inter-group differences\footnote{See Table \ref{tab:imp.samp} for Pearson’s $\chi^2$ test for independence between the groups and for z-test for proportions for each group against \gls{gb} nationally representative proportions.}. However, smart meter ownership is over-represented in both the control (52.0\%) and treatment (55.7\%) groups compared to the 44.0\% national rate at the time \citep{BEIS2021d}. Other electricity-related and socio-demographic variables (except tenure and fuel type) align well across groups\footnote{See Tables \ref{tab:samp.add.socio}, \ref{tab:samp.elec}, and \ref{tab:samp.smdet}.}. There is slight over-representation of electricity-only households and under-representation of those on non-standard tariffs, such as \glspl{tvt}.

Most respondents passed at least two manipulation checks, with no significant group differences\footnote{Correct responses: Check 1 - 78.4\%, Check 2 - 84.1\%, Check 3 - 56.5\%. The lower performance on the final check likely reflects ambiguous wording rather than an understanding of the tasks. Full results in Table \ref{tab:samp.manip}.}. Structured feedback indicates 67.8\% understood all choice tasks, 60.5\% found them realistic, and 60.9\% found them easy to complete, comparable to \citet{Richter2018}\footnote{See Table \ref{tab:samp.feed}.}. Feedback did not differ significantly by group, suggesting the treatment information did not induce cognitive overload \citep{Waerdt2020}.

\section{Results and Discussion}\label{sec:results}

This section presents the findings from the \gls{dce} and \gls{wts} analyses, supported by open-ended responses\footnote{Quotes coded as: ID, group (TR or C), gender, age, \gls{seg}, smart meter (SM), \gls{dsa}, \gls{iwts}. Missing entries denoted as R.}. We first establish baseline preferences in the control group, and then assess the effect of the information treatment and explore socio-demographic heterogeneity.

\subsection{Baseline Demand for Anonymisation}
\begin{figure*}[t]
    \centering
    \subfloat[\gls{iwts}\label{fig:wts}]{\includegraphics[width=0.148\linewidth,valign=t]{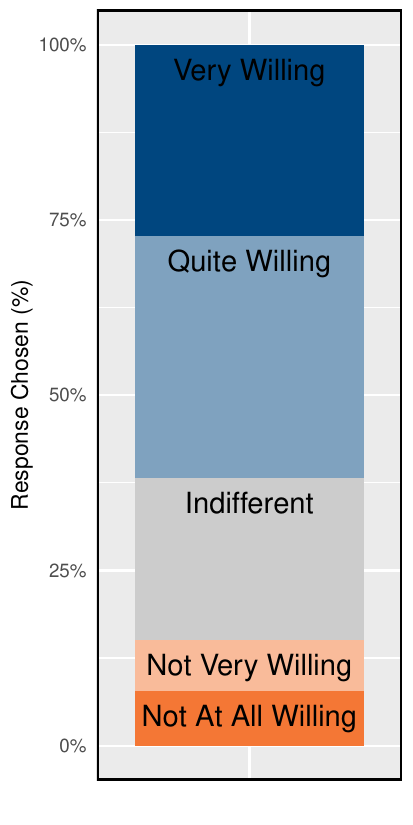}}
    \hfil
    \subfloat[Change in \gls{wts}\label{fig:wtsna}]{\includegraphics[width=0.3\linewidth,valign=t]{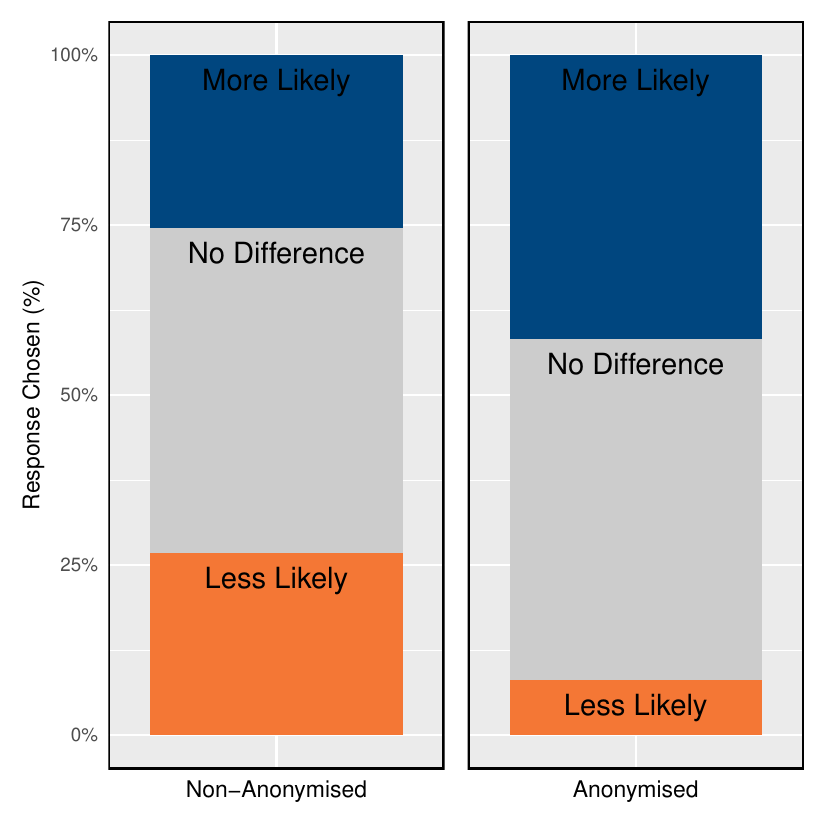}}
    \hfil
    \subfloat[Mean \gls{wtpa}\label{fig:wtpa}] {\includegraphics[width=0.4\linewidth,valign=t]{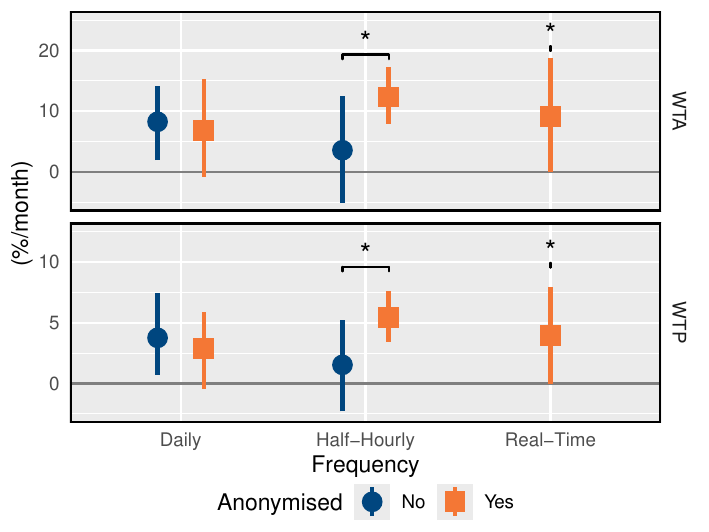}}
    \hfil
    \subfloat[\gls{smd}\label{fig:wsm}]{\includegraphics[width=0.148\linewidth,valign=t]{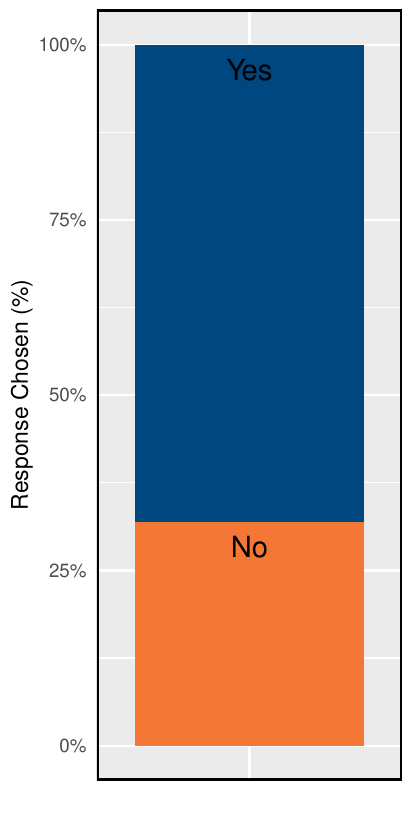}}
    \caption{Baseline Results for Control Group (n=477).  (a) Initial \gls{wts} half-hourly data. (b) Change in \gls{wts} half-hourly data after being given the option to anonymise data for both anonymised and non-anonymised data. (c) Mean \gls{wtpa} generated from \gls{mxl} with 95\% confidence intervals. Reference option: non-anonymised real-time data sharing. Significance labels for difference between anonymised and non-anonymised data at each frequency. See Table \ref{tab:mixl.split} for model parameters, Table \ref{tab:wtpa.conf} for \gls{wtpa} estimates, and Table \ref{tab:poe.freq} for $p$-values of tests. (d) \gls{smd} within control group.}
    \label{fig:demand}
\end{figure*}

Among the control group, 61.8\% were initially willing to share half-hourly smart meter data (Figure \ref{fig:wts})\footnote{Likert plots with hypothesis test can be found in Table \ref{fig:wtsg_het}.}. When anonymisation was introduced, 41.7\% reported increased \gls{wts}, aligning closely with prior work \citep{Knight2018a}\footnote{Average \gls{wts} of 61.7\% to share half-hourly data with suppliers for operational improvements and 40.1\% being more likely to share if data were anonymised. See Tables T249 \& T291 in \citep{Knight2018b}.}. Simultaneously, 26.8\% were less willing to share non-anonymised data, indicating the salience of privacy concerns. In total, over half the sample adjusted their willingness depending on anonymisation (52.2\% for non-anonymised; 49.9\% for anonymised data). As shown in Figure \ref{fig:wtsna}, these shifts are statistically significant, evidencing a clear preference for anonymisation and supporting \textbf{H4}\footnote{We see respondents shift from being more likely or indifferent to sharing anonymised data to being less likely to share non-anonymised data. See Table \ref{tab:wtsna.sig} for results of Wilcoxon Signed-Rank and McNemar-Bowker tests.}.

Figure \ref{fig:wtpa} shows the mean \gls{wtp} as a percentage of monthly bills (fee) to avoid sharing real-time non-anonymised data and the mean \gls{wta} (discount) required to share real-time non-anonymised data. For non-anonymised half-hourly data, the \gls{wtp} is just 1.53\% of respondents' monthly bill (£0.87 for the average bill of £57) and statistically insignificant. This lack of differentiation between half-hourly and real-time data is also observed \citet{Citizen2019}. In the absence of anonymisation there is a clear preference for daily sharing with a \gls{wtp} of 3.76\% (£2.14) and a \gls{wta} of 8.26\% (£4.71), supporting \textbf{H1}.  

Anonymisation significantly increased valuations. For real-time data, the mean \gls{wtp} rose to 3.98\% (£2.27), and the \gls{wta} to 9.16\% (£5.22). For half-hourly anonymised data, the \gls{wtp} was 5.42\% (£3.09) and the \gls{wta} 12.40\% (£7.04). Anonymisation thus elicited markedly higher valuations at both frequencies. For daily anonymised sharing, however, valuations were lower (\gls{wtp} - 2.91\% (£1.66), \gls{wta} - 6.87\% (£3.92)) and statistically indistinguishable from non-anonymised sharing, suggesting daily frequency may already offer sufficient privacy. This attenuation supports \textbf{H2}\footnote{See Table \ref{tab:poe.freq} for $p$-values from complete combinatorial tests.}.

Overall, \gls{wtp} estimates align with prior studies \citep{Richter2018,Skatova2019}. The ratio of \gls{wta} to \gls{wtp}, 3.21 [2.27, 4.47], indicates a strong endowment effect, supporting \textbf{H3}. This was echoed in qualitative responses, where respondents emphasised the intrinsic value of anonymisation and expressed discomfort with paying for privacy:

\blockquote[10071, C, M, 18-34, DE, Yes, MR, QW]{Anonymisation has a big value, bigger than a discount.} 

while a fee elicits moral outrage:

\blockquote[11045, C, M, 35-54, DE, No, MR, NVW]{[...]I wouldn't pay more for my electricity to stay anonymous. That should be free.} 

Interestingly, some also revealed limited understanding of privacy risks:

\blockquote[10001, C, F, 55-64, AB, Yes, BA, QW]{I don't really understand what a hacker will gain from learning about my electricity consumption}

As shown by the wide confidence intervals in Figure \ref{fig:wtpa}, substantial heterogeneity exists across the sample. Finally, Figure \ref{fig:wsm} shows that 68.1\% of the control group expressed interest in having a smart meter when anonymisation was available. While this reflects a positive baseline, a sizeable minority remain hesitant, even with enhanced privacy options.

\subsection{The Effect of Information Asymmetry}
We assess the effect of the information treatment across three post-treatment measures, examining both the full sample and subgroups defined by \glspl{dsa} and \gls{iwts}. Table \ref{tab:samp.dsa} confirms no significant differences in pre-treatment attitudes across experimental groups, validating the randomised design. The sample displays relatively relaxed \gls{dsa}, with most respondents comfortable sharing personal information with third parties, likely lower than population-level privacy concerns\footnote{For instance, \citet{Which2018}, which was a face-to-face survey, found 81\% of UK consumers were concerned about data being sold to third parties.}.

\begin{table}[htb]
\centering
\caption{Split of Data Sharing Attitudes} 
\label{tab:samp.dsa}
\setlength\targetwidth{\linewidth}
\begin{tabularx}{\targetwidth}{@{}rCCCC@{}}
  \toprule
   \multirow{2}{*}{Grouping}&    \multicolumn{2}{c}{Control}& \multicolumn{2}{c}{Treatment}  \\ 
  \cmidrule(lr){2-3} \cmidrule(lr){4-5}
  & n & \% & n & \% \\
  \midrule
  \multicolumn{5}{l}{\textbf{General Data Sharing Attitudes} ($p$ = 0.313)}\\
  \midrule
 Basic Information (BA) & 144 & 30.2 & 136 & 27.9 \\
 Marketing \& Research (MR) & 90 & 18.9 & 80 & 16.4 \\
   Third-Party Access (TP) & 243 & 50.9 & 272 & 55.7 \\ 
   \midrule
  \multicolumn{5}{l}{\textbf{Initial \gls{wts} Half-Hourly Data} ($p$ = 0.828)}\\
  \midrule
   Very Willing (VW) & 130 & 27.3 & 146 & 29.9 \\ 
   Quite Willing (QW) & 165 & 34.6 & 154 & 31.6 \\ 
   Indifferent (IND) & 110 & 23.1 & 118 & 24.2 \\ 
   Not very willing (NVW) & 35 & 7.3 & 34 & 7.0 \\ 
   Not at all willing (NAW) & 37 & 7.8 & 36 & 7.4 \\ 
   \bottomrule
   \multicolumn{5}{@{}P@{}}{Note: Control - n = 477, Treatment - n = 488. $p$-values for Pearson’s Chi-Squared Test for independence between experimental groups.}
\end{tabularx}
\end{table}

\subsubsection{Willingness-to-Share}
\begin{figure}
    \centering
    \includegraphics[width=\linewidth]{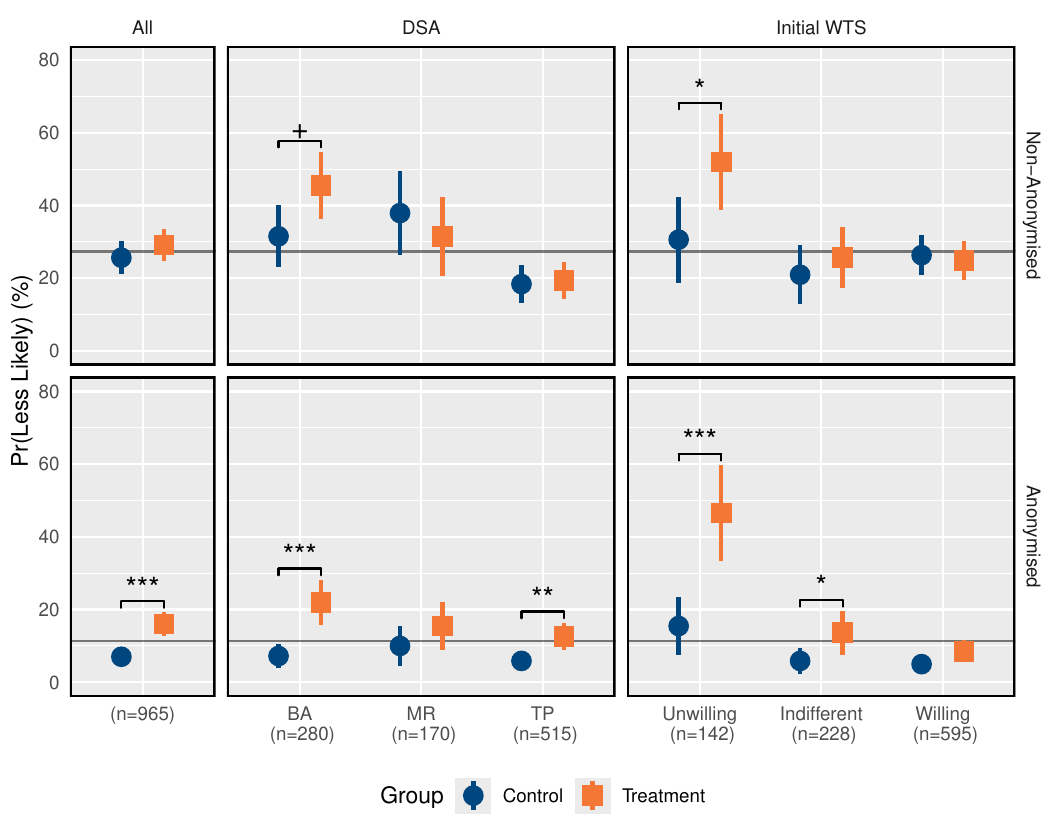}
    \caption{Effect of Treatment on Probability of being Less Likely to Share Half-Hourly Data. Estimated marginal probabilities of change in \gls{wts} based on \gls{mnl} with random effects (see Table \ref{tab:wtsna.ord}). Significance levels indicate results of z-tests between treatment and control group with: + $p<$0.1, * $p<$ 0.05, ** $p<$ 0.01, *** $p<$ 0.001. $p$-values adjusted with Holm correction for multiple comparisons across subgroups. Grey lines indicate marginal mean probability across sample under each framing.}
    \label{fig:wtsna.treat}
\end{figure}

For the change in \gls{wts} half-hourly data a significant treatment effect on the probability of being less likely to share half-hourly data is observed under the anonymised framing, as shown in Figure \ref{fig:wtsna.treat}\footnote{Likert plots with hypothesis test can be found in Table \ref{fig:wtsg_het}.}. This provides partial support for \textbf{H6}. 

The effect is more pronounced among subgroups with higher privacy concerns. Specifically, those who generally only share basic information (BA) and those initially unwilling to share show the largest increases in reluctance to share data when presented with the anonymised scenario (14.8\% [8.5\%, 21.1\%]\footnote{Difference in estimated marginal mean probability and 95\% confidence interval between treatment and control group.} and 21.4\% [5.3\%, 37.5\%], respectively). In contrast, for those already comfortable sharing data with third parties (TP) and those initially indifferent, the effect is weaker (6.7\% [2.6\%, 10.9\%] and 7.8\% [1.8\%, 13.7\%], respectively) but still significant. Under the non-anonymised framing, effects are only significant for the more privacy-concerned subgroups. These patterns support \textbf{H7}, suggesting the treatment’s impact is moderated by pre-existing attitudes.

Contrary to expectations (\textbf{H8}), the treatment effect is not stronger for the non-anonymised framing. This may be due to ordering effects; respondents were first introduced to anonymisation before evaluating the non-anonymised scenario. As such, their judgement may have been influenced by a sense of moral unfairness \citep{Winegar2019}, even in the control group, muting the effect of the treatment. Overall, the observed treatment effects on \gls{wts} support the presence of information asymmetries in smart meter data sharing.

\subsubsection{Willingness-to-Pay/Accept}
We now examine the mean \gls{wtpa} estimates for the different data sharing options. When considering the full sample, only marginal treatment effects are observed in mean \gls{wtpa} estimates, specifically for non-anonymised daily sharing and anonymised half-hourly sharing ($p<0.1$), offering limited support for \textbf{H6}. However, the subgroup analysis reveals clear and divergent treatment effects based on general \gls{dsa}, namely: BM - those who only provide basic information or allow their data to be used for marketing and research, and TP - those who generally share data with third parties. 

\begin{figure}
    \centering
    \includegraphics[width=\linewidth]{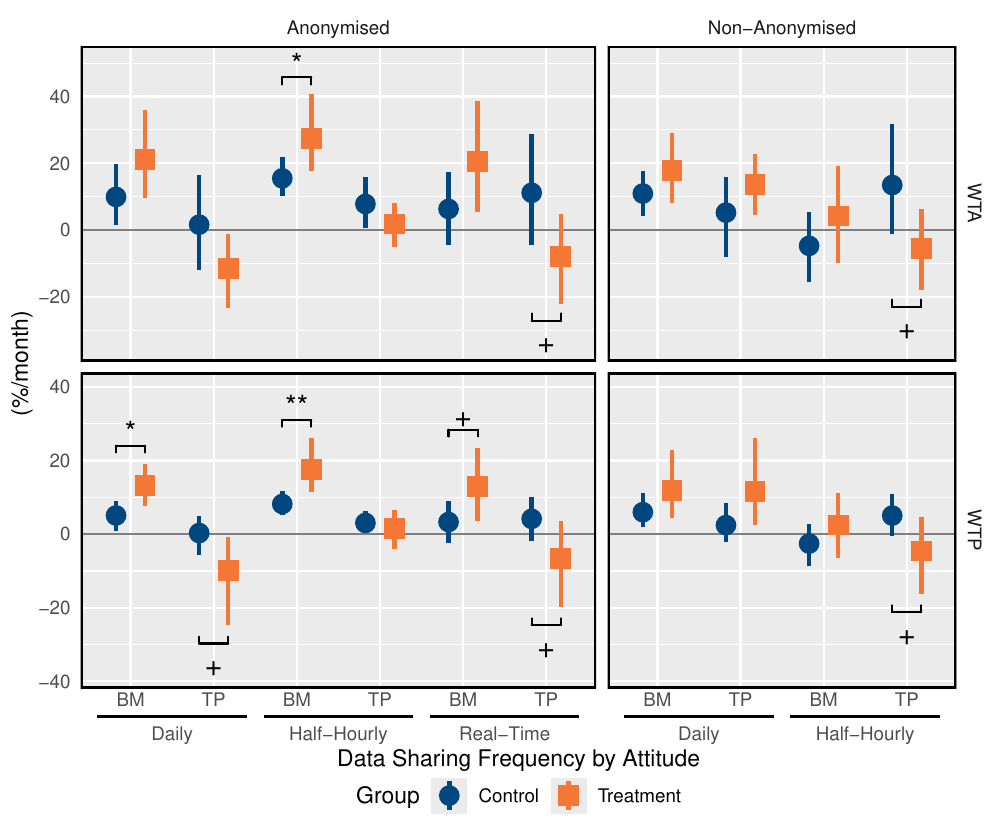}
    \caption{Effect of Treatment on mean \gls{wtpa} for anonymisation and data sharing frequency by \gls{dsa} (BM - Basic sharing or for marketing \& research, TP - sharing with third parties). Reference option: non-anonymised real-time data. See Table \ref{tab:wtpa.conf} for \gls{wtpa} estimates and Table \ref{tab:mixl.split} for underlying \glspl{mxl}. Significance levels indicate results of one-sided complete combinatorial test between treatment and control group with: + $p<$0.1, * $p<$ 0.05, ** $p<$ 0.01, *** $p<$ 0.001. $p$-values adjusted with Holm correction for multiple comparisons across subgroups. See Table \ref{tab:poe.freq} for details.}
    \label{fig:wtpas.treat}
\end{figure}

Among BM respondents, the treatment significantly increased valuations of anonymisation. For instance, the \gls{wtp} for anonymised half-hourly sharing more than doubled from 8.19\% (£4.67) in the control group to 17.60\% (£10.00) in the treatment group (Figure \ref{fig:wtpas.treat})\footnote{Discount valuations (\gls{wta}) were more uncertain due to higher variability in the estimated parameter.}. Open-ended responses reinforce this heightened sensitivity:

\blockquote[10244, TR, F, 55-64, C1, Yes, MR, QW]{I found the minute by minute example frightening to know that so much can be known about what goes on in your home [...].}

\blockquote[10492, TR, F, 65+, C1, Yes, MR, QW]{[...]I had thought I would be influenced entirely by cost, but when asked to choose, I found I didn't like the idea of that quantity of information being readily available when it could be identified directly to me/us as a household.}

In contrast, the TP subgroup showed either no change or negative treatment effects. For anonymised real-time sharing, \gls{wtp} fell from 4.17\% (£2.38) in the control group to -6.50\% (-£3.71) in the treatment group, suggesting a preference for non-anonymised sharing. This may indicate perceived irrelevance of anonymisation\footnote{This potential attribute non-attendance may not be captured due to our \gls{mxl} specification with normally distributed parameters \citep{Rigby2006ModelingFood}.} or a belief it impedes system operations.

These findings support \textbf{H7}, indicating stronger effects among more privacy-concerned respondents. The negative treatment effect within the TP group, however, was unexpected. As highlighted in the open response, some respondents may have been reassured by the additional information, similar to findings in \citet{Loessl2023}:

\blockquote[10230, TR, F, 65+, C1, No, TP, QW]{I would prefer to save money and maybe being anonymised is not so necessary as I first thought.} 

Unlike \citet{Loessl2023}, where treatment effects were consistent across groups, our study shows increased concern primarily among those already cautious. These results highlight not only the importance of accounting for heterogeneity in effects, but also the wording of the information treatment itself. While \citet{Loessl2023} relies on respondents' own perceptions of the sensitivity and potential privacy risks of sharing smart meter data, our study outlines these in a concrete manner based on the current state-of-the-art literature \citep{Teng2022}. As a result, our respondents may have made more informed decisions, better reflecting true preferences and highlighting persistent information asymmetries. Respondents themselves questioned whether such information is adequately disclosed under current consent regimes:

\blockquote[10777, TR, F, 65+, DE, No, MR, VW]{[...] I do not believe these things are shared with people when they opt for a smart meter.}

\subsubsection{Access to High-Resolution Data}
A central aim of the \gls{mhhs} programme is to increase the use of half-hourly smart meter data for settlement. To support this, \gls{ofgem} has proposed a shift to an opt-out model and forgone the use of \glspl{ppt} \citep{OFGEM2019}. To assess how policy framing and anonymisation affect the availability of high-resolution data, we simulate expected market shares using utility distributions from the split-sample \gls{mxl} analysis\footnote{We apply parametric bootstrapping with 1,000 coefficient draws used to simulate choices over 10,000 random draws. Market shares represent average choices weighted across BM and TP subgroups.}. We compare two policy scenarios: (1) opt-in, where daily sharing is the default and consumers receive a discount for higher-resolution sharing; and (2) opt-out, where real-time sharing is default and a fee is imposed to opt for lower-resolution sharing\footnote{We assume that half-hourly data carries half the monetary incentive of real-time or daily data, depending on the scenario.}.

\begin{figure}
    \centering
    \includegraphics[width=\linewidth]{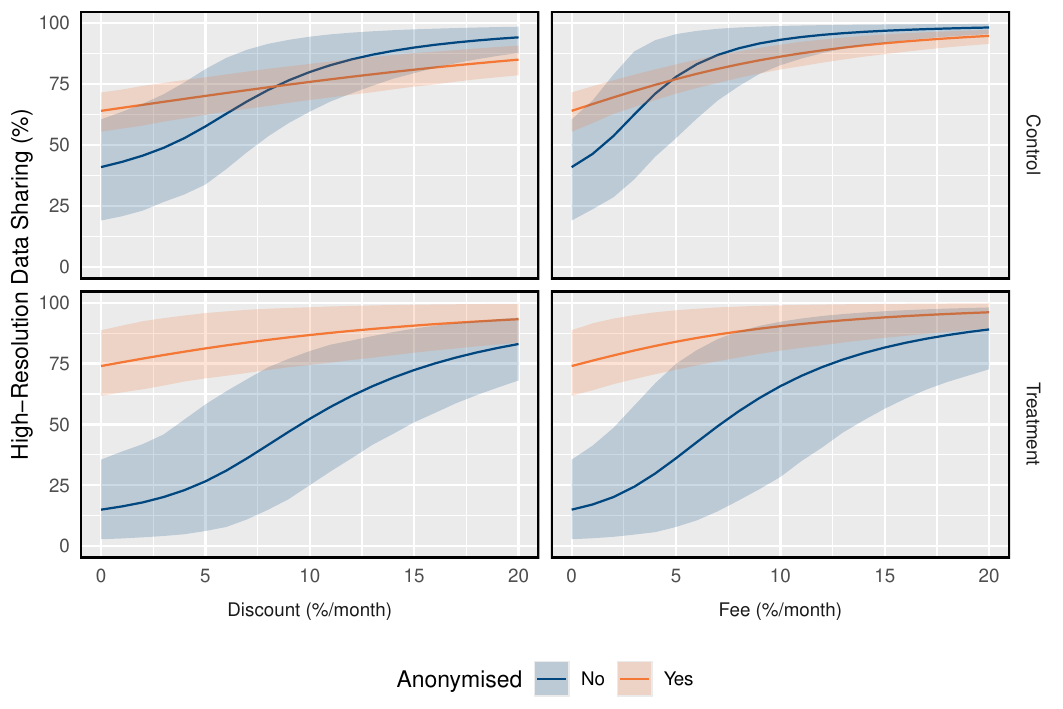}
    \caption{Expected Proportion of Consumers Sharing High-Resolution Data under Different Framing Options. High-Resolution data includes real-time and half-hourly data. Shaded regions represent 95\% confidence intervals. Generated from split-sample \glspl{mxl} in Table \ref{tab:mixl.split}.}
    \label{fig:market_shares}
\end{figure}

Figure \ref{fig:market_shares} illustrates how the proportion of consumers sharing high-resolution data (real-time and half-hourly) changes with increasing incentives. In the control group, without incentives or anonymisation, 41.0\% [19.1\%, 60.6\%] would opt to share high-resolution data. This increases to 64.1\% [55.5\%, 71.5\%] when anonymisation is introduced, underscoring its role in facilitating data sharing. Policy framing also matters: in the opt-out scenario, a 9\% fee suffices to achieve 90\% high-resolution sharing, whereas the opt-in model requires a 15\% discount to reach the same level. As monetary incentives increase, the relative importance of anonymisation diminishes, suggesting substitution between financial and privacy incentives.

The treatment group shows a markedly different pattern. In the absence of anonymisation, only 15.0\% [2.8\%, 35.7\%] are willing to share high-resolution data, far below the control group. Anonymisation increases this to 74.0\% [61.8\%, 88.6\%], in line with the control group. Without anonymisation, even a 20\% fee or discount fails to produce a 90\% market share. With anonymisation, however, similar incentive thresholds as in the control group emerge (10\% fee or 15\% discount). This reinforces the value of a Privacy by Design approach under an opt-in regime, particularly when informed consent is prioritised.

We note that these simulations do not account for status quo bias \citep{Kahneman1991}, which could increase default-option uptake, especially relevant in retail energy markets marked by consumer inertia \citep{Sovacool2017}. In our sample, 32.3\% of smart meter owners were unaware of their current data sharing settings\footnote{See Table \ref{tab:samp.smdet}. Similar levels of consumer unawareness were also observed in \citet{Citizen2019}.}. Additionally, 24.4\% did not own, or were unsure whether they owned, an \gls{ihd}\footnote{Given the central role of \glspl{ihd} in achieving behavioural energy savings, this gap has implications for the net benefits of the smart meter rollout \citep{BEIS2019b}. See Table \ref{tab:samp.smdet}.}.

\subsubsection{Adoption of Smart Metering}
\begin{figure}
    \centering
    \includegraphics[width=0.95\linewidth,valign=t]{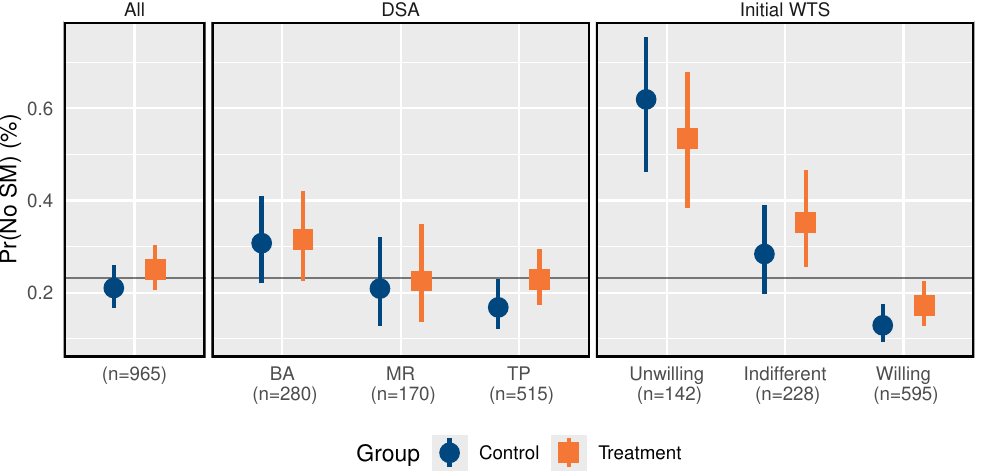}
    \caption{Effect of Treatment on \gls{smd}. Estimated marginal probabilities of change in demand based on binary logistic regression (see Table \ref{tab:wsm.lr}). Significance levels indicate results of z-tests between treatment and control group with: + p $<$ 0.1, * p $<$ 0.05, ** p $<$ 0.01, *** p $<$ 0.001. $p$-values adjusted with Holm correction for multiple comparisons across subgroups. Grey line indicates marginal mean probability across sample.}
    \label{fig:wsm.treat}
\end{figure}

Given the voluntary nature of the smart meter roll-out in \gls{gb}, we assess whether the information treatment influences \gls{smd}. As shown in Figure \ref{fig:wsm.treat}, no significant difference is observed between the treatment and control groups. The estimated probability of not wanting a smart meter is 21.0\% [16.8\%, 26.0\%] in the control group and 25.1\% [20.5\%, 30.3\%] in the treatment group, contradicting \textbf{H6}. Likewise, subgroup analyses show no significant effects, rejecting \textbf{H7}. However, \gls{smd} is correlated with privacy attitudes. Respondents who only share basic information or were initially unwilling to share half-hourly data were also less likely to want a smart meter, consistent with previous studies highlighting the role of privacy in smart meter acceptance \citep{GERPOTT2013483, Gosnell2023}.

The absence of a treatment effect suggests that factors beyond privacy, such as trust, may be shaping smart meter adoption. Among those not wanting a smart meter, only 28.7\% said they would be more willing to share data if it were anonymised, compared to 47.2\% among those who did want a smart meter, indicating deeper scepticism. Trust in energy suppliers has been identified as a key determinant of \gls{wts} in previous work \citep{Grunewald2020c, Maidment2020}, and is reflected in our open-ended responses:

\blockquote[10498, C, M, 55-64, C1, No, BA, IND]{[...]if it saves money and I can trust the company it's ok if we benefit.}

\blockquote[10231, TR, M, 55-64, C1, No, TP, VW]{[...]Most of utility providers treat the public as semi-literate \& think they can be bought for a song.}

\blockquote[10479, TR, M, 65+, AB, No, BA, VW]{[...] I have long suspected that these companies [...] have been buying and selling our information [...].}

These sentiments point to persistent concerns about data governance and corporate intentions, suggesting that improving transparency and public trust may be as critical as addressing privacy through design.

\subsection{Drivers of Heterogeneity}
The previous subgroup analyses revealed significant heterogeneity based on respondents’ \glspl{dsa}. To further explore these patterns, we examine variation across socio-demographic and electricity supply characteristics for the four key measures. Figure \ref{fig:het} presents marginal mean values across groups, while statistical test results are summarised in Tables \ref{tab:robust}, \ref{tab:wtpa.het.tests} and \ref{tab:wts.het.tests}.

\subsubsection{Socio-Demographics}
\begin{figure*}[t]
\centering
\sidesubfloat[]{\includegraphics[width=0.9\textwidth,keepaspectratio]{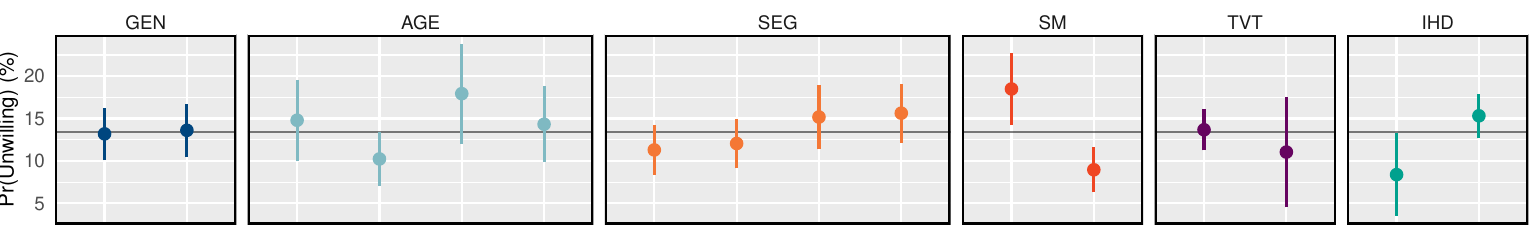}\label{fig:het.wtsg}}
\hfil
\sidesubfloat[]{\includegraphics[width=0.9\textwidth,keepaspectratio]{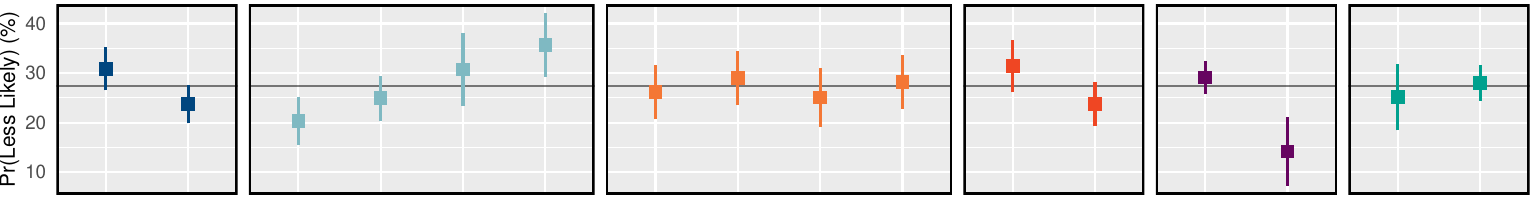}\label{fig:het.cwts}}
\hfil
\sidesubfloat[]{\includegraphics[width=0.9\textwidth,keepaspectratio]{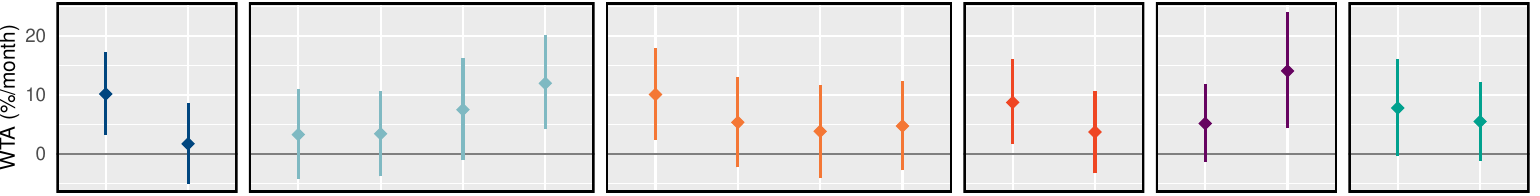}\label{fig:het.wta}}
\hfil
\sidesubfloat[]{\includegraphics[width=0.9\textwidth,keepaspectratio]{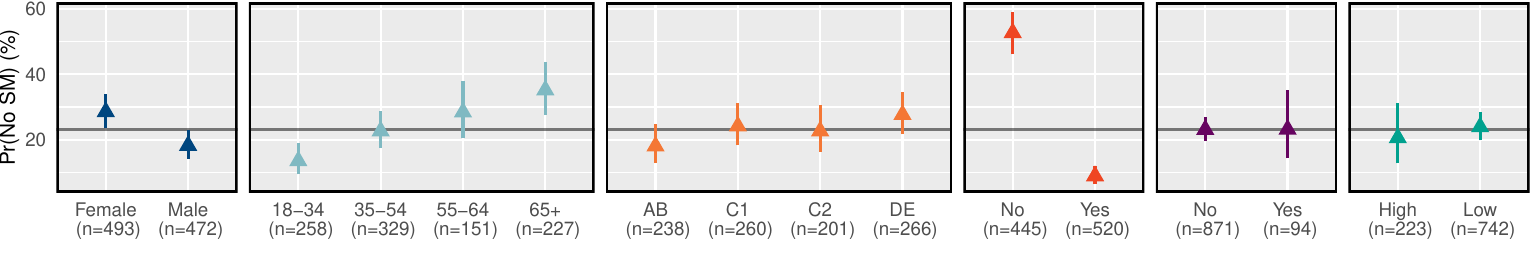}\label{fig:het.wsm}}
\caption{Heterogeneity in Measures. (a) \gls{iwts} for half-hourly data. Marginal mean probabilities of being unwilling based on partial proportional odds model in Table \ref{tab:wtsg.lr}. (b) Change in \gls{wts} for non-anonymised data. Marginal mean probabilities for being less likely based on \gls{mnl} with random effects in Table \ref{tab:wtsna.ord}. (c) Mean \gls{wta} for anonymised real-time data. Marginal means based on \gls{mxl} with interactions in Table \ref{tab:mixl.het}. (d) \gls{smd}. Marginal mean probabilities based on the binary logit model in Table \ref{tab:wsm.lr}. For (a), (b), and (d) grey lines indicate marginal mean probability across sample. Results of pairwise comparisons can be found in Table \ref{tab:wts.het.tests} and \ref{tab:wtpa.het.tests}. Raw response distributions for \gls{iwts} and change in \gls{wts} can be found in Figure \ref{fig:wtsg.sig}.}
\label{fig:het}
\end{figure*}

\paragraph{Gender} While no significant gender difference is observed in being initially unwilling to share half-hourly data (Figure \ref{fig:het.wtsg}), women are significantly more likely to become less willing to share post-information (Figure \ref{fig:het.cwts}), have a higher \gls{wta} for anonymised real-time data (Figure \ref{fig:het.wta}), and are more likely to reject smart meters (Figure \ref{fig:het.wsm}). These results support \textbf{H5} and align with existing literature indicating greater privacy concerns among women \citep{Knight2018a,Richter2018}.

\paragraph{Age} A similar pattern emerges across age groups. While \gls{iwts} does not vary significantly, older respondents exhibit greater reluctance post-information, marginally higher \gls{wtpa}, and lower \gls{smd}, again supporting \textbf{H5}. Prior studies attribute this to reduced trust and lower digital literacy among older adults \citep{Knight2018a,Citizen2019}.

\paragraph{\gls{seg}} Lower \glspl{seg} (C2 and DE) are marginally more likely to be initially unwilling to share data and less likely to want a smart meter, consistent with \textbf{H5}. However, post-information change in \gls{wts} differences are not significant, and conversely, \gls{wta} is slightly higher among higher \glspl{seg} (though not significant). This apparent contradiction has also been observed in \citep{Richter2018}, with the authors concluding that this group were less concerned about data privacy\footnote{\textit{Open Data} cluster, which had lower average privacy valuations had a higher proportion of \gls{seg} DE (37\%).}. This is sometimes attributed to a lack of knowledge of the information embedded within smart meter data \citep{Citizen2019}. However, the open responses put forward a different explanation. As one respondent put it:

\blockquote[10580, TR, F, 65+, DE, Yes, BA, IND]{My options were based on costs, I can't afford to pay more.}

This, together with the non-monetary measures suggests the low \gls{wtpa} likely reflects affordability rather than a lack of privacy concerns.

\subsubsection{Electricity Supply Characteristics}

\paragraph{Smart Meter Ownership} All measures vary significantly with smart meter ownership. Respondents without a smart meter are more likely to be initially unwilling, show greater reluctance post-information, have higher \gls{wtpa}, and are less likely to want a smart meter. These findings robustly support \textbf{H5}, confirm previous research linking smart meter resistance with privacy concerns \citep{GERPOTT2013483,Gosnell2023}, and are echoed by respondents:

\blockquote[11012, TR, M, 35-54, C2, No, BA, NAW]{This is extremely sinister[...]. Under no circumstances should anyone ever have a so-called smart meter.}

\paragraph{Time-Varying Tariffs}  No significant differences are observed in \gls{iwts} or \gls{smd} across tariff types. However, respondents on \glspl{tvt} are less likely to reduce their \gls{wts} post-information and have significantly higher \gls{wtpa} for daily data sharing. This may reflect heightened awareness of the connection between consumption patterns and personal data, as well as greater digital engagement and trust, traits associated with early adopters of such tariffs \citep{Richter2018,Loessl2023}.

\paragraph{IHD Engagement} Respondents who use their \gls{ihd} more than once a week are less likely to be initially unwilling to share data and have a marginally higher \gls{wtpa} for anonymisation. This suggests greater engagement may be associated with both digital literacy and potentially a more informed valuation of privacy.

Overall, the heterogeneity analysis supports many socio-demographic patterns identified in prior studies while offering new insights into the role of electricity supply characteristics. Importantly, each of the four privacy-related measures varies across groupings in distinct ways, reinforcing the importance of context and framing. This suggests that a single metric is insufficient to capture the full spectrum of privacy concerns or demand for anonymisation. A holistic approach is therefore needed to understand consumer preferences and design equitable policy interventions\footnote{Including socio-demographic interactions in the \gls{mxl} reduced, but did not eliminate, preference heterogeneity. See Table \ref{tab:mixl.het} for full estimates.}.

\subsection{Policy Implications}
The results of this study has several implications for smart meter data governance and the design of \glspl{ppt}.

\paragraph{Fostering Informed Consent} The evidence of information asymmetries suggests that existing consent mechanisms fail to ensure informed decision-making, leaving consumers exposed. This reinforces proposals by Citizens Advice \citep{CitizensAdvice2018a} and the Energy Digitalisation Taskforce \citep{EnergyDigitalisationTaskforce2022} for a user-facing data dashboard. Such tools could provide clarity on data sharing options, increase transparency, and promote informed consent by clearly explaining what personal information is shared through smart meter data.

\paragraph{Privacy by Design} While presenting full privacy implications risks overwhelming users \citep{Waerdt2020}, the evolving nature of data re-identification threats \citep{Teng2022} underscores the limitations of consent-based regimes like \gls{gdpr} and the \gls{dapf}. A proactive, Privacy by Design approach, emphasising strong privacy defaults and user-centric design, is more appropriate \citep{Kingsmill2015}. For \gls{gb}, this implies a shift to an opt-in model with daily data sharing as the default, combined with \glspl{ppt}. Such a model would align with public expectations, reduce moral resistance to privacy costs, and mitigate distributional inequities, particularly for lower \gls{seg} consumers constrained by affordability rather than preference. Our findings also suggest that anonymisation becomes more valuable in this context, especially given the observed endowment effect and information asymmetries.

\paragraph{Leveraging Heterogeneity} The significant variation in privacy preferences indicates that uniform privacy solutions are likely inefficient. Flexible, tuneable methods, such as differential privacy, could better accommodate heterogeneity than static approaches like aggregation or pseudonymisation, which have been the focus of the \gls{mhhs} \citep{Teng2022}. Furthermore, the lack of strong preferences between real-time and half-hourly data suggests that including higher-resolution sharing options in the \gls{dapf} could be beneficial.

\section{Conclusions}\label{sec:conc}
This study examined consumer demand for anonymisation of smart meter data in the \gls{gb} context, with a particular focus on the effects of information asymmetries and question framing. Using a mixed-methods approach, we combined estimates of monetary (\gls{wtpa}) and non-monetary (\gls{wts}, \gls{smd}) preferences with qualitative analysis of open responses.

To the best of our knowledge, this is the first study to estimate consumers’ \gls{wtpa} for anonymisation specifically, rather than to avoid data sharing altogether \citep{Richter2018,Skatova2019,Loessl2023}. We show that, on average, consumers are willing to pay or require compensation for anonymisation, and many are more willing to share if anonymisation is available. At the same time, a substantial share is less willing to share non-anonymised data when the anonymised option is presented. Yet, anonymisation alone was insufficient to convince some respondents to accept smart meters.

We also identify a significant endowment effect, with \gls{wta} values exceeding \gls{wtp}, and confirm the presence of information asymmetries: consumers informed about privacy risks showed significantly higher \gls{wtpa} and reduced \gls{wts}, particularly those less comfortable with third-party data sharing. In addition, demand for anonymisation and privacy concerns vary by age, gender, \gls{seg}, smart meter ownership, and tariff type, though the influence of these factors differed between monetary and non-monetary framings. Qualitative responses offered additional depth, revealing concerns over data misuse, distrust in suppliers, and discomfort with the idea of paying to protect one’s privacy. These narratives also highlight the need for clearer communication and stronger trust-building mechanisms.

Our findings underscore the critical role of informed consent in smart meter data governance. We propose several policy interventions that would more accurately reflect consumer preferences, including an opt-in approach, greater transparency, and privacy-by-design principles.

Future research could address the limitations of this study. Our online sample displayed relatively low privacy concerns compared to \gls{gb} benchmarks \citep[e.g.][]{Which2018}. Qualitative methods such as interviews or focus groups could offer a more representative understanding of privacy attitudes and allow richer exploration of contextual framing. Finally, the role of trust, which emerged consistently in qualitative feedback, warrants further investigation, particularly to engage the substantial minority resistant to smart meters regardless of privacy safeguards.

\section{CRediT authorship contribution statement}
\textbf{Saurab Chhachhi}: Conceptualisation, Data curation, Methodology, Formal analysis, Software, Investigation, Project Administration, Visualisation, Writing - original draft, Writing - review \& editing, Funding acquisition. \textbf{Fei Teng}: Conceptualisation, Supervision, Funding Acquisition, Writing - review and editing.

\section{Declaration of Competing Interests}
The authors declare that they have no known competing financial interests or personal relationships that could have appeared to influence the work reported in this paper.

\section{Data Availability}
All code required to reproduce the analysis can be found on \href{https://github.com/saurabac/Smart-Meter-Data-Privacy-Survey}{Github}. Raw data will be made available on request for research purposes, in line with participant consent forms.

\section{Acknowledgements}
We thank Debbie Shuttlewood and the Accent team for programming and conducting the survey. We would also like to thank Dr. Paul J. Metcalfe of PJM Economics for reviewing the experimental design of the discrete choice experiment and statistical analysis plan. In addition, we thank Pudong Ge for contributing to the literature review on information embedded within smart meter data for the experimental design. Finally, we are grateful to Prof. Pierre Pinson, Dr. Philip Grünewald, Dr. Chien-Fei Chen, Dr. Aruna Sivakumar, Prof. Amrita Chhachhi, Dr. Nicole Watson and Dr. Anna Gorbatcheva for their comments and suggestions. This research was partly funded by Research England (Strategic Priorities Fund 2021-22) and the Economic and Social Research Council (grant ES/P000703/1:2113082). 

\bibliography{references}
\bibliographystyle{elsarticle-harv}

\onecolumn
\newpage
\clearpage
\pagenumbering{arabic}
\setcounter{page}{1}
\appendix
\makeatletter
\def\@seccntformat#1{\@ifundefined{#1@cntformat}%
   {\csname the#1\endcsname\space}
   {\csname #1@cntformat\endcsname}}
\newcommand\section@cntformat{\appendixname\thesection.\space} 
\makeatother
\renewcommand{\thesection}{\Alph{section}}
\counterwithin{equation}{section}
\counterwithin{figure}{section}
\counterwithin{table}{section}

\section{Survey Questions}\label{app:survey}
{\tiny\tabcolsep=3pt  
 \begin{longtable}{cll}
 \captionsetup{font=scriptsize}
    \caption{Survey Questions}
       \label{tab:survey_qs} \tabularnewline
    \toprule
     & Question & Options \\
    \midrule
    \endfirsthead
    \multicolumn{3}{@{}l}{\ldots continued}\\
    \toprule
     & Question & Options \\
    \midrule
    \endhead
    \bottomrule
    \multicolumn{3}{@{}r}{continued\ldots}\\
    \endfoot
    \bottomrule
    \endlastfoot
    \multirow{2}{*}{1} & \multirow{2}{*}{\makecell[l]{Are you responsible, either fully or jointly,\\ for paying the electricity bills in your household?}} & Yes \\
          &      & No \\
    \cmidrule(lr){2-3}
    \multirow{3}{*}{2} & \multirow{3}{*}{\makecell[l]{Please select your gender:}} & Male \\
          &    & Female  \\
          &     & Other / refused \\
          \cmidrule(lr){2-3}
    \multirow{5}{*}{3} & \multirow{5}{*}{\makecell[l]{Please select your age bracket:}} & 18-34  \\
          &     & 35-54 \\
          &     & 55-64  \\
          &     & 75+ \\
          &      & Refused \\
          \cmidrule(lr){2-3}
    \multirow{6}{*}{4} & \multirow{6}{*}{\makecell[l]{Choose one option that best describes\\ your ethnic group or background:}}  & White  \\
          &      & Asian/Asian British \\
          &      & Black African/Caribbean/Black British \\
          &      & Mixed/multiple ethnic groups  \\
          &      & Other ethnic group  \\
          &      & Refused \\
          \cmidrule(lr){2-3}
    \multirow{10}{*}{5} & \multirow{10}{*}{\makecell[l]{How would you describe the occupation of\\ the chief income earner in your household?}} & Senior managerial or professional \\
          &      & Intermediate managerial, administrative or professional \\
          &      & Supervisor; clerical; junior managerial, administrative or professional \\
          &      & Manual worker (with industry qualifications) \\
          &      & Manual worker (with no qualifications) \\
          &      & Unemployed due to ill health \\
          &      & Unemployed for another reason \\
          &      & Retired \\
          &      & Student \\
          &      & Prefer not to say \\
          \cmidrule(lr){2-3}
    \multirow{3}{*}{6} & \multirow{3}{*}{\makecell[l]{Does the chief income earner have a state pension,\\ a private pension or both?}} & State only \\
          &      & Private only \\
          &      & Both \\
          \cmidrule(lr){2-3}
    \multirow{6}{*}{7} & \multirow{6}{*}{\makecell[l]{How would you describe the chief income earner`s\\ occupation before retirement?}} & Senior managerial or professional \\
          &      & Intermediate managerial, administrative or professional \\
          &      & Supervisor; clerical; junior managerial, administrative or professional \\
          &      & Manual worker (with industry qualifications) \\
          &      & Manual worker (with no qualifications) \\
          &      & None of these \\
          \cmidrule(lr){2-3}
    \multirow{4}{*}{8} & \multirow{4}{*}{\makecell[l]{Please indicate the first part of your postcode\\ (e.g. for SW7 2AZ enter SW7):}} & Postcode not found on the lookup or not given \\
          &      & England  \\
          &      & Wales  \\
          &      & Scotland  \\
          \cmidrule(lr){2-3}
    \multirow{3}{*}{9} & \multirow{3}{*}{\makecell[l]{Do you have a smart meter?}} &  Yes \\
          &      & No \\
          &      & Don't know \\
          \cmidrule(lr){2-3}
    \multirow{3}{*}{10} & \multirow{3}{*}{\makecell[l]{If you know the amount (excluding gas) please enter either\\ your monthly or yearly electricity bill below.}} & \_\_ monthly \\
        & & \_\_ annually\\ 
        && Don't Know \\
        \cmidrule(lr){2-3}
    \multirow{5}{*}{11} & \multirow{5}{*}{\makecell[l]{Typically, when you share your data in order to\\ use a service or product, for example when you
shop online, \\ use an energy company or book a holiday online,\\ which level of data sharing do you sign up to?\\ Please tick all that apply}} & \makecell[l]{a) Sharing the basic information to enable the company to provide me with \\the service they offer.}\\
     & & \makecell[l]{b) Allowing the company to use my information for marketing, research,\\ forecasting etc.}\\
     & & \makecell[l]{c) I allow my data and information to be passed to third parties.} \\
     \cmidrule(lr){2-3}
    \multirow{5}{*}{12}& \multirow{5}{*}{\makecell[l]{How willing would you be to share your half-hourly electricity \\consumption data with your energy supplier?}} & Very Willing \\
          &      & Quite Willing \\
          &      & Neither willing nor unwilling \\
          &      & Not very willing \\
          &      & Not at all willing \\
          \cmidrule(lr){2-3}
    \multirow{4}{*}{13}&\multirow{4}{*}{\makecell[l]{Considering the information you have just read,\\ would you be more or less likely to\\ share your half-hourly electricity consumption\\ data if it was not anonymised before being shared?}} & More likely \\
          &      & It makes no difference \\
          &      & Less likely \\ \\
          \cmidrule(lr){2-3}
    \multirow{4}{*}{14}&\multirow{4}{*}{\makecell[l]{Considering the information you have just read,\\ would you be more or less likely to\\ share your half-hourly electricity consumption\\ data if it was anonymised before being shared?}} & More likely \\
          &      & It makes no difference \\
          &      & Less likely \\ \\
          \cmidrule(lr){2-3}
    \multirow{4}{*}{15}&\multirow{2}{*}{\makecell[l]{Thinking about the choices you have just made,\\ would you opt to have a smart meter with one\\ of the data sharing options you selected or \\choose not to have a smart meter?}} & Have a smart meter with one of the selected data sharing options \\
          &      & Not have a smart meter \\ \\ \\
          \cmidrule(lr){2-3}
    \multirow{2}{*}{16}&\multirow{2}{*}{\makecell[l]{A negative expected change in your monthly bill\\ (shown in green) indicates a reduction in your electricity bill}} & True \\
          &      & False \\
          \cmidrule(lr){2-3}
    \multirow{3}{*}{17}&\multirow{2}{*}{\makecell[l]{A half-hourly frequency means that your\\ total electricity consumption for each \\half hour is sent to your supplier every 30 minutes}} & True \\
          &      & False \\ \\
          \cmidrule(lr){2-3}
    \multirow{3}{*}{18}&\multirow{2}{*}{\makecell[l]{Anonymisation of your consumption\\ data ensures your information can be linked\\ back to you in the event of data breach}} & True \\
          &      & False \\ \\
          \cmidrule(lr){2-3}
    \multirow{6}{*}{19}&\multirow{6}{*}{\makecell[l]{I was able to understand the choices}} & Strongly disagree  \\
          &      & Disagree  \\
          &      & Neither agree nor disagree  \\
          &      & Somewhat agree  \\
          &      & Strongly agree  \\
          &      & Don't know \\
          \cmidrule(lr){2-3}
    \multirow{6}{*}{20}&\multirow{6}{*}{\makecell[l]{I found the options realistic}} & Strongly disagree  \\
          &      & Disagree \\
          &      & Neither agree nor disagree \\
          &      & Somewhat agree \\
          &      & Strongly agree \\
          &      & Don't know \\
          \cmidrule(lr){2-3}
    \multirow{6}{*}{21}&\multirow{6}{*}{\makecell[l]{I found it easy to choose between the options}} & Strongly disagree  \\
          &      & Disagree  \\
          &      & Neither agree nor disagree  \\
          &      & Somewhat agree  \\
          &      & Strongly agree  \\
          &      & Don't know \\
          \cmidrule(lr){2-3}
          
    \multirow{5}{*}{22} & \multirow{2}{*}{\makecell[l]{Do you have any comments or feedback about the\\ choice task? Were any of the data sharing \\options (frequency, anonymisation) difficult to understand?\\ Were the differences between the options \\in each scenario clear?}} & Open-ended\\ \\ \\ \\ \\
    \cmidrule(lr){2-3}
    
    \multirow{4}{*}{23}&\multirow{3}{*}{\makecell[l]{We would now like to understand more \\about your current electricity supply. Does your home have\\ a dual fuel (both electricity and gas) connection \\or electricity only connection?}} & Dual Fuel (if you have a gas boiler for heating and/or hot water and/or a gas stove) \\
          &      & Electricity Only \\
          &      & Don't know \\ \\
          \cmidrule(lr){2-3}
    \multirow{6}{*}{24}&\multirow{6}{*}{\makecell[l]{What type of electricity tariff are you currently on?}} & Standard Variable Tariff (default tariff if you made no active tariff selection) \\
          &      & Fixed Rate (fixed rate guaranteed for specified time e.g. 12 months) \\
          &      & Pre-payment (if you have a pre-payment meter which you have to top up) \\
          &      & Time-of-Use Tariff (rate varies depending on the time of day e.g. Octopus Agile) \\
          &      & Economy 7 or 10 \\
          &      & Don't know \\
          \cmidrule(lr){2-3}
    \multirow{4}{*}{25}&\multirow{4}{*}{\makecell[l]{Thinking about when you had your smart meter\\ installed, which frequency did you select?}} & Half-hourly \\
          &      & Daily \\
          &      & Monthly \\
          &      & Don't know \\
          \cmidrule(lr){2-3}
    \multirow{3}{*}{26}&\multirow{3}{*}{\makecell[l]{Do you own an In-Home Energy Display?}} & Yes \\
          &      & No \\
          &      & Don't know \\
          \cmidrule(lr){2-3}
    \multirow{6}{*}{27}&\multirow{6}{*}{\makecell[l]{How often do you check your In-Home Display?}} & Daily \\
          &      & 2-3 times a week \\
          &      & Once a week \\
          &      & Once a month \\
          &      & Less than once a month \\
          &      & Never \\
          \cmidrule(lr){2-3}
    \multirow{7}{*}{28}&\multirow{7}{*}{\makecell[l]{Which of the following best describes your household?}} & I/we own my own home (mortgage or outright) \\
          &      & I/we own my own home (through a shared ownership or Keyworker scheme) \\
          &      & I/we rent from a private landlord \\
          &      & I/we live in Student Accommodation \\
          &      & I/we rent from a Housing Association/Council \\
          &      & I/we live with my parents \\
          &      & Prefer not to say \\
          \cmidrule(lr){2-3}
    \multirow{7}{*}{29}&\multirow{7}{*}{\makecell[l]{Please select your pre-tax household income bracket:}} & Less than £20,000 \\
          &      & Between £20,000 and £39,999 \\
          &      & Between £40,000 and £59,999 \\
          &      & Between £60,000 and £79,999 \\
          &      & Between £80,000 and £99,999 \\
          &      & More than £100,000 \\
          &      & Prefer not to say 
    \end{longtable}%
  }

\newpage 
\begin{table}[h]
    \centering
    \caption{Attribute Restrictions and Privacy Implications}
     \resizebox{0.8\textwidth}{!}{
    \begin{tabular}{ccccccc}
    \toprule
        {} & Option 1 & Option 2 & Option 3 & Option 4 & Option 5 & Option 6 \\ 
        \midrule
        Anonymised & Yes & Yes &Yes& No & No & No \\ 
        Frequency & Real-Time & Half-Hourly & Daily & Real-Time & Half-Hourly & Daily \\ 
        \makecell{Expected Change\\ in Monthly Bill} & [-20\%, 20\%] & [-20\%, 20\%] & [0\%,20\%] & [-20\%,-0\%] & [-20\%,-0\%] & 0\% \\ 
        No. of Profiles & 9 & 9 & 5 & 5 & 5 & 1 \\ 
        \makecell{Large Appliance\\ Ownership} &  &  &  & \checkmark & \checkmark &  \\
        \makecell{Small Appliance\\ Ownership} &  &  &  & \checkmark &  &  \\ 
        \makecell{Appliance Usage\\ and Routines} &  &  &  & real-time & per half-hour &  \\
        Occupancy &  &  &  & real-time & per half-hour & per day \\ 
        Household Details &  &  &  & \checkmark & \checkmark & \checkmark \\
        Income Level &  &  &  & \checkmark & \checkmark & \checkmark \\
        \makecell{Marital \\ \& Employment Status}  &  &  &  & \checkmark & \checkmark &  \\
        Housing Details &  &  &  & \checkmark & \checkmark & \checkmark \\ 
        \bottomrule
    \end{tabular}
    }
    \label{tab:sv_restrict}
\end{table}

\newpage
\begin{longtable}{ccc}
    \caption{Choice Task Introduction and Treatment}
     \label{tab:choice_screen} \tabularnewline
    \toprule
        No. & Group & Screen \\
    \midrule
    \endfirsthead
    \multicolumn{3}{@{}l}{\ldots continued}\\
    \toprule
        No. & Group & Screen \\
    \midrule
    \endhead
    \bottomrule
        \multicolumn{3}{r@{}}{continued \ldots}\\
    \endfoot
    \bottomrule
    \multicolumn{3}{l}{Note: TR - Treatment group, C - Control group.}\\
    \endlastfoot
        1 & TR \& C &\adjustbox{valign=c}{\includegraphics[width=0.7\textwidth]{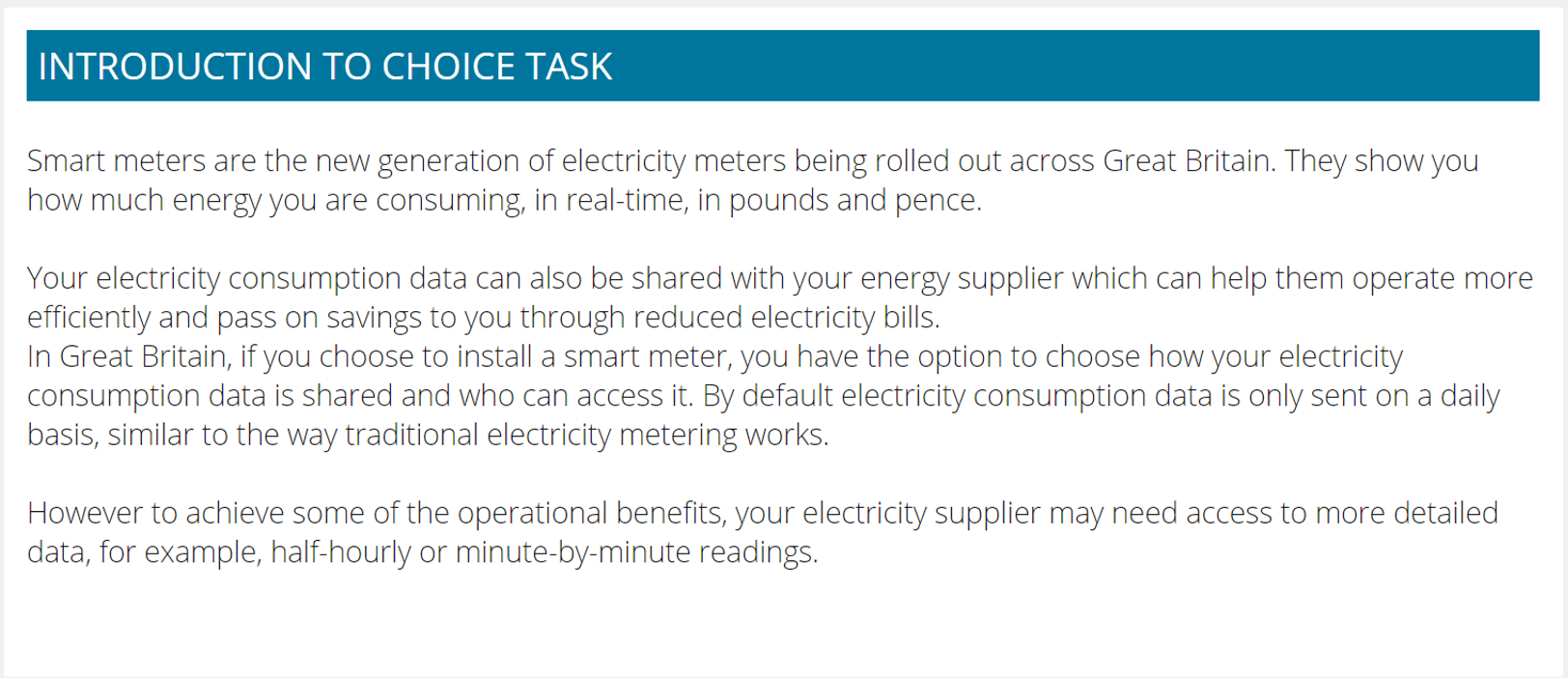}}\\
        2 & TR \& C &  \adjustbox{valign=c}{\includegraphics[width=0.7\textwidth]{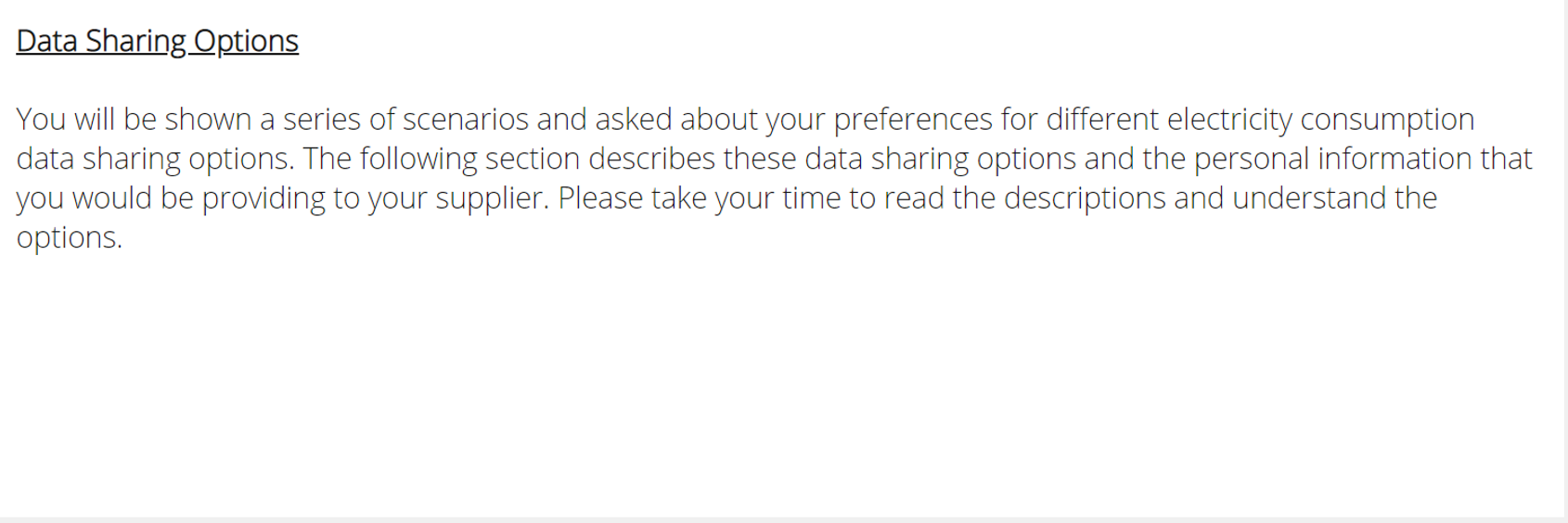}}\\
        3 & TR \& C &  \adjustbox{valign=c}{\includegraphics[width=0.7\textwidth]{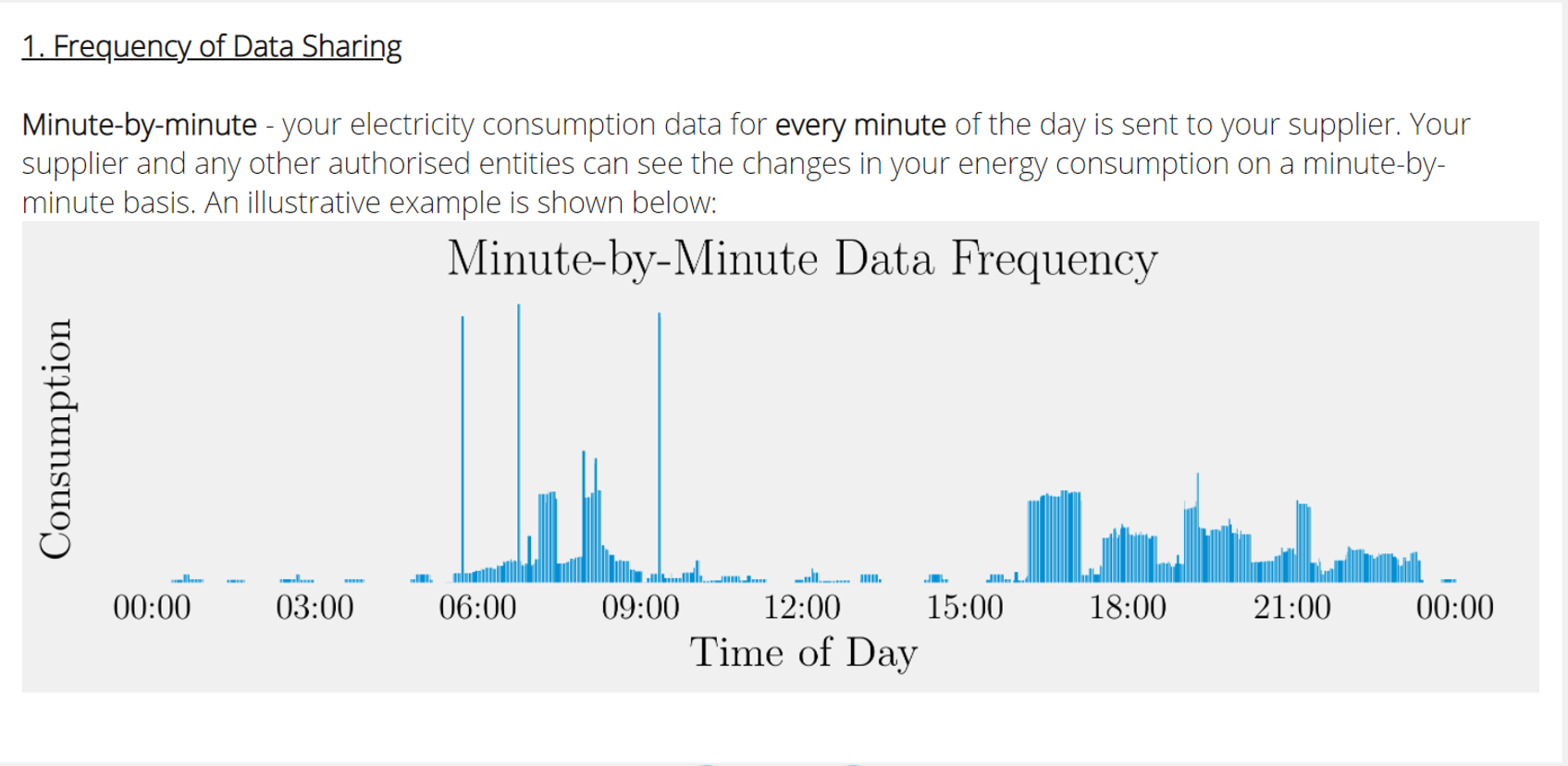}}\\
        4 & TR \& C &  \adjustbox{valign=c}{\includegraphics[width=0.7\textwidth]{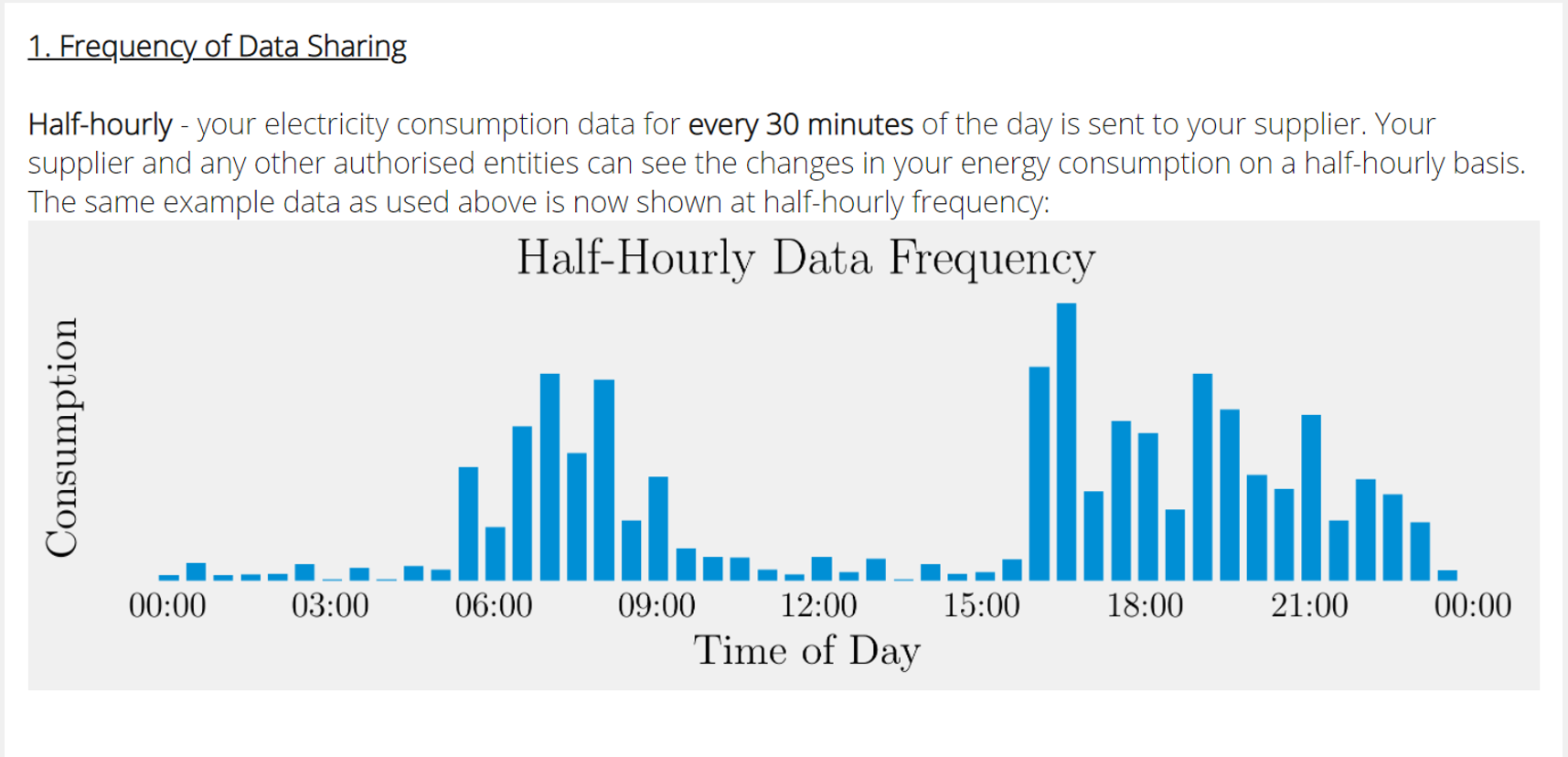}}\\
        5 & TR \& C &  \adjustbox{valign=c}{\includegraphics[width=0.7\textwidth]{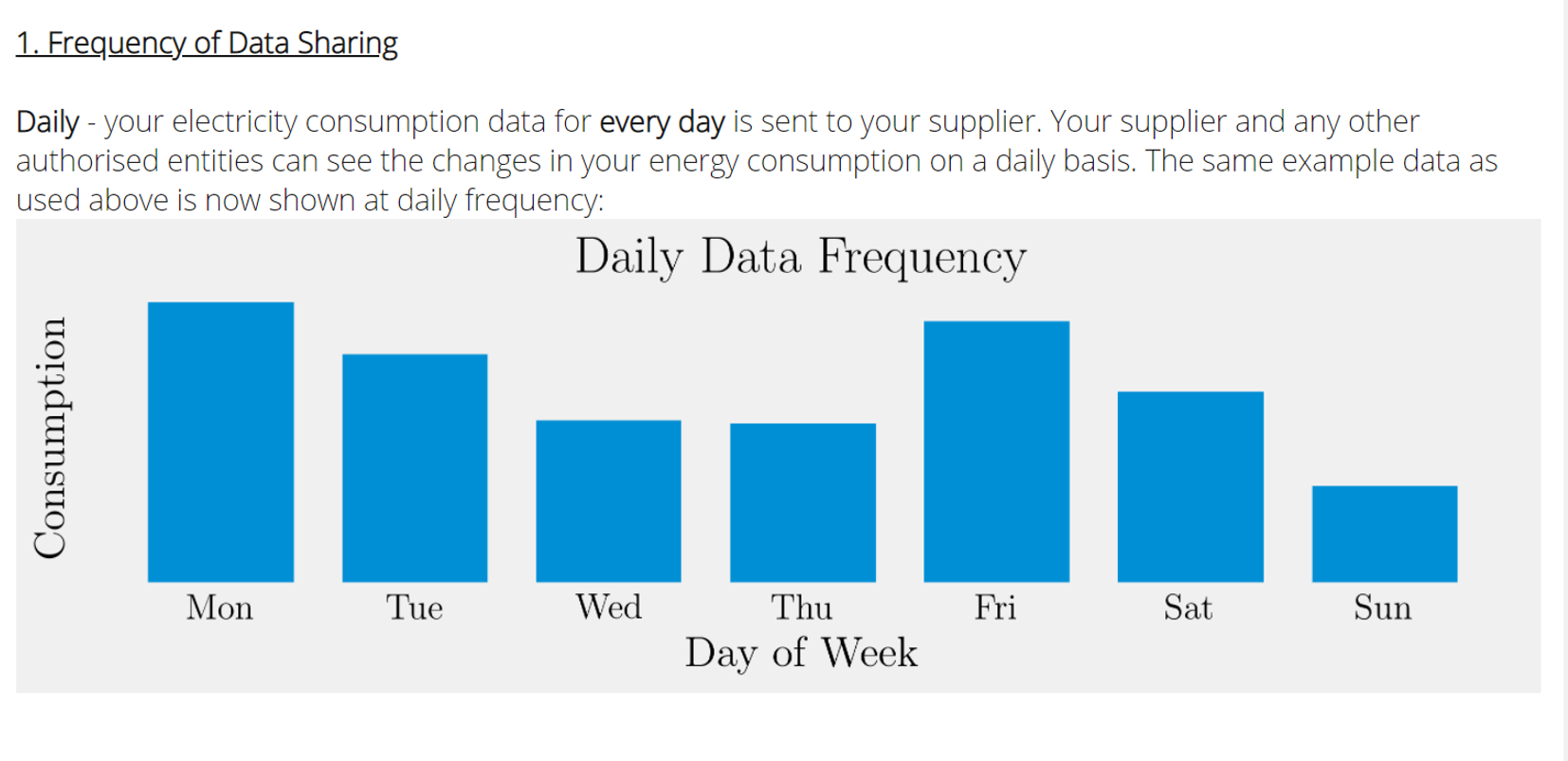}}\\
        6 & TR \& C &  \adjustbox{valign=c}{\includegraphics[width=0.7\textwidth]{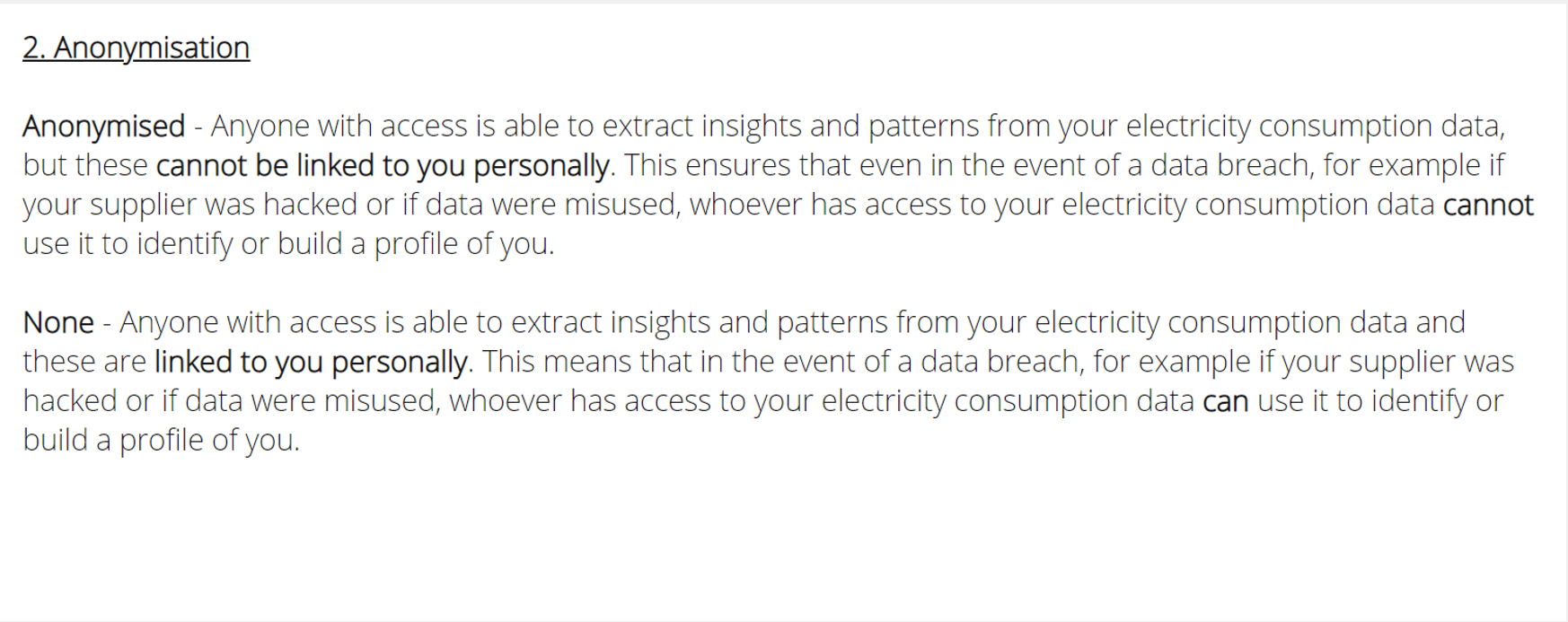}}\\
        7 & TR  &  \adjustbox{valign=c}{\includegraphics[width=0.7\textwidth]{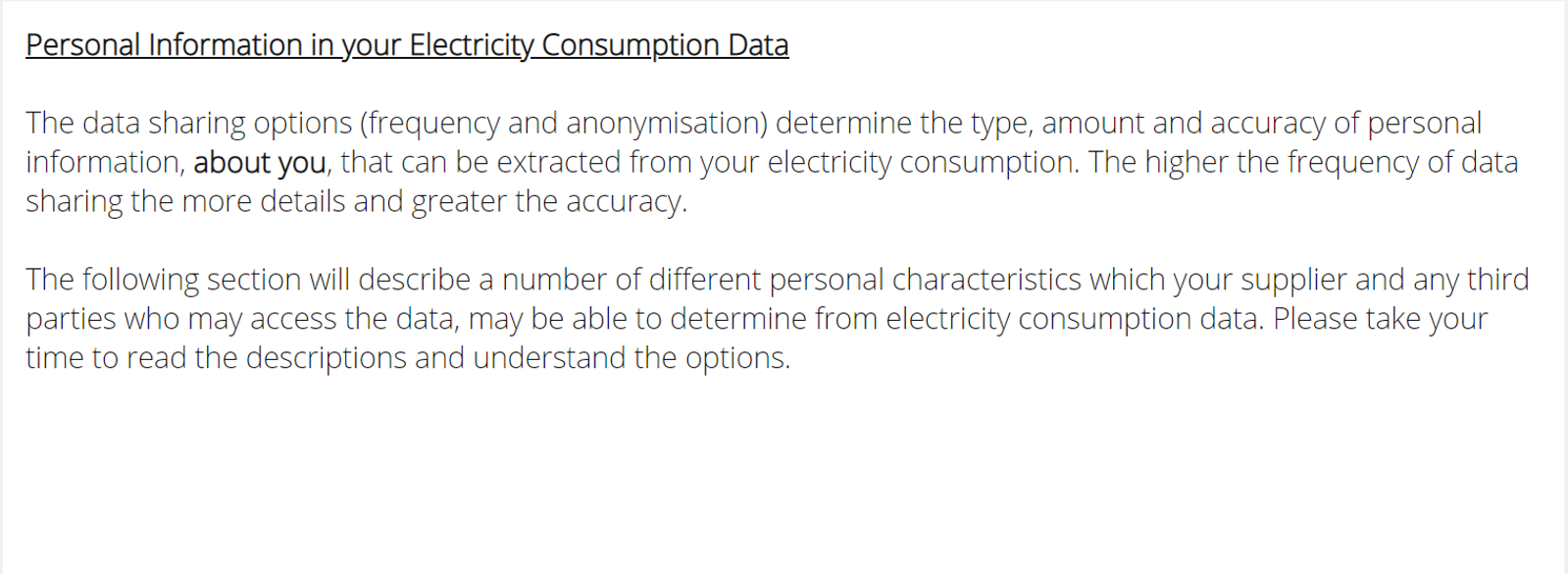}}\\
        8 & TR &  \adjustbox{valign=c}{\includegraphics[width=0.7\textwidth]{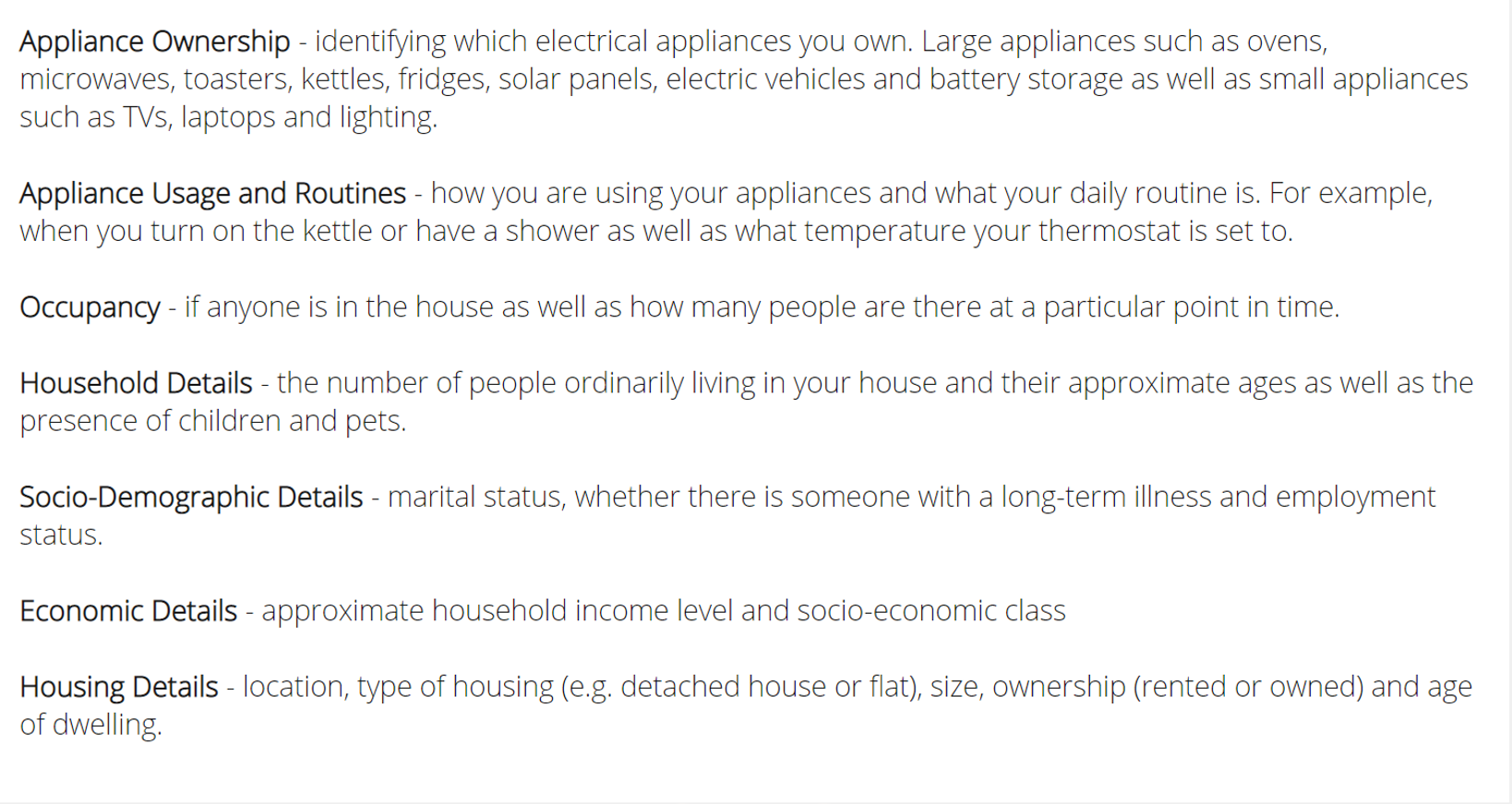}}\\
        9 & TR &  \adjustbox{valign=c}{\includegraphics[width=0.7\textwidth]{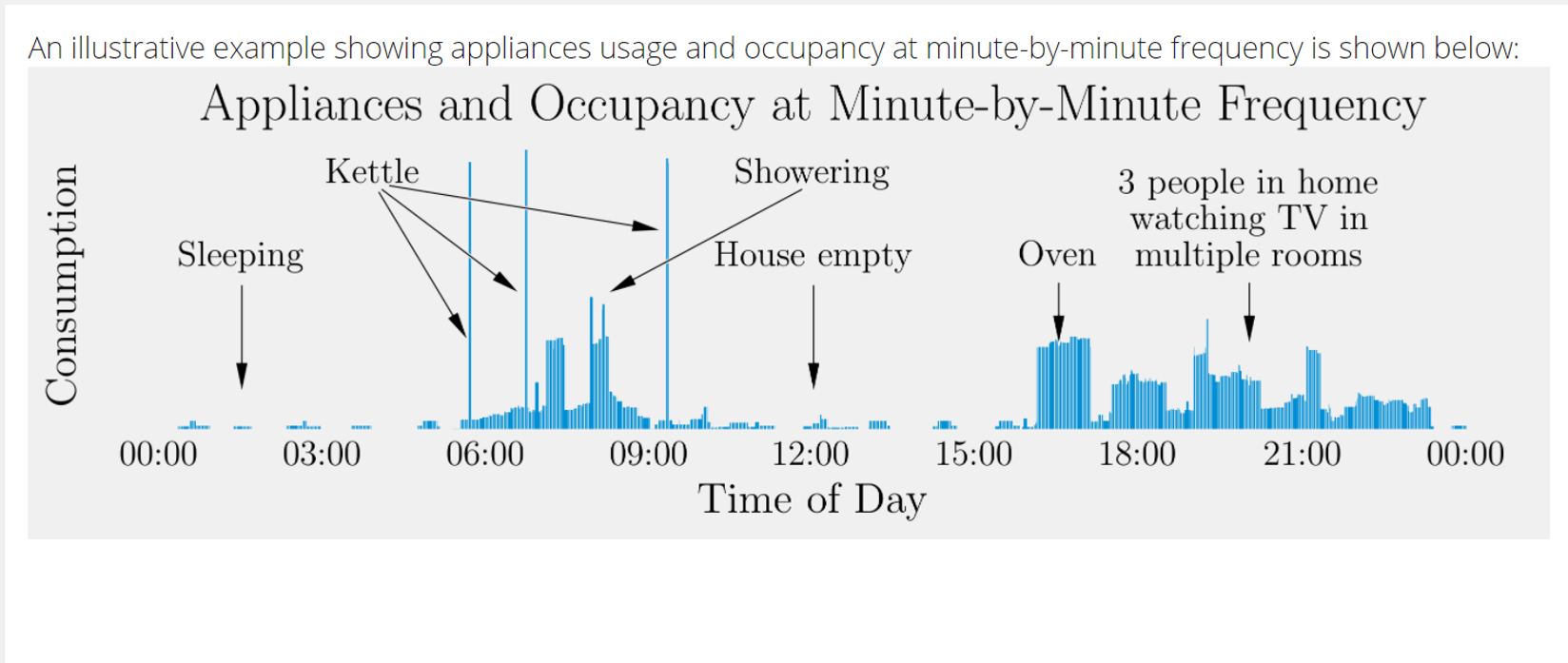}}\\
        10 & TR &  \adjustbox{valign=c}{\includegraphics[width=0.7\textwidth]{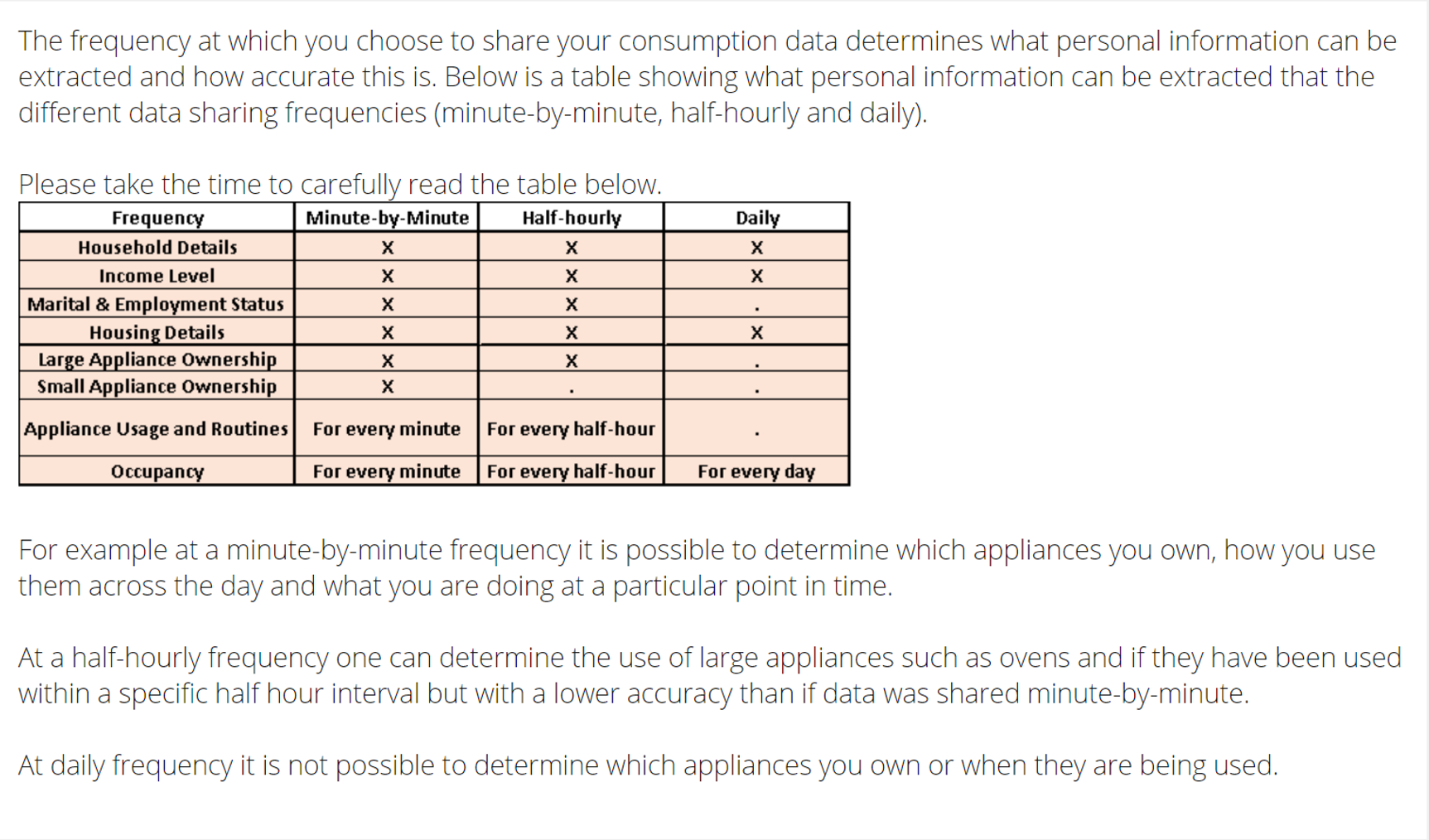}}
    \end{longtable}

\newpage
\section{Sample Statistics and Data Imputation}
\begin{table}[H]
\centering
\caption{Sample Statistics} 
\label{tab:full.samp}
\setlength\targetwidth{\textwidth}
\begin{tabularx}{\targetwidth}{@{}CCCCCCCl@{}}
    \toprule
    \multicolumn{2}{c}{ }  & \multicolumn{2}{c}{\makecell{Control (n=477)}} & \multicolumn{2}{c}{\makecell{Treatment (n=488)}} & \multicolumn{1}{c}{GB} & \multirow{2}{*}{\makecell{$p$-value}} \\
    \cmidrule(l{3pt}r{3pt}){3-4} \cmidrule(l{3pt}r{3pt}){5-6}
    &  & n & \% & n & \% & \% & \\
    \midrule
    \multirow{5}{2cm}{\makecell{Age$^1$}} & 18-34   & 123   & 25.8  & 134   & 27.5  & 28    & C-T: 0.573  \\
                                      & 35-54   & 157   & 32.9  & 172   & 35.2  & 34    & C-GB: 0.622	\\
                                      & 55-64   & 75    & 15.7  & 76    & 15.6  & 15    & T-GB: 0.888  \\
                                      & 65+     & 119   & 24.9  & 105   & 21.5  & 23    &           \\
                                      & Refused & 3     & 0.6   & 1     & 0.2   &       &           \\
    \cmidrule(lr){2-8} 
    \multirow{3}{2cm}{\makecell{Gender$^1$}} & Male     & 232   & 48.6  & 237   & 48.6  & 49    & C-T: 0.918\\
                                         & Female   & 241   & 50.5  & 248   & 50.8  & 51    & C-GB: 0.981\\
                                         & Refused  & 4     & 0.8   & 3     & 0.6   &       & T-GB: 0.989\\
    \cmidrule(lr){2-8}
    \multirow{6}{2cm}{\makecell{Ethnicity$^2$}} & White     & 421   & 88.3  & 425   & 87.1  & 86    &  C-T: 0.808  \\
                                            & Asian     & 30    & 6.3   & 33    & 6.8   & 8     &  C-GB: 0.521  \\
                                            & Black     & 13    & 2.7   & 11    & 2.3   & 3     &  T-GB: 0.484      \\
                                            & Mixed     & 9     & 1.9   & 14    & 2.9   & 2     &               \\
                                            & Other     & 4     & 0.8   & 4     & 0.8   & 1     &               \\
                                            & Refused   & 0     & 0.0   & 1     & 0.2   &       &               \\
    \cmidrule(lr){2-8}
    \multirow{5}{2cm}{\makecell{Socio-\\Economic\\ Group\\ (SEG)$^3$}}  & AB        & 120   & 25.2  & 116   & 23.8  & 27    &  C-T: 0.516\\
                                                                & C1        & 127   & 26.6  & 131   & 26.8  & 28    &  C-GB: 0.352 \\
                                                                & C2        & 90    & 18.9  & 110   & 22.5  & 20    &  T-GB: 0.299\\
                                                                & DE        & 135   & 28.3  & 123   & 25.2  & 25    &           \\
                                                                & Refused   & 5     & 1.0   & 8     & 1.6   &       &           \\
    \cmidrule(lr){2-8}
    \multirow{4}{2cm}{\makecell{Region$^1$}}& England   & 410   & 86.0  & 388   & 79.5  & 87    &  C-T: 0.0377\\
                                        & Wales     & 20    & 4.2   & 27    & 5.5   & 5     &  C-GB: 0.533 \\
                                        & Scotland  & 32    & 6.7   & 42    & 8.6   & 8     &  T-GB: 0.000\\
                                        & Refused   & 15    & 3.1   & 31    & 6.4   &       &           \\
    \cmidrule(lr){2-8}
    \multirow{3}{2.1cm}{\makecell{Smart\\ Meter \\ Ownership$^4$}} & Yes           & 248   & 52.0  & 272   & 55.7  & 44    &  C-T: 0.059\\
                                                                & No            & 226   & 47.4  & 206   & 42.2  & 56    &  C-GB: 0.000\\
                                                                & DK    & 3     & 0.6   & 10    & 2.0   &       &  T-GB: 0.000\\
    \bottomrule
    \multicolumn{8}{@{}P@{}}{Note: $p$-values for Pearson's Chi-Squared Test for Independence between the Control and Treatment group (C-T), and for z-test for proportions for each group against the GB nationally representative proportions. Percentages may not add up due to rounding. $^1$2018 \gls{ons} Population Projections \citep{ONS2019}. $^2$2011 UK Government Ethnicity Facts and Figures \citep{gov2011}. $^3$National Readership Survey Social Grades \citep{NRS2016}. $^4$December 2020 Quarterly Smart Meter Statistics \citep[Table 5a]{BEIS2021d}. Includes smart meters in smart and traditional mode.}
\end{tabularx}
\end{table}
\begin{table}[H]
\centering
\caption{Sample Statistics with Data Imputation} 
\label{tab:imp.samp}
\setlength\targetwidth{\textwidth}
\begin{tabularx}{\targetwidth}{@{}CCCCCCCl@{}}
\toprule
\multicolumn{2}{c}{ } & \multicolumn{2}{c}{\makecell{Control (n=477)}} & \multicolumn{2}{c}{\makecell{Treatment (n=488)}} & \multicolumn{1}{c}{GB} & \multirow{2}{*}{$p$-value} \\
\cmidrule(l{3pt}r{3pt}){3-4} \cmidrule(l{3pt}r{3pt}){5-6}
  &    & n & \% & n & \% & \% & \\
\midrule
\multirow{4}{2cm}{\makecell{Age$^1$}} & 18-34 & 123 & 25.8 & 135 & 27.7 & 28 & C-T:  0.494\\
 & 35-54 & 157 & 32.9 & 172 & 35.2 & 34& C-GB: 0.494\\
 & 55-64 & 75 & 15.7 & 76 & 15.6 & 15&  T-GB:  0.895\\
 & 65+ & 122 & 25.6 & 105 & 21.5 & 23\\
  \cmidrule(l{3pt}r{3pt}){2-8}
\multirow{3}{2cm}{\makecell{Gender$^1$}} & Male & 234 & 49.1 & 238 & 48.8 & 49& C-T:0.980\\
 & Female & 243 & 50.9 & 250 & 51.2 & 51& C-GB: 1.000\\
 &  &  &  & & & &  T-GB:  0.997\\
 \cmidrule(l{3pt}r{3pt}){2-8}
\multirow{5}{2cm}{\makecell{Ethnicity$^2$}} & White &421 & 88.3 & 426 & 87.3 & 86&  C-T:0.861 \\
 & Asian &  30 & 6.3 & 33 & 6.8 & 8 & C-GB: 0.521\\
 & Black & 13 & 2.7 & 11 & 2.3 & 3&  T-GB: 0.458\\
 & Mixed & 9 & 1.9 & 14 & 2.9 & 2\\
 & Other & 4 & 0.8 & 4 & 0.8 & 1\\
  \cmidrule(l{3pt}r{3pt}){2-8}
\multirow{4}{2cm}{\makecell{Socio-\\Economic\\ Group\\ (SEG)$^3$}} & AB & 121 & 25.4 & 117 & 24.0 & 27 &  C-T: 0.462\\
 & C1 & 128 & 26.8 & 132 & 27.0 & 28 &  C-GB: 0.260\\
 & C2 & 90 & 18.9 & 111 & 22.7 & 20 &  T-GB: 0.270\\
 & DE & 138 & 28.9 & 128 & 26.2 & 25\\
  \cmidrule(l{3pt}r{3pt}){2-8}
\multirow{3}{2cm}{\makecell{Region$^1$}} & England & 425 & 89.1 & 417 & 85.5 & 87&  C-T: 0.235\\
 & Wales & 20 & 4.2 & 28 & 5.7 & 5 &  C-GB: 0.309\\
 & Scotland & 32 & 6.7 & 43 & 8.8 & 8 &  T-GB: 0.566\\
  \cmidrule(l{3pt}r{3pt}){2-8}
\multirow{3}{2.1cm}{\makecell{Smart\\ Meter\\ Ownership$^4$}} & Yes & 248 & 52.0 & 272 & 55.7 & 44 &  C-T: 0.270\\
 & No & 229 & 48.0 & 216 & 44.3 & 56  &  C-GB: 0.000\\
 &  &  &  & &  & &  T-GB: 0.000\\
\bottomrule
\multicolumn{8}{@{}P@{}}{Note: $p$-values for Pearson's Chi-Squared Test for Independence between the Control and Treatment group (C-T), and for z-test for proportions for each group against the GB nationally representative proportions. Percentages may not add up due to rounding. $^1$2018 \gls{ons} Population Projections \citep{ONS2019}. $^2$2011 UK Government Ethnicity Facts and Figures \citep{gov2011}. $^3$National Readership Survey Social Grades \citep{NRS2016}. $^4$December 2020 Quarterly Smart Meter Statistics \citep[Table 5a]{BEIS2021d}. Includes smart meters in smart and traditional modes.}
\end{tabularx}
\end{table}
\begin{table}[H]
\centering
\caption{Additional Socio-Demographic Characteristics} 
\label{tab:samp.add.socio}
\setlength\targetwidth{\textwidth}
\begin{tabularx}{\targetwidth}{@{}CCCCCCC@{}}
  \toprule
  & &   \multicolumn{2}{c}{Control (n=477)}& \multicolumn{2}{c}{Treatment (n=488)}  & \multirow{2}{*}{$p$-value} \\ 
  \cmidrule(lr){3-4} \cmidrule(lr){5-6}
  & & n & \% & n & \% \\
  \midrule
\multirow{7}{*}{\makecell{Household\\Income}} & $<$£20k & 138 & 28.9 & 152 & 31.1 & \multirow{7}{*}{0.470} \\ 
   & 20-40k & 163 & 34.2 & 166 & 34.0 \\ 
   & 40-60k & 82 & 17.2 & 87 & 17.8 \\ 
   & 60-80k & 30 & 6.3 & 26 & 5.3 \\ 
   & 80-100k & 19 & 4.0 & 9 & 1.8 \\ 
   & $>$100k & 6 & 1.3 & 10 & 2.0 \\ 
   & Refused & 39 & 8.2 & 38 & 7.8 \\
   \cmidrule(lr){2-7}
  \multirow{3}{*}{Tenure} & Owner & 281 & 58.9 & 243 & 49.8 & \multirow{3}{*}{0.000}\\ 
   & Tenant & 195 & 40.9 & 230 & 47.1 \\ 
   & Refused & 1 & 0.2 & 15 & 3.1 \\ 
   \midrule
   \multicolumn{7}{l}{With Data Imputation}\\
   \midrule
   \multirow{6}{*}{\makecell{Household\\Income}} & $<$20k & 155 & 32.5 & 169 & 34.6 & \multirow{6}{*}{0.507} \\ 
   & 20-40k & 173 & 36.3 & 178 & 36.5 \\ 
   & 40-60k & 91 & 19.1 & 90 & 18.4 \\ 
   & 60-80k & 32 & 6.7 & 30 & 6.1 \\ 
   & 80-100k & 19 & 4.0 & 10 & 2.0 \\ 
   & $>$100k & 7 & 1.5 & 11 & 2.3 \\ 
   \cmidrule(lr){2-7}
   \multirow{2}{*}{Tenure} & Owner & 281 & 58.9 & 247 & 50.6 & \multirow{2}{*}{0.012}\\ 
   & Tenant & 196 & 41.1 & 241 & 49.4 \\ 
   \bottomrule
   \multicolumn{7}{@{}P@{}}{Note: $p$-values from Pearson’s Chi-Squared Test for independence between control and treatment groups.}
\end{tabularx}
\end{table}

\begin{table}[H]
\centering
\caption{Electricity Supply Characteristics} 
\label{tab:samp.elec}
\setlength\targetwidth{\textwidth}
\begin{tabularx}{\targetwidth}{@{}ccCCCCCl@{}}
  \toprule
  &    & \multicolumn{2}{c}{Control (n=477)} & \multicolumn{2}{c}{Treatment (n=488)} & GB & \multirow{2}{*}{$p$-value} \\ 
  \cmidrule(lr){3-4} \cmidrule(lr){5-6}
  & & n & \% & n & \% & \% \\
  \midrule
\multirow{3}{*}{Fuel Type$^1$} & Dual Fuel & 357 & 74.8 & 325 & 66.6 & 84.9 & C-T: 0.018 \\ 
   & Elec Only & 102 & 21.4 & 136 & 27.9 & 15.1 & C-GB: 0.000\\ 
   & Don't Know & 18 & 3.8 & 27 & 5.5 &  & T-GB: 0.000\\ 
   \cmidrule(lr){2-8}
  \multirow{6}{*}{Tariff} & SVT & 106 & 22.2 & 108 & 22.1 & 36.3 & C-T: 0.337\\ 
   & Fixed & 221 & 46.3 & 205 & 42.0 & 28.7 & C-GB: 0.000\\ 
   & Pre-Pay$^2$ & 51 & 10.7 & 71 & 14.5 & 16.7 & T-GB: 0.000\\ 
   & ToU$^3$ & 19 & 4.0 & 17 & 3.5 & \multirow{2}{*}{18.3} \\ 
   & Eco7/10$^3$ & 24 & 5.0 & 34 & 7.0 &  \\ 
   & Don't Know & 56 & 11.7 & 53 & 10.9 &  \\ 
   \midrule
   \multicolumn{7}{l}{with Data Imputation}\\
   \midrule
   \multirow{3}{*}{Fuel Type$^1$} & Dual Fuel &375 & 78.6 & 352 & 72.1 & 84.9 & C-T: 0.0237 \\ 
   & Elec Only & 102 & 21.4 & 136 & 27.9 & 15.1& C-GB: 0.000\\  
   & & & & & & & T-GB: 0.000\\
   \cmidrule(lr){2-8}
  \multirow{6}{*}{Tariff} & SVT & 162 & 34.0 & 161 & 33.0 & 36.3& C-T: 0.2316\\ 
   & Fixed & 221 & 46.3 & 205 & 42.0 & 28.7 & C-GB: 0.000\\ 
   & Pre-Pay$^2$ & 51 & 10.7 & 71 & 14.5 & 16.7& T-GB: 0.000\\ 
   & ToU$^3$ & 19 & 4.0 & 17 & 3.5 & \multirow{2}{*}{18.3}\\ 
   & Eco7/10$^3$ & 24 & 5.0 & 34 & 7.0 &  \\ 
   \bottomrule
   \multicolumn{8}{@{}P@{}}{Note: $p$-values for Pearson’s Chi-Squared Test for Independence between the Control and Treatment group (C-T), and for z-test for proportions for each group against GB nationally representative proportions. 
   $^1$Fuel type data based on proportion of domestic customers connected to the gas grid in 2021 \citep{BEIS2024}. Tariff data from OFGEM for April 2021 \citep[Tab 1]{OFGEM2021c}. 
   $^2$Pre-payment meters estimated at 4.4 million based on latest available OFGEM data \citep[p. 49]{OFGEM2019b}. 
   $^3$Category 'Other non-standard variable tariffs'. Split between Time-of-Use (ToU) and Economy 7/10 not available. }
\end{tabularx}
\end{table}

\begin{table}[H]
\centering
\caption{Smart Meter Characteristics} 
\label{tab:samp.smdet}
\setlength\targetwidth{\textwidth}
\begin{tabularx}{\targetwidth}{@{}CCCCCCC@{}}
  \toprule
  & &   \multicolumn{2}{c}{Control (n=477)}& \multicolumn{2}{c}{Treatment (n=488)}  & \multirow{2}{*}{$p$-value} \\ 
  \cmidrule(lr){3-4} \cmidrule(lr){5-6}
  & & n & \% & n & \% \\
  \midrule
\multirow{5}{*}{\makecell{Data Sharing\\ Resolution}} & NA & 229 & 48.0 & 216 & 44.3 & \multirow{5}{*}{0.396}\\ 
   & Half-Hourly & 41 & 8.6 & 58 & 11.9 \\ 
   & Daily & 74 & 15.5 & 83 & 17.0 \\ 
   & Monthly & 46 & 9.6 & 50 & 10.2 \\ 
   & DK & 87 & 18.2 & 81 & 16.6 \\ 
   \cmidrule(lr){2-7}
  \multirow{4}{*}{IHD} & NA & 229 & 48.0 & 216 & 44.3 &\multirow{4}{*}{0.444}\\ 
   & Yes & 182 & 38.2 & 211 & 43.2 \\ 
   & No & 55 & 11.5 & 52 & 10.7 \\ 
   & DK & 11 & 2.3 & 9 & 1.8 \\ 
   \cmidrule(lr){2-7}
  \multirow{7}{*}{\makecell{IHD \\Engagement}} & NA & 295 & 61.8 & 277 & 56.8 &\multirow{7}{*}{0.317}\\ 
   & Daily & 63 & 13.2 & 70 & 14.3 \\ 
   & 2-3/week & 39 & 8.2 & 51 & 10.5 \\ 
   & 1/week & 31 & 6.5 & 33 & 6.8 \\ 
   & 1/month & 9 & 1.9 & 19 & 3.9 \\ 
   & $<$ 1/month & 18 & 3.8 & 13 & 2.7 \\ 
   & Never & 22 & 4.6 & 25 & 5.1 \\ 
   \bottomrule
   \multicolumn{7}{@{}P@{}}{Note: $p$-values for Pearson’s Chi-Squared Test for Independence between the Control and Treatment group. NA - Not applicable (i.e. do not have a smart meter/IHD), DK - Don't know.}
\end{tabularx}
\end{table}

\begin{table}[H]
\centering
\caption{Structured Feedback} 
\label{tab:samp.feed}
\setlength\targetwidth{\textwidth}
\begin{tabularx}{\targetwidth}{@{}CCCCCCC@{}}
\toprule
  & &   \multicolumn{2}{c}{Control (n=477)}& \multicolumn{2}{c}{Treatment (n=488)}  & \multirow{2}{*}{$p$-value} \\ 
  \cmidrule(lr){3-4} \cmidrule(lr){5-6}
  & & n & \% & n & \% \\
  \midrule
\multirow{6}{*}{\makecell{I was able to\\ understand \\the choices}} & SD & 15 & 3.1 & 17 & 3.5 & \multirow{6}{*}{0.831} \\ 
   & DA & 27 & 5.7 & 29 & 5.9 \\ 
   & NAD & 96 & 20.1 & 113 & 23.2 \\ 
   & SWA & 169 & 35.4 & 173 & 35.5 \\ 
   & SA & 163 & 34.2 & 149 & 30.5 \\ 
   & DK & 7 & 1.5 & 7 & 1.4 \\ 
   \cmidrule(lr){2-7}
  \multirow{6}{*}{\makecell{I found the\\ options realistic}} & SD & 22 & 4.6 & 14 & 2.9 & \multirow{6}{*}{\makecell{0.139}}\\ 
   & DA & 38 & 8.0 & 37 & 7.6 \\ 
   & NAD & 123 & 25.8 & 121 & 24.8 \\ 
   & SWA & 188 & 39.4 & 172 & 35.2 \\ 
   & SA & 94 & 19.7 & 130 & 26.6 \\ 
   & DK & 12 & 2.5 & 14 & 2.9 \\ 
   \cmidrule(lr){2-7}
  \multirow{6}{*}{\makecell{I found it \\easy to choose \\between \\the options}} & SD & 25 & 5.2 & 25 & 5.1 & \multirow{6}{*}{\makecell{0.609}}\\ 
   & DA & 38 & 8.0 & 53 & 10.9 \\ 
   & NAD & 110 & 23.1 & 109 & 22.3 \\ 
   & SWA & 164 & 34.4 & 151 & 30.9 \\ 
   & SA & 133 & 27.9 & 140 & 28.7 \\ 
   & DK & 7 & 1.5 & 10 & 2.0 \\ 
   \bottomrule
   \multicolumn{7}{@{}P@{}}{Note: $p$-values for Pearson’s Chi-Squared Test for Independence between the Control and Treatment group. SD - Strongly Disagree, DA - Disagree, NAD - Neither agree nor disagree, SWA - Somewhat agree, SA - Strongly agree, DK - Don't know.}
\end{tabularx}
\end{table}

\begin{table}[H]
\centering
\caption{Manipulation Checks} 
\label{tab:samp.manip}
\setlength\targetwidth{\textwidth}
\begin{tabularx}{\targetwidth}{@{}cCCCCCC@{}}
\toprule
   \multirow{2}{*}{Statement} & \multirow{2}{*}{Response} &   \multicolumn{2}{c}{\makecell{Control\\(n=477)}}& \multicolumn{2}{c}{\makecell{Treatment\\(n=488)}}  & \multirow{2}{*}{$p$-value} \\ 
  \cmidrule(lr){3-4} \cmidrule(lr){5-6}
  & & n & \% & n & \% \\
  \midrule
\multirow{2}{*}{\makecell{A negative expected change\\ in your monthly bill \\(shown in green)\\ indicates a reduction in\\ your electricity bill.}} & \textbf{True} & 373 & 78.2 & 384 & 78.7 & \multirow{2}{*}{0.915} \\ 
& False & 104 & 21.8 & 104 & 21.3 \\  \\ \\ \\
   \cmidrule(lr){1-7}
 \multirow{2}{*}{\makecell{A half-hourly frequency \\means that your total\\ electricity consumption\\ for each half-hour\\ is sent to your electricity\\ supplier every 30 minutes.}} & \textbf{True} & 407 & 85.3 & 405 & 83.0 & \multirow{2}{*}{0.366}\\
   & False & 70 & 14.7 & 83 & 17.0 \\ \\ \\ \\ \\
   \cmidrule(lr){1-7}
  \multirow{2}{*}{\makecell{Anonymisation of your\\ consumption data\\ ensures your information \\can be linked back to you\\ in the event of a data breach.}} & True & 204 & 42.8 & 216 & 44.3 & \multirow{2}{*}{0.687}\\ 
   & \textbf{False} & 273 & 57.2 & 272 & 55.7 \\ \\ \\ \\
   \bottomrule
   \multicolumn{7}{@{}P@{}}{Note: $p$-values for Pearson’s Chi-Squared Test for Independence between the Control and Treatment group. Correct response in bold.}
\end{tabularx}
\end{table}

\begin{figure}[H]
    \centering
    \subfloat[Completion Time\label{fig:samp_time}]{
    \includegraphics[width=0.45\linewidth]{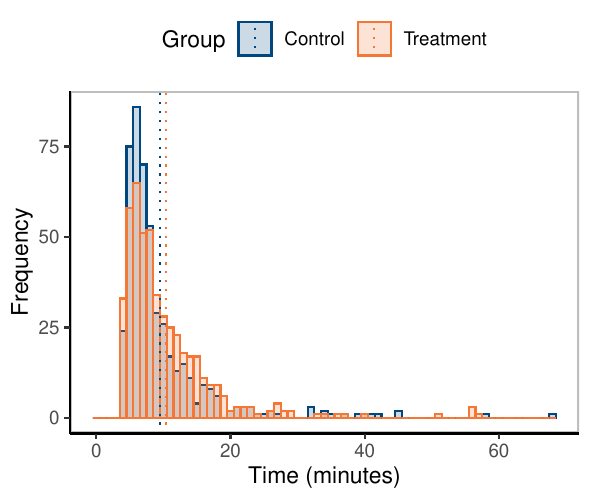}}
    \hfil
    \subfloat[Monthly Bills\label{fig:bill}]{
    \includegraphics[width=0.45\linewidth]{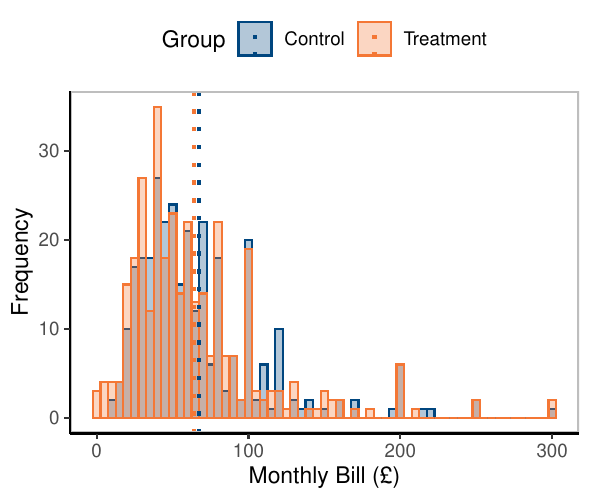}}
    \caption{Distribution of Survey Completion Time and Provided Monthly Bills by Control and Treatment Groups. Dotted Lines Indicate Sample Means.}
    \label{fig:enter-label}
\end{figure}

\newpage
\section{Robustness Analysis}
\begin{landscape}
\begin{longtable}{cccp{9.5cm}p{4.5cm}p{4cm}}
    \caption{Summary of Hypotheses and Robustness Tests}
    \label{tab:robust}\tabularnewline
    \toprule
        No. & Measure &  Group &  Summary & Conclusion & Source \\
        \midrule
        \endfirsthead
        \multicolumn{5}{@{}l}{\ldots continued}\\
        \toprule
        No. & Measure & Group  & Summary & Conclusion & Source \\
        \midrule
        \endhead
        \bottomrule
        \multicolumn{6}{@{}r}{continued\ldots}\\
        \endfoot
        \bottomrule
        \endlastfoot
         \textbf{H1}& \makecell{\gls{wtpa} for\\ Frequency} & - & Daily sharing has significantly higher value (both \gls{wtp} and \gls{wta}) than real-time sharing for all \gls{mxl} specifications with control group and higher value than half-hourly sharing for the subgroup models (BM and TP). For the treatment group models, we see a significantly higher value for daily sharing for the full sample and BM but not for TP. Results using mixed logit specifications without price mis-interpretation correction similar dynamics are observed. In the control group daily sharing has significantly higher value than real-time sharing for the full sample and BM. While for the treatment group daily sharing has higher value compared to both half-hourly and real-time sharing. & \textbf{H1} is supported. &  \multirow{2}{4cm}{Table \ref{tab:mixl.split} for \glsxtrlongpl{mxl} and Table \ref{tab:wtpa.conf} for \gls{wtpa} estimates and Table \ref{tab:poe.freq} for $p$-values for combinatorial tests. For $p$-values for combinatorial tests without price mis-interpretation correction see Table \ref{tab:poe.freq.robust}.}\\
         
         \cmidrule(lr){1-5}
         \textbf{H2}& \makecell{\gls{wtpa} for\\ Anonymisation} & - & No significant differences between data sharing frequencies once anonymisation is introduced. Similarly, anonymised data sharing is significantly higher for real-time and half-hourly sharing but not for daily sharing. These results show anonymisation has less additive value at lower frequencies. Result holds for all \gls{mxl} specifications except for a marginally significant difference between anonymised half-hourly and anonymised real-time sharing in the treatment group. & \textbf{H2} is supported. &\\
         
         \cmidrule(lr){1-5}
         \textbf{H3}& \makecell{\gls{wtpa} for\\ Anonymisation} & - & Ratio of \gls{wta} to \gls{wtp} greater than 1 ($p < 0.05$) for all \gls{mxl} specifications except subgroup of those who share data with third parties in the treatment group ($p < 0.10$). Combinatorial tests show slightly more mixed results with \gls{wta} significantly greater than \gls{wtp} for daily and anonymised half-hourly sharing for control group and but only marginally significant (before adjustment for multiple comparisons) for treatment group. & \textbf{H3} is supported. & \\
         
         \cmidrule(lr){1-6}
         \textbf{H4} & Change in \gls{wts} & - & Wilcoxon Signed-Rank Test shows significant difference in distribution of responses between anonymised and non-anonymised framing. McNemar-Bowker test shows significant change in response from less likely or indifferent to more likely when shifting from non-anonymised to anonymised framing. Question framing (Anon) coefficient for the probability of being less likely in the partial proportional odds, binary logistic regression and \gls{mnl} are significant and negative ($p < 0.01$).& \textbf{H4} is supported. & Table \ref{tab:wtsna.sig} for non-parametric test results and \ref{tab:wtsna.ord} for logistic regression model coefficients.\\
         
         \pagebreak
         \textbf{H5}& \gls{iwts} & AGE & No significant difference in estimated marginal mean probabilities of being willing or unwilling after adjustment for multiple comparisons. No significant difference in response distribution based on the Kruskal-Wallis test. &\textbf{H5} for differences in \gls{iwts} across age is not supported. & Table \ref{tab:wtsg.sig} for non-parametric tests. Table \ref{tab:wts.het.tests} for results of z-tests on estimated marginal mean probabilities from partial proportional odds model in Table \ref{tab:wtsg.lr}. \\ 
         \cmidrule(lr){3-5}
         & &\gls{seg} & Marginally significant difference in estimated marginal mean probabilities of being unwilling before adjustment for multiple comparisons. Significant differences were found in the Kruskal-Wallis test, with follow-up Dunn’s pairwise comparisons showing significance for AB vs. C2, AB vs. DE, and C1 vs. DE. & \textbf{H5} for differences in \gls{iwts} across \gls{seg} is marginally supported with more deprived respondents having lower \gls{iwts}. \\ 
         \cmidrule(lr){3-5}
         & &GEN& No significant difference in estimated marginal mean probabilities of being willing or unwilling after adjustment for multiple comparisons. No significant difference in response distribution based on Mann-Whitney U test. &  \textbf{H5} for differences in \gls{iwts} across gender is not supported.  \\
         \cmidrule(lr){3-5}
         & &SM & Significant difference in estimated marginal mean probabilities of being willing or unwilling after adjustment for multiple comparisons. Significant difference in Mann-Whitney U test. & \textbf{H5} for differences in \gls{iwts} across smart meter ownership is supported with those who do not have a smart meter having a lower \gls{iwts}. \\
         \cmidrule(lr){3-5}
         & &\gls{dsa} & Significant difference in all response categories between those sharing with third parties (TP) and those only sharing basic information (BA) after adjustment for multiple comparisons. Significant differences were found in the Kruskal-Wallis test, with follow-up Dunn’s pairwise comparisons showing significance for BA vs. TP. & \textbf{H5} for differences in \gls{iwts} across \gls{dsa} is supported with those who share less information having a lower \gls{iwts}.\\
         \cmidrule(lr){3-5}
         & &\gls{ihd} & Significant difference in estimated marginal mean probabilities of being willing or unwilling after adjustment for multiple comparisons. Significant difference in Mann-Whitney U test. & \textbf{H5} for differences in \gls{iwts} across \gls{ihd} engagement is supported with those who engage with their \gls{ihd} more having a higher \gls{iwts}. \\
         \cmidrule(lr){3-5}
         & &\gls{tvt} & No significant difference in estimated marginal mean probabilities of being willing or unwilling after adjustment for multiple comparisons. No significant difference in response distribution based on Mann-Whitney U test. & \textbf{H5} for differences in \gls{iwts} across \gls{tvt} is not supported.  \\
         \pagebreak
         \textbf{H5}& \makecell{Change in \gls{wts}\\ Non-anonymised} & AGE   & Significant difference in estimated marginal mean probabilities of being more likely to share with younger respondents having a higher probability. Significant differences were found in the Kruskal-Wallis test, with follow-up Dunn’s pairwise comparisons showing significance for all age groups except 55-64 vs. 65+. & \textbf{H5} is supported. & Table \ref{tab:wts.het.tests} for $p$-values of z-tests, Table \ref{tab:wtsna.ord} for \gls{mnl} for random-effects model, Table \ref{tab:wtsn.ord} for partial proportional odds model, and Table \ref{tab:wtsn.sig} for non-parametric tests.\\ 
         \cmidrule(lr){3-5}
         & &\gls{seg} & Significant difference in estimated marginal mean probabilities of being more likely to share with DE having a lower probability than AB. Significant differences were found in the Kruskal-Wallis test, with follow-up Dunn’s pairwise comparisons showing significance for AB vs. DE. &  \textbf{H5} is supported for being more likely but not less likely. &\\
         \cmidrule(lr){3-5}
         & &GEN & Significant difference in estimated marginal mean probabilities of being more likely to share with women having a lower probability than men. Significant difference in Mann-Whitney U test. & \textbf{H5} is supported. &\\
         \cmidrule(lr){3-5}
         & &SM & Significant difference in estimated marginal mean probabilities of being less likely to share with smart meter owners having a lower probability than those who do not have one. Significant difference in Mann-Whitney U test. & \textbf{H5} is supported. & \\
         \cmidrule(lr){3-5}
         & &\gls{dsa} & Significant difference in estimated marginal mean probabilities of being less likely to share with BA and MR having a higher probability than TP. Significant differences were found in the Kruskal-Wallis test, with follow-up Dunn’s pairwise comparisons showing significance for BA and MR vs. TP. & \textbf{H5} is supported. &\\
         \cmidrule(lr){3-5}
         & &\gls{ihd}& No significant difference in estimated marginal mean probabilities. Significant difference in Mann-Whitney U test with those who engage with their \gls{ihd} having a higher probability of being more likely. & \textbf{H5} is not supported when accounting for other socio-demographic differences. & \\
         \cmidrule(lr){3-5}
         & &\gls{tvt}& Significant difference in estimated marginal mean probabilities of being less likely to share with those in \gls{tvt} having a lower probability. Significant difference in Mann-Whitney U test. & \textbf{H5} is supported. & \\
         \pagebreak
         \textbf{H5}& \makecell{Change in \gls{wts}\\ Anonymised} & AGE   & Significant difference in estimated marginal mean probabilities of being more likely to share with younger respondents having a higher probability. Significant differences were found in the Kruskal-Wallis test, with follow-up Dunn’s pairwise comparisons showing significance for all age groups except 55-64 vs. 65+ and 18-34 vs. 35-54. & \textbf{H5} is supported. & Table \ref{tab:wts.het.tests} for $p$-values of z-tests, Table \ref{tab:wtsna.ord} for \gls{mnl} for random-effects model, Table \ref{tab:wtsa.ord} for partial proportional odds model, and Table \ref{tab:wtsa.sig} for non-parametric tests.\\ 
         \cmidrule(lr){3-5}
         & &\gls{seg} & Significant difference in estimated marginal mean probabilities of being more likely to share with DE having a lower probability than AB and C1. Significant differences were found in the Kruskal-Wallis test, with follow-up Dunn’s pairwise comparisons showing significance for AB and C1 vs. DE. & \textbf{H5} is supported for being more likely but not less likely. &\\
         \cmidrule(lr){3-5}
         & &GEN & Significant difference in estimated marginal mean probabilities of being more likely to share with women having a lower probability than men. No significant difference in Mann-Whitney U test. & \textbf{H5} is supported. &\\
         \cmidrule(lr){3-5}
         & &SM & Significant difference in estimated marginal mean probabilities of being less likely to share with smart meter owners having a lower probability than those who do not have one. Significant difference in Mann-Whitney U test. & \textbf{H5} is supported. & \\
         \cmidrule(lr){3-5}
         & &\gls{dsa} & No significant difference in estimated marginal mean probabilities. No significant difference in the Kruskal-Wallis test. & \textbf{H5} is not supported. &\\
         \cmidrule(lr){3-5}
         & &\gls{ihd} & No significant difference in estimated marginal mean probabilities. Significant difference in Mann-Whitney U test with those who engage with their \gls{ihd} having a higher probability of being more likely. & \textbf{H5} is not supported when accounting for other socio-demographic differences. & \\
         \cmidrule(lr){3-5}
         & &\gls{tvt}& Significant difference in estimated marginal mean probabilities of being less likely to share with those in \gls{tvt} having a lower probability. No significant difference in Mann-Whitney U test. & \textbf{H5} is supported. & \\
         \pagebreak
         \textbf{H5} & \makecell{\gls{wtpa} for\\ Anonymisation} & AGE  & \gls{mxl} with interactions shows marginally significant difference in \gls{wtpa} for anonymised real-time data sharing for 65+ vs. 18-34, before adjustment for multiple comparisons. Significant differences for anonymised daily sharing between 65+ and other groups.  Segmented \gls{mxl} analysis results in larger differences. & \textbf{H5} is marginally supported with older age groups having higher \gls{wtpa} for anonymisation. & \multirow{2}{4cm}{Table \ref{tab:mixl.het} for \gls{mxl} with mean-shifting interaction terms for socio-demographics. Table \ref{tab:wtpa.het.tests} for combinatorial tests on marginal mean \gls{wtpa} distributions and Table \ref{tab:seg.het} for combinatorial tests from segmented analysis.}\\ 
         \cmidrule(lr){3-5}
         & &\gls{seg} & \gls{mxl} with interactions shows no significant differences across \gls{seg}. Segmented \gls{mxl} analysis shows marginally significant difference in \gls{wtpa} for anonymised daily and half-hourly data sharing for AB and C1 vs. C2 and DE, before adjustment for multiple comparisons. Significant differences for anonymised daily sharing between 65+ and other groups. & \textbf{H5} is marginally supported with more deprived groups (DE) having lower \gls{wtpa} for anonymisation. & \\
         \cmidrule(lr){3-5}
         & &GEN & \gls{mxl} with interactions shows significant difference in \gls{wtpa} for anonymised real-time and daily data sharing, with women having higher valuations than men. Significant differences for all anonymised sharing resolutions from segmented \gls{mxl} analysis. & \textbf{H5} is supported with women having higher \gls{wtpa} for anonymisation. & \\
         \cmidrule(lr){3-5}
         & &SM & \gls{mxl} with interactions shows significant difference in \gls{wtpa} for anonymised real-time and daily data sharing, with smart meter owners having a lower valuation. Significant differences for all anonymised sharing resolutions from segmented \gls{mxl} analysis. & \textbf{H5} is supported with smart meter owners having lower \gls{wtpa} for anonymisation. & \\
         \cmidrule(lr){3-5}
         & &\gls{dsa} & \gls{mxl} with interactions shows significant difference in \gls{wtpa} for anonymised real-time and daily data sharing, with BA and MR having higher valuations than TP. Significant differences for all anonymised sharing resolutions from segmented \gls{mxl} analysis. & \textbf{H5} is supported with those sharing data with third parties having lower \gls{wtpa} for anonymisation. & \\
         \cmidrule(lr){3-5}
         & &\gls{ihd} & \gls{mxl} with interactions shows marginally significant difference in \gls{wtpa} for anonymised real-time data sharing, with those engaging more with their \gls{ihd} having higher valuations. Segmented \gls{mxl} analysis shows no significant differences. & \textbf{H5} is marginally supported with those engaging regularly with their \gls{ihd} having higher \gls{wtpa} for anonymisation. & \\
         \cmidrule(lr){3-5}
         & &\gls{tvt}& \gls{mxl} with interactions shows significant difference in \gls{wtpa} for anonymised daily data sharing, with those on \glspl{tvt} having higher valuations. Segmented \gls{mxl} analysis show no significant difference. & \textbf{H5} is marginally supported with those on \glspl{tvt} having higher \gls{wtpa} for anonymisation. & \\
         \pagebreak
         \textbf{H5} & \makecell{Demand for \\Smart metering} & AGE  & Significant differences in estimated marginal mean probability of wanting a smart meter with younger respondents being more likely. Significant differences were found in the Kruskal-Wallis test, with follow-up Dunn’s pairwise comparisons showing significance for all age groups except 55-64 vs. 65+ and 35-54 vs. 55-64.  & \textbf{H5} is supported. & Table \ref{tab:wsm.lr} for binary logistic regression model, Table \ref{tab:wts.het.tests} for $p$-values of z-tests, and Table \ref{tab:wsm.sig} for non-parametric tests.\\ 
         \cmidrule(lr){3-5}
         & &\gls{seg}& Significant differences in estimated marginal mean probability of wanting a smart meter with DE being less likely than AB. Significant differences were found in the Kruskal-Wallis test, with follow-up Dunn’s pairwise comparisons showing significance for DE vs. all other groups.  & \textbf{H5} is marginally supported.\\
         \cmidrule(lr){3-5}
         & &GEN & Significant differences in estimated marginal mean probability of wanting a smart meter with women being less likely than men. No significant difference observed in Mann-Whitney U test. & \textbf{H5} is supported.\\
         \cmidrule(lr){3-5}
         & &SM & Significant difference in estimated marginal mean probability of wanting a smart meter with smart meter owners having a lower probability than those who do not have one. Significant difference in Mann-Whitney U test. & \textbf{H5} is supported. & \\
         \cmidrule(lr){3-5}
         & &\gls{dsa} & Significant difference in estimated marginal mean probability of wanting a smart meter with BA having a lower probability than TP. Significant differences were found in the Kruskal-Wallis test, with follow-up Dunn’s pairwise comparisons showing significance for BA vs. MR and TP. & \textbf{H5} is supported. &\\
         \cmidrule(lr){3-5}
         & &\gls{ihd} & No significant difference in estimated marginal mean probability. Significant difference in Mann-Whitney U test. & \textbf{H5} is not supported when accounting for other socio-demographic differences. & \\
         \cmidrule(lr){3-5}
         & &\gls{tvt} & No significant differences. & \textbf{H5} is not supported. & \\
         
         \pagebreak
         \textbf{H6} & \makecell{Change in \gls{wts}\\ Non-Anonymised} & - & No significant difference in estimated marginal mean probabilities or non-parametric tests. & \textbf{H6} is not supported. & Tables \ref{tab:wtsn.gp.sig} and \ref{tab:wtsn.ll.sig} for non-parametric tests and Table \ref{tab:wtsna.ord} for \gls{mnl}. \\
         \cmidrule(lr){2-6}
          &  \makecell{Change in \gls{wts}\\ Anonymised} & - & Significant difference in estimated marginal mean probabilities and Mann-Whitney U tests. &\textbf{H6} is supported. &Tables \ref{tab:wtsa.gp.sig} and \ref{tab:wtsa.ll.sig} for non-parametric tests and Table \ref{tab:wtsna.ord} for \gls{mnl}. \\
          \cmidrule(lr){2-6}
          & \makecell{\gls{wtpa} for\\ Anonymisation} & - & Marginally significant difference for daily and anonymised half-hourly sharing for \gls{wtp} and daily sharing for \gls{wta} in the full sample analysis. Similar results observed when excluding the price mis-interpretation correction with only the \gls{wtp} for daily sharing being marginally significant. & Limited support for \textbf{H6}. & Table \ref{tab:poe.freq} and \ref{tab:poe.freq.robust} for $p$-values of combinatorial tests.\\
          \cmidrule(lr){2-6}
          & \makecell{Demand for \\Smart Metering} & - & No significant treatment coefficient in binary logistic regression or non-parametric tests. & \textbf{H6} is not supported. & Table \ref{tab:wsm.gp.sig} for non-parametric tests and Table \ref{tab:wsm.lr} for binary logistic regression model.\\
         \cmidrule(lr){1-6}

         \textbf{H7} & \multirow{2}{*}{\makecell{Change in \gls{wts}\\ Non-Anonymised}} & \gls{dsa} & Significant difference in estimated marginal mean probabilities for BA with smaller insignificant differences for MR and TP. No significant difference in pairwise Dunn's tests. & \textbf{H7} is supported with the treatment effect reducing as general data privacy concerns reduce. & Table \ref{tab:wtsn.ll.sig} for non-parametric tests and Table \ref{tab:wtsna.ord} for \gls{mnl}. \\
         \cmidrule(lr){3-5}
          & & \gls{iwts} & Significant difference in estimated marginal mean probabilities for those initially unwilling with smaller insignificant differences for those indifferent or willing. No significant difference in pairwise Dunn's tests. & \textbf{H7} is supported with a lower \gls{iwts} resulting in a higher treatment effect. & \\
         \cmidrule(lr){2-6}
          &  \multirow{2}{*}{\makecell{Change in \gls{wts}\\ Anonymised}} &\gls{dsa} & Significant difference in estimated marginal mean probabilities for BA with smaller significant differences for TP. Significant difference in pairwise Dunn's tests for BA and TP. & \textbf{H7} is supported with the treatment effect reducing as general data privacy concerns reduce. & Table \ref{tab:wtsn.ll.sig} for non-parametric tests and Table \ref{tab:wtsna.ord} for \gls{mnl}. \\
          \cmidrule(lr){3-5}
          & & \gls{iwts} & Significant difference in estimated marginal mean probabilities for those initially unwilling and those who were indifferent with smaller insignificant differences for those initially willing. Significant difference in pairwise Dunn's tests for those unwilling and indifferent. &  \textbf{H7} is supported with a lower \gls{iwts} resulting in a higher treatment effect.\\
          \cmidrule(lr){2-6}
          & \multirow{2}{*}{\makecell{\gls{wtpa} for\\ Anonymisation}}& \gls{dsa} & A significant positive treatment effect for BM (all resolutions) and a negative effect for TP (real-time and daily) were observed for anonymised sharing. The treatment effect is greater for BM at all resolutions expect non-anonymised daily sharing. The same results are observed with the model without price mis-interpretation correction. & \textbf{H7} is supported. &  Table \ref{tab:wtpa.conf} for \gls{wtpa} estimates and Table \ref{tab:poe.freq} and \ref{tab:poe.freq.robust} for $p$-values of combinatorial tests.\\
          \cmidrule(lr){2-6}
          & \multirow{2}{*}{\makecell{Demand for \\Smart Metering}} &\gls{dsa} & No significant difference in pairwise Dunn's tests and no significant treatment interaction coefficients in binary logistic regression. & \textbf{H7} is not supported. & Table \ref{tab:wsm.gp.sig} for non-parametric tests and Table \ref{tab:wsm.lr} for binary logistic regression model.\\
          \cmidrule(lr){3-5}
          & & \gls{iwts} & No significant difference in pairwise Dunn's tests and no significant treatment interaction coefficients in binary logistic regression. & \textbf{H7} is not supported. & \\
         \cmidrule(lr){1-6}
         \textbf{H8} & \makecell{Change in \gls{wts}} & - &  Treatment effect for non-anonymised sharing is insignificant with a mean difference in the estimated marginal mean probability being 3.5\% [-2.7\%, 9.7\%], whereas, for anonymised sharing it is 9.0\% [5.1\%, 12.9\%]. & \textbf{H8} is not supported. & See Figure \ref{fig:wtsna.treat}.\\
\end{longtable}
\end{landscape}

\newpage
\section{Supporting Tables for Willingness-to-Share Analysis}\label{app:wts}
\begin{table}[H]
\centering
\caption{Non-Parametric Testing of Willingness-to-Share Responses} 
\label{tab:wtsna.sig}
\setlength\targetwidth{\textwidth}
\begin{tabularx}{\targetwidth}{@{}CCCCCCCCCC@{}}
    \toprule
    \multicolumn{2}{c}{\multirow{2}{*}{Grouping}} & \multirow{2}{*}{\makecell{Test\\Stat}} & \multicolumn{3}{c}{$p$} & \multicolumn{4}{c}{Effect} \\ 
    \cmidrule(lr){4-6} \cmidrule(lr){7-10}
     &  &  & unadj. & adj. & sig. & Est. & Low & High & Size \\ 
    \midrule
    \multicolumn{10}{l}{\textbf{Equality of Proportions:} Pearson's Chi Square Goodness-of-Fit Test}\\
    \cmidrule(lr){1-10}
    \multicolumn{2}{c}{WTSG(4)}  & 278.58 & 0.000 &  & *** & 0.47 & 0.43 & 0.51  & v. large\\ 
      \multicolumn{2}{c}{WTSN(2)}  & 65.37 & 0.000 &  & *** & 0.25 & 0.19 & 0.31 & medium\\ 
      \multicolumn{2}{c}{WTSA(2)}  & 199.54 & 0.000 &  & *** & 0.41 & 0.36 & 0.46 & v. large\\ 
      \cmidrule(lr){1-10}
      \multicolumn{9}{l}{\textbf{Difference in Response:} Wilcoxon Signed-Rank Test}\\
      \cmidrule(lr){1-10}
       WTSG & 0& 54519.5 & 0.000 &  & *** & -0.60 & -0.65 & -0.54 & large \\ 
       WTSN & 0 & 69958 & 0.760 &  & n.s. &  &  &  & \\ 
       WTSA &0  & 30186 & 0.000 &  & *** & -0.55 & -0.61 & -0.47 & large\\ 
       WTSA & WTSN & 6710 & 0.000 &  & *** & -0.71 & -0.74 & -0.67 & large \\ 
      \cmidrule(lr){1-10}
      \multicolumn{10}{l}{\textbf{Change in Response due to Anonymisation}: McNemar-Bowker Test}\\
      \cmidrule(lr){1-10}
      \multicolumn{2}{c}{Omnibus(3)}  & 131.51 & 0.000 &  & *** & 0.28 & 0.23 & 0.32 & large\\ 
      More  & No Diff. & 2.84 & 0.092 & 0.092 & + & 0.07 & -0.01 & 0.16 &  small\\ 
      More & Less & 110.77 & 0.000 & 0.000 & *** & 0.46 & 0.41 & 0.48 & large\\ 
      No Diff. & Less & 17.89 & 0.000 & 0.000 & *** & 0.31 & 0.17 & 0.40 & large\\ 
    \bottomrule  \multicolumn{10}{@{}P@{}}{
  Note: + $p<$ 0.1, * $p<$ 0.05, ** $p<$ 0.01, *** $p<$ 0.001.
  For Goodness-of-fit tests the $\chi^2_P$ is reported and the effect size is Pearson's $C$ with interpretation based on \cite{Funder2019}. For Wilcoxon signed-rank tests the effect sizes are given by the biserial rank correlation $r^{rb}=1-\frac{2W}{n_1*n_2}$ with interpretation based on \cite{Vargha2000}. Omnibus McNemar-Bowker test for symmetry on 3x3 table is followed by pairwise 2x2 McNemar tests with Holm correction for multiple comparisons. The $\chi^2_{MN}$ is reported and effect size is given by Cohen's $g$ with interpretation based on \cite{Cohen2013}.} 
\end{tabularx}
\end{table}

\begin{table}[H]
\centering
\caption{Non-Parametric Testing of Heterogeneity in Initial WTS} 
\label{tab:wtsg.sig}
\setlength\targetwidth{\textwidth} 
\begin{tabularx}{\targetwidth}{@{}CCCCCCCCCC@{}}
  \toprule
\multicolumn{2}{c}{\multirow{2}{*}{Grouping}} & \multirow{2}{*}{\makecell{Test\\Stat}} & \multicolumn{3}{c}{$p$} & \multicolumn{4}{c}{Effect} \\ 
\cmidrule(lr){4-6} \cmidrule(lr){7-10}
 &  &  & unadj. & adj. & sig. & Est. & Low & High & Size \\ 
  \midrule
  \multicolumn{10}{l}{\textbf{Age (AGE)}} \\
  \cmidrule(lr){1-10}
\multicolumn{2}{c}{Omnibus (3)}  & 2.78 & 0.427 &  & n.s. &  &  &  &  \\ 
  \cmidrule(lr){1-10}
  \multicolumn{10}{l}{\textbf{Gender (GEN)}}\\
  \cmidrule(lr){1-10}
   M & F & 109016 & 0.078 &  & + & -0.06 & -0.14 & 0.01 & neg. \\ 
  \cmidrule(lr){1-10}
  \multicolumn{10}{l}{\textbf{Socio-Economic Group (SEG)}}\\
  \cmidrule(lr){1-10}
  \multicolumn{2}{c}{Omnibus (3)}  & 15.13 & 0.002 &  & ** & 0.02 & 0.01 & 1.00 & small \\ 
  AB & C1 & 1.33 & 0.183 & 0.549 & n.s. &  &  &  &  \\ 
  AB & C2 & 2.39 & 0.017 & 0.077 & + & -0.13 & -0.23 & -0.02 & small \\ 
  AB & DE & 3.71 & 0.000 & 0.001 & ** & -0.18 & -0.28 & -0.08 & small \\ 
  C1 & C2 & 1.17 & 0.244 & 0.549 & n.s. &  &  &  &  \\ 
  C1 & DE & 2.42 & 0.015 & 0.077 & + & -0.12 & -0.21 & -0.02 & small \\ 
  C2 & DE & 1.09 & 0.275 & 0.549 & n.s. &  &  &  &  \\ 
  \cmidrule(lr){1-10}
\multicolumn{10}{l}{\textbf{Smart Meter Ownership (SM)}}\\
  \cmidrule(lr){1-10}
  Yes & No  &  80827 & 0.000 &  & *** & -0.30 & -0.37 & -0.23 & medium \\ 
  \cmidrule(lr){1-10}
\multicolumn{10}{l}{\textbf{Data Sharing Attitude (DSA)}}\\
  \cmidrule(lr){1-10}
  \multicolumn{2}{c}{Omnibus (2)} &  9.38 & 0.009 &  & ** & 0.01 & 0.00 & 1.00 & v. small \\ 
  BA & MR & 1.26 & 0.209 & 0.419 & n.s. &  &  &  &  \\ 
  BA & TP & 3.05 & 0.002 & 0.007 & ** & 0.12 & 0.04 & 0.21 & small \\ 
  MR & TP & 1.18 & 0.238 & 0.419 & n.s. &  &  &  &  \\ 
    \cmidrule(lr){1-10}
  \multicolumn{10}{l}{\textbf{IHD Engagement (IHD)}}\\
  \cmidrule(lr){1-10}
    Low & High & 108340 & 0.000 &  & *** & 0.31 & 0.23 & 0.39 & medium \\ 
      \cmidrule(lr){1-10}
  \multicolumn{10}{l}{\textbf{Time-Varying Tariff (TVT)}}\\
  \cmidrule(lr){1-10}
  No & Yes & 41391.5 & 0.854 &  & n.s. &  &  &  &  \\ 
   \bottomrule  \multicolumn{10}{@{}P@{}}{
   Note: + $p<$ 0.1, * $p<$ 0.05, ** $p<$ 0.01, *** $p<$ 0.001. Kruskal-Wallis (KW) rank-sum omnibus test for groupings with $\geq $3 (df in brackets). Effect size given by $\hat\epsilon^2_{ord} = \frac{\chi^2_{KW}(n+1)}{(n^2-1)}$, where, $\chi^2_{KW}$ is the test statistic. Effect size interpretation is based on \cite{Field2013}. Dunn's test for post-hoc pairwise comparisons with Holm correction for multiple comparisons (adj.). Here the $z$ test statistic is reported and effect size is the biserial rank correlation $r^{rb}=1-\frac{2W_{MW}}{n_1*n_2}$, where, $W_{MW}$ is the test statistic for corresponding Mann-Whitney (MW) U Test. Effect size interpretation based on \cite{Cohen2013}. For dichotomous groupings $W_{MW}$, and $r^{rb}$ are reported.}
\end{tabularx}
\end{table}

\begin{table}[H]
\centering
\caption{Non-Parametric Testing of Heterogeneity in Change in WTS for Non-Anonymised Data} 
\label{tab:wtsn.sig}
\setlength\targetwidth{\textwidth} 
\resizebox{\textwidth}{!}{
\begin{tabularx}{\targetwidth}{@{}CCCCCCCCCC@{}}
  \toprule
\multicolumn{2}{c}{\multirow{2}{*}{Grouping}} & \multirow{2}{*}{\makecell{Test\\Stat}} & \multicolumn{3}{c}{$p$} & \multicolumn{4}{c}{Effect} \\ 
\cmidrule(lr){4-6} \cmidrule(lr){7-10}
 &  &  & unadj. & adj. & sig. & Est. & Low & High & Size \\ 
  \midrule
  \multicolumn{10}{l}{\textbf{Age (AGE)}} \\
  \cmidrule(lr){1-10}
\multicolumn{2}{c}{Omnibus (3)} & 91.37 & 0.000 &  & *** & 0.09 & 0.07 & 1.00 & medium \\ 
  18-34 & 35-54 & 3.84 & 0.000 & 0.000 & *** & -0.17 & -0.26 & -0.08 & small \\ 
  18-34 & 55-64 & 6.63 & 0.000 & 0.000 & *** & -0.37 & -0.47 & -0.27 & medium \\ 
  18-34 & 65+ & 8.78 & 0.000 & 0.000 & *** & -0.42 & -0.50 & -0.33 & medium \\ 
  35-54 & 55-64 & 3.67 & 0.000 & 0.000 & *** & -0.20 & -0.30 & -0.09 & small \\ 
  35-54 & 65+ & 5.56 & 0.000 & 0.000 & *** & -0.26 & -0.35 & -0.17 & small \\ 
  55-64 & 65+ & 1.13 & 0.257 & 0.257 & n.s. &  &  &  &  \\ 
  \cmidrule(lr){1-10}
  \multicolumn{10}{l}{\textbf{Gender (GEN)}}\\
  \cmidrule(lr){1-10}
  M & F & 107074 & 0.021 &  & * & -0.08 & -0.15 & -0.01 & neg. \\ 
  \cmidrule(lr){1-10}
  \multicolumn{10}{l}{\textbf{Socio-Economic Group (SEG)}}\\
  \cmidrule(lr){1-10}
  \multicolumn{2}{c}{Omnibus (3)}   & 9.49 & 0.023 &  & * & 0.01 & 0.00 & 1.00 & v. small \\ 
  AB & C1 & 0.96 & 0.337 & 1.000 & n.s. &  &  &  &  \\ 
  AB & C2 & 0.38 & 0.701 & 1.000 & n.s. &  &  &  &  \\ 
  AB & DE & 2.84 & 0.005 & 0.027 & * & -0.14 & -0.23 & -0.04 & small \\ 
  C1 & C2 & 0.52 & 0.600 & 1.000 & n.s. &  &  &  &  \\ 
  C1 & DE & 1.92 & 0.055 & 0.220 & n.s. &  &  &  &  \\ 
  C2 & DE & 2.32 & 0.020 & 0.102 & n.s. &  &  &  &  \\ 
  \cmidrule(lr){1-10}
\multicolumn{10}{l}{\textbf{Smart Meter Ownership (SM)}}\\
  \cmidrule(lr){1-10}
  Yes & No & 92952.5 & 0.000 &  & *** & -0.20 & -0.27 & -0.13 & small \\ 
  \cmidrule(lr){1-10}
\multicolumn{10}{l}{\textbf{Data Sharing Attitude (DSA)}}\\
  \cmidrule(lr){1-10}
  \multicolumn{2}{c}{Omnibus (2)}  & 53.94 & 0.000 &  & *** & 0.06 & 0.03 & 1.00 & small \\ 
  BA & MR & 0.51 & 0.609 & 0.609 & n.s. &  &  &  &  \\ 
  BA & TP & 6.62 & 0.000 & 0.000 & *** & 0.26 & 0.18 & 0.34 & small \\ 
  MR & TP & 4.99 & 0.000 & 0.000 & *** & 0.24 & 0.14 & 0.33 & small \\ 
  \cmidrule(lr){1-10}
  \multicolumn{10}{l}{\textbf{IHD Engagement (IHD)}}\\
  \cmidrule(lr){1-10}
    Low & High & 98391.5 & 0.000 &  & *** & 0.19 & 0.10 & 0.27 & small \\ 
  \cmidrule(lr){1-10}
  \multicolumn{10}{l}{\textbf{Time-Varying Tariff (TVT)}}\\
  \cmidrule(lr){1-10}
  No & Yes & 51010 & 0.000 &  & *** & 0.25 & 0.13 & 0.36 & small \\ 
   \bottomrule  \multicolumn{10}{@{}P@{}}{
 Note: + $p<$ 0.1, * $p<$ 0.05, ** $p<$ 0.01, *** $p<$ 0.001. Kruskal-Wallis (KW) rank-sum omnibus test for groupings with $\geq $3 (df in brackets). Effect size given by $\hat\epsilon^2_{ord} = \frac{\chi^2_{KW}(n+1)}{(n^2-1)}$, where, $\chi^2_{KW}$ is the test statistic. Effect size interpretation is based on \cite{Field2013}. Dunn's test for post-hoc pairwise comparisons with Holm correction for multiple comparisons (adj.). Here the $z$ test statistic is reported and effect size is the biserial rank correlation $r^{rb}=1-\frac{2W_{MW}}{n_1*n_2}$, where, $W_{MW}$ is the test statistic for corresponding Mann-Whitney (MW) U Test. Effect size interpretation based on \cite{Cohen2013}. For dichotomous groupings $W_{MW}$, and $r^{rb}$ are reported.} 
\end{tabularx}
}
\end{table}

\begin{table}[H]
\centering
\caption{Non-Parametric Testing of Heterogeneity in Change in WTS for Anonymised Data} 
\label{tab:wtsa.sig}
\setlength\targetwidth{\textwidth} 
\resizebox{\textwidth}{!}{
\begin{tabularx}{\targetwidth}{@{}CCCCCCCCCC@{}}
   \toprule
\multicolumn{2}{c}{\multirow{2}{*}{Grouping}} & \multirow{2}{*}{\makecell{Test\\Stat}} & \multicolumn{3}{c}{$p$} & \multicolumn{4}{c}{Effect} \\ 
\cmidrule(lr){4-6} \cmidrule(lr){7-10}
 &  &  & unadj. & adj. & sig. & Est. & Low & High & Size \\ 
  \midrule
\multicolumn{10}{l}{\textbf{Age (AGE)}} \\
  \cmidrule(lr){1-10}
\multicolumn{2}{c}{Omnibus (3)}  & 28.64 & 0.000 &  & *** & 0.03 & 0.02 & 1.00 & small \\ 
  18-34 & 35-54 & 1.25 & 0.210 & 0.248 & n.s. &  &  &  &  \\ 
  18-34 & 55-64 & 4.72 & 0.000 & 0.000 & *** & -0.26 & -0.36 & -0.15 & small \\ 
  18-34 & 65+ & 3.53 & 0.000 & 0.002 & ** & -0.17 & -0.27 & -0.07 & small \\ 
  35-54 & 55-64 & 3.86 & 0.000 & 0.001 & *** & -0.20 & -0.30 & -0.09 & small \\ 
  35-54 & 65+ & 2.52 & 0.012 & 0.035 & * & -0.11 & -0.21 & -0.02 & small \\ 
  55-64 & 65+ & 1.54 & 0.124 & 0.248 & n.s. &  &  &  &  \\ 
  \cmidrule(lr){1-10}
  \multicolumn{10}{l}{\textbf{Gender (GEN)}}\\
  \cmidrule(lr){1-10}
  M & F & 110488.5 & 0.137 &  & n.s. &  &  &  &  \\ 
  \cmidrule(lr){1-10}
  \multicolumn{10}{l}{\textbf{Socio-Economic Group (SEG)}}\\
  \cmidrule(lr){1-10}
  \multicolumn{2}{c}{Omnibus (3)}   & 13.79 & 0.003 &  & ** & 0.01 & 0.00 & 1.00 & small \\ 
  AB & C1 & 0.75 & 0.451 & 0.591 & n.s. &  &  &  &  \\ 
  AB & C2 & 1.94 & 0.053 & 0.211 & n.s. &  &  &  &  \\ 
  AB & DE & 3.43 & 0.001 & 0.004 & ** & -0.16 & -0.26 & -0.06 & small \\ 
  C1 & C2 & 1.26 & 0.209 & 0.591 & n.s. &  &  &  &  \\ 
  C1 & DE & 2.74 & 0.006 & 0.031 & * & -0.13 & -0.22 & -0.03 & small \\ 
  C2 & DE & 1.29 & 0.197 & 0.591 & n.s. &  &  &  &  \\ 
 \cmidrule(lr){1-10}
\multicolumn{10}{l}{\textbf{Smart Meter Ownership (SM)}}\\
  \cmidrule(lr){1-10}
  Yes & No  & 103285 & 0.002 &  & ** & -0.11 & -0.18 & -0.03 & neg. \\ 
  \cmidrule(lr){1-10}
\multicolumn{10}{l}{\textbf{Data Sharing Attitude (DSA)}}\\
  \cmidrule(lr){1-10}
  \multicolumn{2}{c}{Omnibus (2)}   & 2.78 & 0.249 &  & n.s. &  &  &  &  \\ 
  \cmidrule(lr){1-10}
  \multicolumn{10}{l}{\textbf{IHD Engagement (IHD)}}\\
  \cmidrule(lr){1-10}
    Low & High & 92376.5 & 0.004 &  & ** & 0.12 & 0.03 & 0.20 & small \\ 
  \cmidrule(lr){1-10}
  \multicolumn{10}{l}{\textbf{Time-Varying Tariff (TVT)}}\\
  \cmidrule(lr){1-10}
  No & Yes & 43696.5 & 0.237 &  & n.s. &  &  &  &  \\ 
  \cmidrule(lr){1-10}
  \multicolumn{10}{l}{\textbf{Want a Smart Meter (WSM)}}\\
  \cmidrule(lr){1-10}
  No & Yes & 72742 & 0.000 &  & *** &  -0.284 & -0.354 & -0.210 &  medium\\ 
   \bottomrule  \multicolumn{10}{@{}P@{}}{
 Note: + $p<$ 0.1, * $p<$ 0.05, ** $p<$ 0.01, *** $p<$ 0.001. Kruskal-Wallis (KW) rank-sum omnibus test for groupings with $\geq $3 (df in brackets). Effect size given by $\hat\epsilon^2_{ord} = \frac{\chi^2_{KW}(n+1)}{(n^2-1)}$, where, $\chi^2_{KW}$ is the test statistic. Effect size interpretation is based on \cite{Field2013}. Dunn's test for post-hoc pairwise comparisons with Holm correction for multiple comparisons (adj.). Here the $z$ test statistic is reported and effect size is the biserial rank correlation $r^{rb}=1-\frac{2W_{MW}}{n_1*n_2}$, where, $W_{MW}$ is the test statistic for corresponding Mann-Whitney (MW) U Test. Effect size interpretation based on \cite{Cohen2013}. For dichotomous groupings $W_{MW}$, and $r^{rb}$ are reported.} 
\end{tabularx}
}
\end{table}

\begin{table}[H]
\centering
\caption{Non-Parametric Testing of Treatment Effect on WTS for Non-Anonymised Data} 
\label{tab:wtsn.gp.sig}
\setlength\targetwidth{\textwidth} 
\begin{tabularx}{\targetwidth}{@{}CCCCCCCCCCC@{}}
  \toprule
\multicolumn{2}{c}{Grouping} & \multirow{2}{*}{\makecell{Test\\Stat}} & \multicolumn{4}{c}{$p$-value} & \multicolumn{4}{c}{Effect}\\
\cmidrule(lr){4-7} \cmidrule(lr){8-11}
 &  &  & unadj. & adj. & full & sig. & Est. & Low & High & Size \\ 
  \midrule
Control & Treatment & 117613 & 0.761 &  &  & n.s. &  &  &  &  \\ 
\cmidrule(lr){1-11}
\multicolumn{11}{l}{\textbf{Data Sharing Attitudes (DSA)}}\\
\cmidrule(lr){1-11}
\multicolumn{2}{c}{Omnibus (5)}  & 54.36 & 0.000 &  &  & *** & 0.06 & 0.04 & 1.00 & small \\ 
  \multicolumn{2}{c}{Basic} & 0.30 & 0.761 & 1.000 & 1.000 & n.s. &  &  &  &  \\ 
 \multicolumn{2}{c}{Marketing} & 0.55 & 0.582 & 1.000 & 1.000 & n.s. &  &  &  &  \\ 
  \multicolumn{2}{c}{Third Party}& 0.15 & 0.881 & 1.000 & 1.000 & n.s. &  &  &  &  \\ 
  \cmidrule(lr){1-11}
\multicolumn{11}{l}{\textbf{Initial Willingness-to-Share (IWTS)}}\\
\cmidrule(lr){1-11}
  \multicolumn{2}{c}{Omnibus (9)} & 54.88 & 0.000 &  &  & *** & 0.06 & 0.04 & 1.00 & small \\ 
  \multicolumn{2}{c}{Not at all willing} & 0.87 & 0.383 & 1.000 & 1.000 & n.s. &  &  &  &  \\ 
  \multicolumn{2}{c}{Not very willing} & 0.86 & 0.391 & 1.000 & 1.000 & n.s. &  &  &  &  \\ 
  \multicolumn{2}{c}{Indifferent} & 0.05 & 0.964 & 1.000 & 1.000 & n.s. &  &  &  &  \\ 
  \multicolumn{2}{c}{Quite willing} & 0.63 & 0.527 & 1.000 & 1.000 & n.s. &  &  &  &  \\ 
  \multicolumn{2}{c}{Very willing} & 0.59 & 0.557 & 1.000 & 1.000 & n.s. &  &  &  &  \\ 
   \bottomrule  \multicolumn{11}{@{}P@{}}{
  Note: + $p<$ 0.1, * $p<$ 0.05, ** $p<$ 0.01, *** $p<$ 0.001. Kruskal-Wallis (KW) rank-sum omnibus test for groupings with $\geq $3 (df in brackets). Effect size given by $\hat\epsilon^2_{ord} = \frac{\chi^2_{KW}(n+1)}{(n^2-1)}$, where, $\chi^2_{KW}$ is the test statistic. Effect size interpretation is based on \cite{Field2013}. Dunn's test for post-hoc pairwise comparisons with Holm correction for multiple comparisons across relevant comparisons (adj.) and all comparisons (full). Here the $z$ test statistic is reported and effect size is the biserial rank correlation $r^{rb}=1-\frac{2W_{MW}}{n_1*n_2}$, where, $W_{MW}$ is the test statistic for corresponding Mann-Whitney (MW) U Test. Effect size interpretation based on \cite{Cohen2013}. For dichotomous groupings $W_{MW}$, and $r^{rb}$ are reported.} 
\end{tabularx}
\end{table}

\begin{table}[H]
\centering
\caption{Non-Parametric Test of Treatment Effect on being Less Likely to Share for Non-Anonymised Data} 
\label{tab:wtsn.ll.sig}
\setlength\targetwidth{\textwidth} 
\begin{tabularx}{\targetwidth}{@{}CCCCCCCCCCC@{}}
 \toprule
\multicolumn{2}{c}{Grouping} & \multirow{2}{*}{\makecell{Test\\Stat}} & \multicolumn{4}{c}{$p$-value} & \multicolumn{4}{c}{Effect}\\
\cmidrule(lr){4-7} \cmidrule(lr){8-11}
 &  &  & unadj. & adj. & full & sig. & Est. & Low & High & Size \\ 
  \midrule
Control & Treatment & 0.25 & 0.616 &  &  & n.s. &  &  &  &  \\ 
   \cmidrule(lr){1-11}
\multicolumn{11}{l}{\textbf{Data Sharing Attitudes (DSA)}}\\
\cmidrule(lr){1-11}
\multicolumn{2}{c}{Omnibus (5)} & 48.91 & 0.000 &  &  & *** & 0.21 & 0.13 & 0.27 & medium \\ 
  \multicolumn{2}{c}{Basic} & 1.86 & 0.173 & 0.518 & 1.000 & n.s. &  &  &  &  \\ 
  \multicolumn{2}{c}{Marketing} & 0.05 & 0.821 & 1.000 & 1.000 & n.s. &  &  &  &  \\ 
  \multicolumn{2}{c}{Third Party} & 0.04 & 0.851 & 1.000 & 1.000 & n.s. &  &  &  &  \\  \cmidrule(lr){1-11}
\multicolumn{11}{l}{\textbf{Initial Willingness-to-Share (IWTS)}}\\
\cmidrule(lr){1-11}
  \multicolumn{2}{c}{Omnibus (9)}& 38.98 & 0.000 &  &  & *** & 0.18 & 0.05 & 0.23 & small \\ 
  \multicolumn{2}{c}{Not at all willing} & 3.13 & 0.077 & 0.385 & 1.000 & n.s. &  &  &  &  \\ 
  \multicolumn{2}{c}{Not very willing} & 0.72 & 0.395 & 1.000 & 1.000 & n.s. &  &  &  &  \\ 
  \multicolumn{2}{c}{Indifferent}  & 0.21 & 0.646 & 1.000 & 1.000 & n.s. &  &  &  &  \\ 
  \multicolumn{2}{c}{Quite willing} & 0.23 & 0.632 & 1.000 & 1.000 & n.s. &  &  &  &  \\ 
  \multicolumn{2}{c}{Very willing} & 0.03 & 0.852 & 1.000 & 1.000 & n.s. &  &  &  &  \\ 
   \bottomrule  \multicolumn{11}{@{}P@{}}{
  Note: + $p<$ 0.1, * $p<$ 0.05, ** $p<$ 0.01, *** $p<$ 0.001. Pearson's Chi-squared tests for independence with post-hoc pairwise testing of subgroups across control and treatment groups. $p$-values adjusted for multiple comparisons across relevant groups (adj.) and all possible comparisons (full). Effect size given by Cramer's V, $V = \sqrt{\frac{\chi^2}{n(k-1)}}$, where $\chi^2$ is the test statistic, $n$ is the sample size, and $k$ is the number of groupings. Interpretation based on \cite{Funder2019}.}
\end{tabularx}
\end{table}

\begin{table}[H]
\centering
\caption{Non-Parametric Testing of Treatment Effect on WTS for Anonymised Data} 
\label{tab:wtsa.gp.sig}
\setlength\targetwidth{\textwidth} 
\begin{tabularx}{\targetwidth}{@{}CCCCCCCCCCC@{}}
   \toprule
\multicolumn{2}{c}{Grouping} & \multirow{2}{*}{\makecell{Test\\Stat}} & \multicolumn{4}{c}{$p$-value} & \multicolumn{4}{c}{Effect}\\
\cmidrule(lr){4-7} \cmidrule(lr){8-11}
 &  &  & unadj. & adj. & full & sig. & Est. & Low & High & Size \\ 
  \midrule
Control & Treatment & 110087 & 0.109 &  &  & n.s. &  &  &  &  \\ 
  \cmidrule(lr){1-11}
\multicolumn{11}{l}{\textbf{Data Sharing Attitudes (DSA)}}\\
\cmidrule(lr){1-11}
\multicolumn{2}{c}{Omnibus (5)} &  6.50 & 0.260 &  &  & n.s. &  &  &  &  \\ 
  \cmidrule(lr){1-11}
\multicolumn{11}{l}{\textbf{Initial Willingness-to-Share (IWTS)}}\\
\cmidrule(lr){1-11}
  \multicolumn{2}{c}{Omnibus (9)} & 118.15 & 0.000 &  &  & *** & 0.12 & 0.10 & 1.00 & medium \\ 
  \multicolumn{2}{c}{Not at all willing} & 1.94 & 0.052 & 0.262 & 0.833 & n.s. &  &  &  &  \\ 
  \multicolumn{2}{c}{Not very willing} & 1.81 & 0.071 & 0.282 & 0.918 & n.s. &  &  &  &  \\ 
  \multicolumn{2}{c}{Indifferent} & 1.15 & 0.249 & 0.747 & 1.000 & n.s. &  &  &  &  \\ 
  \multicolumn{2}{c}{Quite willing} & 0.20 & 0.839 & 1.000 & 1.000 & n.s. &  &  &  &  \\ 
  \multicolumn{2}{c}{Very willing} & 0.41 & 0.680 & 1.000 & 1.000 & n.s. &  &  &  &  \\ 
   \bottomrule  \multicolumn{11}{@{}P@{}}{
  Note: + $p<$ 0.1, * $p<$ 0.05, ** $p<$ 0.01, *** $p<$ 0.001. Kruskal-Wallis (KW) rank-sum omnibus test for groupings with $\geq $3 (df in brackets). Effect size given by $\hat\epsilon^2_{ord} = \frac{\chi^2_{KW}(n+1)}{(n^2-1)}$, where, $\chi^2_{KW}$ is the test statistic. Effect size interpretation is based on \cite{Field2013}.  Dunn's test for post-hoc pairwise comparisons with Holm correction for multiple comparisons across relevant comparisons (adj.) and all comparisons (full). Here the $z$ test statistic is reported and effect size is the biserial rank correlation $r^{rb}=1-\frac{2W_{MW}}{n_1*n_2}$, where, $W_{MW}$ is the test statistic for corresponding Mann-Whitney (MW) U Test. Effect size interpretation based on \cite{Cohen2013}. For dichotomous groupings $W_{MW}$, and $r^{rb}$ are reported.}
\end{tabularx}
\end{table}

\begin{table}[H]
\centering
\caption{Non-Parametric Test Treatment Effect on being Less Likely to Share for Anonymised Data} 
\label{tab:wtsa.ll.sig}
\setlength\targetwidth{\textwidth} 
\begin{tabularx}{\targetwidth}{@{}CCCCCCCCCCC@{}}
   \toprule
\multicolumn{2}{c}{\multirow{2}{*}{Grouping}} & \multirow{2}{*}{\makecell{Test\\Stat}} & \multicolumn{4}{c}{$p$-value} & \multicolumn{4}{c}{Effect}\\
\cmidrule(lr){4-7} \cmidrule(lr){8-11}
 &  &  & unadj. & adj. & full & sig. & Est. & Low & High & Size \\ 
  \midrule
Control & Treatment & 13.80 & 0.000 &  &  & *** & 0.12 & 0.05 & 0.18 & small \\ 
  \cmidrule(lr){1-11}
\multicolumn{11}{l}{\textbf{Data Sharing Attitudes (DSA)}}\\
\cmidrule(lr){1-11}
\multicolumn{2}{c}{Omnibus (5)} & 27.26 & 0.000 &  &  & *** & 0.15 & 0.05 & 0.21 & small \\ 
  \multicolumn{2}{c}{Basic} & 10.90 & 0.001 & 0.003 & 0.013 & ** & 0.19 & 0.05 & 0.31 & small \\ 
  \multicolumn{2}{c}{Marketing} & 0.57 & 0.452 & 0.452 & 1.000 & n.s. &  &  &  &  \\ 
  \multicolumn{2}{c}{Third Party} & 5.39 & 0.020 & 0.040 & 0.222 & * & 0.09 & 0.00 & 0.18 & v. small \\ 
   \cmidrule(lr){1-11}
\multicolumn{11}{l}{\textbf{Initial Willingness-to-Share (IWTS)}}\\
\cmidrule(lr){1-11}
  \multicolumn{2}{c}{Omnibus (9)} & 132.66 & 0.000 &  &  & *** & 0.36 & 0.28 & 0.41 & large \\ 
  \multicolumn{2}{c}{Not at all willing}& 6.03 & 0.014 & 0.056 & 0.337 & + & 0.26 & 0.00 & 0.51 & medium \\ 
  \multicolumn{2}{c}{Not very willing} & 7.24 & 0.007 & 0.036 & 0.193 & * & 0.30 & 0.00 & 0.55 & large \\ 
  \multicolumn{2}{c}{Indifferent}  & 4.61 & 0.032 & 0.095 & 0.646 & + & 0.13 & 0.00 & 0.26 & small \\ 
  \multicolumn{2}{c}{Quite willing}  & 1.01 & 0.316 & 0.628 & 1.000 & n.s. &  &  &  &  \\ 
 \multicolumn{2}{c}{Very willing} & 1.01 & 0.314 & 0.628 & 1.000 & n.s. &  &  &  &  \\ 
   \bottomrule  \multicolumn{11}{@{}P@{}}{
  Note: + $p<$ 0.1, * $p<$ 0.05, ** $p<$ 0.01, *** $p<$ 0.001. Pearson's Chi-squared tests for independence with post-hoc pairwise testing of subgroups across control and treatment groups. $p$-values adjusted for multiple comparisons across relevant groups (adj.) and all possible comparisons (full). Effect size given by Cramer's V, $V = \sqrt{\frac{\chi^2}{n(k-1)}}$, where $\chi^2$ is the test statistic, $n$ is the sample size, and $k$ is the number of groupings. Interpretation based on \cite{Funder2019}.}
\end{tabularx}
\end{table}

\begin{landscape}
\begin{table}[H]
\centering
\caption{Logistic Regression Models for Initial Willingness-to-Share} 
\label{tab:wtsg.lr}
\setlength\targetwidth{1.5\textwidth} 
\resizebox{0.9\textwidth}{!}{
\begin{tabularx}{\targetwidth}{@{} rCCCCCCCCCCCCC @{}}
    \toprule
    & \multirow{2}{*}{\makecell{PO \\ Model}} & \multicolumn{4}{c}{Partial PO Model} & \multicolumn{4}{c}{Binary Logistic Models} & \multicolumn{4}{c}{Multinomial Logit Model} \\
    \cmidrule(lr){3-6} \cmidrule(lr){7-10} \cmidrule(lr){11-14}
    &           & NVW       & IND       & QW        & VW        & $>$ NAW   & $>$ NVW & $>$ IND & $>$ QW   & NVW     & IND     & QW      & VW \\ 
    \midrule
    \multicolumn{14}{l}{Covariates}\\
    \midrule
    GEN Female & -0.251*   & 0.086     & 0.037     & -0.188    & -0.463**  & 0.142     & 0.087   & -0.188  & -0.482** & 0.108   & 0.342   & 0.253   & -0.244  \\ 
               & (0.121)   & (0.254)   & (0.189)   & (0.141)   & (0.152)   & (0.264)   & (0.193) & (0.143) & (0.154)  & (0.353) & (0.288) & (0.280) & (0.289) \\ 
   
    SM No      & -0.819*** & -1.178*** & -0.838*** & -0.817*** & -0.698*** & -1.293*** & -0.902*** & -0.809*** & -0.746*** & -0.809+ & -0.892* & -1.363*** & -1.746*** \\ 
               & (0.141)   & (0.342)   & (0.230)   & (0.162)   & (0.179)   & (0.348)   & (0.231)   & (0.163)   & (0.183)   & (0.445) & (0.375) & (0.364)   & (0.374)   \\ 
   
    SEG C1     & -0.079    & \multicolumn{4}{c}{-0.076}  & 0.372   & -0.073  & -0.156   & -0.061  & 0.756   & 0.466   & 0.287   & 0.294   \\ 
               & (0.166)   & \multicolumn{4}{c}{(0.167)} & (0.391) & (0.278) & (0.202)  & (0.204) & (0.521) & (0.431) & (0.409) & (0.418) \\ 
    SEG C2     & -0.363*   & \multicolumn{4}{c}{-0.348*} & -0.239  & -0.242  & -0.442*  & -0.319  & -0.048  & 0.098   & -0.317  & -0.469  \\  
               & (0.176)   & \multicolumn{4}{c}{(0.177)} & (0.373) & (0.289) & (0.211)  & (0.221) & (0.536) & (0.415) & (0.395) & (0.408) \\
    SEG DE     & -0.389*   & \multicolumn{4}{c}{-0.384*} & 0.049   & -0.165  & -0.623** & -0.284  & 0.371   & 0.534   & -0.247  & -0.204  \\ 
               & (0.167)   & \multicolumn{4}{c}{(0.167)} & (0.348) & (0.267) & (0.198)  & (0.209) & (0.491) & (0.385) & (0.372) & (0.381) \\
  
    AGE 35-54  & 0.411**   & 0.372     & 0.431   & 0.240   & 0.658**  & 0.359   & 0.378   & 0.247   & 0.699*** & 0.069  & 0.288    & 0.164   & 0.882*  \\ 
               & (0.152)   & (0.379)   & (0.262) & (0.180) & (0.203)  & (0.391) & (0.264) & (0.183) & (0.207)  & (0.496) & (0.413) & (0.408) & (0.426) \\ 
    AGE 55-64  & 0.338+    & -0.485    & -0.236  & 0.359   & 0.685**  & -0.600  & -0.328  & 0.357   & 0.733**  & -0.493  & -1.105* & -0.636  & 0.042   \\ 
               & (0.193)   & (0.385)   & (0.287) & (0.225) & (0.245)  & (0.395) & (0.286) & (0.227) & (0.250)  & (0.525) & (0.446) & (0.420) & (0.442) \\ 
    AGE 65+    & 0.508**   & -0.241    & 0.039   & 0.361+  & 0.920*** & -0.245  & -0.004  & 0.398+  & 1.003*** & -0.396  & -0.580  & -0.450  & 0.555   \\ 
               & (0.174)   & (0.373)   & (0.275) & (0.203) & (0.220)  & (0.385) & (0.276) & (0.207) & (0.227)  & (0.515) & (0.421) & (0.409) & (0.425) \\ 
 
    DSA MR     & 0.186     & \multicolumn{4}{c}{0.176}   & 0.747+  & 0.188   & 0.267   & 0.100   & 1.056*  & 0.524   & 0.810+  & 0.782+  \\ 
               & (0.175)   & \multicolumn{4}{c}{(0.176)} & (0.404) & (0.280) & (0.210) & (0.237) & (0.526) & (0.438) & (0.426) & (0.446) \\ 
    DSA TP     & 0.409**   & \multicolumn{4}{c}{0.401**} & 0.485+  & 0.202   & 0.373*  & 0.562** & 0.622   & 0.195   & 0.398   & 0.878** \\
               & (0.138)   & \multicolumn{4}{c}{(0.138)} & (0.276) & (0.214) & (0.162) & (0.181) & (0.397) & (0.305) & (0.298) & (0.309) \\ 
   
    IHD High   & 0.531**   & 1.093+    & 0.696*  & 0.632** & 0.506** & 1.095+  & 0.718*  & 0.621** & 0.534** & 0.605   & 0.678   & 1.033   & 1.387*  \\ 
               & (0.163)   & (0.651)   & (0.350) & (0.211) & (0.189) & (0.661) & (0.356) & (0.212) & (0.191) & (0.771) & (0.690) & (0.671) & (0.671) \\ 
 
    TVT Yes    & 0.005     & 1.070     & 0.249   & -0.207  & 0.103   & 1.184   & 0.285   & -0.238  & 0.133   & 1.305   & 1.458+  & 0.914   & 1.232   \\ 
               & (0.199)   & (0.730)   & (0.355) & (0.231) & (0.248) & (0.737) & (0.358) & (0.233) & (0.251) & (0.821) & (0.755) & (0.757) & (0.759) \\ 
    \midrule
    \multicolumn{14}{l}{Intercepts}\\
    \midrule
    NAW $\mid$ NVW & -2.684*** & -3.063*** &           &          &          & 2.745*** &          &         &           &  -0.247  &         &          &         \\ 
                   & (0.244)   & (0.427)   &           &          &          & (0.501)  &          &         &           &  (0.675) &         &          &         \\ 
    NVM $\mid$ IND & -1.915*** &           & -1.962*** &          &          &          & 2.012*** &         &           &          & 1.211*  &          &         \\ 
                   & (0.230)   &           & (0.305)   &          &          &          & (0.356)  &         &           &          & (0.545) &          &         \\ 
    IND $\mid$ QW  & -0.549*   &           &           & -0.614*  &          &          &          & 0.709** &           &          &         & 1.983*** &         \\ 
                   & (0.219)   &           &           & (0.239)  &          &          &          & (0.257) &           &          &         & (0.524)  &         \\ 
    QW  $\mid$ VW  & 0.978***  &           &           &          & 1.163*** &          &          &         & -1.305*** &          &         &          & 1.241*  \\ 
                   & (0.221)   &           &           &          & (0.257)  &          &          &         & (0.282)   &          &         &          & (0.545) \\  
    \midrule
    $n_{ind}$/$n_{obs}$          & 965       & \multicolumn{4}{c}{965}     & 965     & 965     & 965     & 965     & \multicolumn{4}{c}{965}      \\ 
    df         & 16        & \multicolumn{4}{c}{37}      & 13      & 13      & 13      & 13      & \multicolumn{4}{c}{52}       \\ 
    AIC        & 2713      & \multicolumn{4}{c}{2705}    & 479     & 778     & 1217    & 1092    & \multicolumn{4}{c}{2711}     \\ 
    BIC        & 2791      & \multicolumn{4}{c}{2885}    & 543     & 841     & 1280    & 1155    & \multicolumn{4}{c}{2964}     \\ 
    logLik         & -1340.4   & \multicolumn{4}{c}{-1315.4} & -226.72 & -376.01 & -595.26 & -532.91 & \multicolumn{4}{c}{-1303.52} \\ 
    $R_{MF}^2$ & 0.041     & \multicolumn{4}{c}{0.059}   & 0.123   & 0.067   & 0.073   & 0.077   & \multicolumn{4}{c}{0.068}    \\ 
    $R_{CS}^2$ & 0.113     & \multicolumn{4}{c}{0.157}   & 0.064   & 0.055   & 0.093   & 0.088   & \multicolumn{4}{c}{0.178}    \\ 
    $R_{NK}^2$ & 0.119     & \multicolumn{4}{c}{0.167}   & 0.154   & 0.096   & 0.127   & 0.127   & \multicolumn{4}{c}{0.188}    \\ 
    \bottomrule  
    \multicolumn{14}{@{}P@{}}{
    Note: + $p<$ 0.1, * $p<$ 0.05, ** $p<$ 0.01, *** $p<$ 0.001. Note: + $p <$ 0.1, * $p <$ 0.05, ** $p <$ 0.01, *** $p <$ 0.001. Standard errors in parentheses. McFadden (MF), Cox \& Snell (CS) and Nagelkerke (NK) pseudo-$R^2$. PO - Proportional Odds model, NAW - Not at all willing, NVW - Not very willing, IND - Indifferent, QW - Quite willing, VW - Very willing, GEN - Gender, SM - Smart Meter Ownership, SEG - Socio-Economic Group, DSA - Data Sharing Attitude, MR - Marketing \& Research, TP - Third Party, IHD - Engagement with In-Home Display more than once a week, TVT - Time-Varying Tariff. Base categories: GEN=Male, SM=Yes, SEG=AB, AGE=18-34, DSA=Basic, IHD=Low, TVT=No. }
\end{tabularx}
}
\end{table}
\end{landscape}
\begin{table}[H]
\centering
\caption{Logistic Regression Models for Change in WTS for Non-Anonymised Data}
\label{tab:wtsn.ord}
\setlength\targetwidth{1.1\textwidth} 
\resizebox{0.85\textwidth}{!}{
\begin{tabularx}{\targetwidth}{@{} rCCCCCCC @{}}
    \toprule
    & \multirow{2}{*}{\makecell{PO \\ Model}} & \multicolumn{2}{c}{Partial PO Model} 
    & \multicolumn{2}{c}{Binary Logistic Models} & \multicolumn{2}{c}{Multinomial Logit Model} \\
    \cmidrule(lr){3-4} \cmidrule(lr){5-6} \cmidrule(lr){7-8}
               &   &  ND & LL & $>$ ML   & $>$ ND & ND  & LL \\ 
    \midrule
    \multicolumn{8}{l}{Covariates}\\
    \cmidrule(lr){1-8}
    GEN Female &  0.495*** & \multicolumn{2}{c}{0.492***}  & 0.660*** & 0.408*  & 0.579** & 0.827*** \\
               & (0.129)   & \multicolumn{2}{c}{(0.132)}   & (0.173)  & (0.161) & (0.181) & (0.209)  \\ 
               
    No SM      & 0.235     & \multicolumn{2}{c}{0.265+}  & 0.100   & 0.373*  & -0.019  & 0.369   \\ 
               & (0.151)   & \multicolumn{2}{c}{(0.153)} & (0.201) & (0.187) & (0.211) & (0.241) \\
               
    SEG C1     & 0.132     & 0.082  & 0.220  & 0.027   & 0.188   & -0.043  & 0.159   \\ 
               & (0.178)   & (0.214) & (0.215) & (0.221) & (0.217) & (0.235) & (0.269) \\ 
    SEG C2     & 0.003     & 0.197  & -0.169   & 0.156   & -0.197  & 0.227   & -0.032  \\ 
               & (0.186)   & (0.233) & (0.237) & (0.239) & (0.240) & (0.250) & (0.298) \\ 
    SEG DE     & 0.262     & 0.502* & 0.072  & 0.582*  & 0.061   & 0.606*  & 0.511+  \\ 
               & (0.177)   & (0.234) & (0.216) & (0.241) & (0.218) & (0.251) & (0.286) \\ 
    
    AGE 35-54  & 0.587***  & 0.679*** & 0.331    & 0.706*** & 0.323    & 0.703*** & 0.731**  \\ 
               & (0.168)   & (0.189)   & (0.217)   & (0.194)  & (0.221)  & (0.208)  & (0.255)  \\ 
    AGE 55-64  & 1.283***  & 1.918*** & 0.699**  & 2.060*** & 0.666**  & 2.043*** & 2.095*** \\ 
               & (0.206)   & (0.301)   & (0.253)   & (0.307)  & (0.256)  & (0.318)  & (0.359)  \\ 
    AGE 65+    & 1.560***  & 2.111*** & 1.061*** & 2.245*** & 1.002*** & 2.124*** & 2.484*** \\ 
               & (0.193)   & (0.272)   & (0.230)   & (0.281)  & (0.232)  & (0.294)  & (0.326)  \\ 
  
    DSA MR     & 0.085     & \multicolumn{2}{c}{0.065}     & 0.177   & -0.007    & 0.212   & 0.147     \\ 
               & (0.190)   & \multicolumn{2}{c}{(0.192)}   & (0.278) & (0.213)   & (0.293) & (0.308)   \\ 
    DSA TP     & -0.629*** & \multicolumn{2}{c}{-0.642***} & -0.512* & -0.841*** & -0.248  & -1.018*** \\ 
               & (0.149)   & \multicolumn{2}{c}{(0.151)}   & (0.203) & (0.177)   & (0.215) & (0.236)   \\ 
  
    IHD High   & -0.295+   & \multicolumn{2}{c}{-0.258}   & -0.334  & -0.203  & -0.308  & -0.384  \\ 
               & (0.174)   & \multicolumn{2}{c}{(0.176)}  & (0.215) & (0.231) & (0.226) & (0.277) \\ 
  
    TVT Yes    & -0.558*   & \multicolumn{2}{c}{-0.536*}  & -0.505+ & -0.801* & -0.341 & -1.004** \\ 
               & (0.218)   & \multicolumn{2}{c}{(0.225)}  & (0.263) & (0.332) & (0.272) & (0.380) \\ 
               
    WTS NVM    & -0.332    & 0.388   & -0.689+   & 0.452     & -0.635+   & 0.776    & 0.020     \\ 
               & (0.330)   & (0.624)  & (0.366)  & (0.637)   & (0.371)   & (0.657)  & (0.678)   \\ 
    WTS IND    & -0.748**  & -0.469    & -1.047*** & -0.515    & -1.048*** & -0.104   & -1.155*   \\ 
               & (0.274)   & (0.478)  & (0.296)  & (0.486)   & (0.299)   & (0.508)  & (0.518)   \\ 
    WTS QW     & -0.950*** & -1.132*   & -0.833**  & -1.285**  & -0.802**  & -1.028*  & -1.579**  \\ 
               & (0.269)   & (0.459)  & (0.284)  & (0.470)   & (0.288)   & (0.494)  & (0.497)   \\ 
    WTS VW     & -1.366*** & -1.574*** & -1.250*** & -1.671*** & -1.196*** & -1.313** & -2.171*** \\ 
               & (0.277)   & (0.464)  & (0.299)  & (0.476)   & (0.304)   & (0.499)  & (0.510)   \\ 
    \cmidrule(lr){1-8}
    \multicolumn{8}{l}{Intercepts}\\
    \cmidrule(lr){1-8}
    ML $\mid$ ND & -1.285*** & -1.093*   &         & 1.037*  &         & 0.225   &          \\
                 &  (0.331)  & (0.488)   &         & (0.523) &         & (0.553) &          \\
    ND $\mid$ LL & 1.075**   &           & 0.611+  &         & -0.499  &         & 0.425    \\
                 & (0.330)   &           & (0.360) &         & (0.379) &         & (0.575)  \\
    \cmidrule(lr){1-8}
    $n_{ind}$/$n_{obs}$         & 965     & \multicolumn{2}{c}{965}     & 965     & 965     & \multicolumn{2}{c}{965}      \\ 
    df         & 18      & \multicolumn{2}{c}{28}      & 17      & 17      & \multicolumn{2}{c}{34}       \\ 
    AIC        & 1873    & \multicolumn{2}{c}{1844}    & 934     & 1049    & \multicolumn{2}{c}{1840}     \\ 
    BIC        & 1961    & \multicolumn{2}{c}{1980}    & 1017    & 1132    & \multicolumn{2}{c}{2006}     \\ 
    logLik         & -918.74 & \multicolumn{2}{c}{-894.03} & -450.01 & -507.54 & \multicolumn{2}{c}{-886.13}  \\ 
    $R_{MF}^2$ & 0.107   & \multicolumn{2}{c}{0.131}   & 0.198   & 0.107   & \multicolumn{2}{c}{0.139}    \\ 
    $R_{CS}^2$ & 0.204   & \multicolumn{2}{c}{0.244}   & 0.206   & 0.118   & \multicolumn{2}{c}{0.256}    \\ 
    $R_{NK}^2$ & 0.232   & \multicolumn{2}{c}{0.277}   & 0.300     & 0.171   & \multicolumn{2}{c}{0.291}    \\ 
    \bottomrule
    \multicolumn{8}{@{}P@{}}{Note: + $p <$ 0.1, * $p <$ 0.05, ** $p <$ 0.01, *** $p <$ 0.001. Standard errors in parentheses. McFadden (MF), Cox \& Snell (CS) and Nagelkerke (NK) pseudo-$R^2$. PO - Proportional Odds model, ML - More Likely, ND - No Difference, LL - Less Likely, NAW - Not at all willing, NVW - Not very willing, IND - Indifferent, QW - Quite willing, VW - Very willing, GEN - Gender, SM - Smart Meter Ownership, DSA - Data Sharing Attitude, MR - Marketing \& Research, TP - Third Party, IHD - Engagement with In-Home Display more than once a week, TVT - Time-Varying Tariff, WTS - Initial Willingness-to-Share. Base categories: GEN=Male, SM=Yes, AGE=18-34, DSA=Basic, IHD=Low, TVT=No, WTS=NAW.}
\end{tabularx}
}
\end{table} 
\begin{table}[H]
\centering
\caption{Logistic Regression Models for Change in WTS for Anonymised Data}
\label{tab:wtsa.ord}
\setlength\targetwidth{1.1\textwidth} 
\resizebox{0.85\textwidth}{!}{
\begin{tabularx}{\targetwidth}{@{} rCCCCCCC @{}}
    \toprule
    & \multirow{2}{*}{\makecell{PO \\ Model}} & \multicolumn{2}{c}{Partial PO Model} 
    & \multicolumn{2}{c}{Binary Logistic Models} & \multicolumn{2}{c}{Multinomial Logit Model} \\
    \cmidrule(lr){3-4} \cmidrule(lr){5-6} \cmidrule(lr){7-8}
               &   &  ND & LL & $>$ ML   & $>$ ND & ND  & LL \\
    \midrule
    \multicolumn{8}{l}{Covariates}\\
    \cmidrule(lr){1-8}
    GEN Female & 0.289*  & \multicolumn{2}{c}{0.297*}   & 0.336*  & 0.322   & 0.304*  & 0.512*  \\ 
               & (0.131) & \multicolumn{2}{c}{(0.131)}  & (0.144) & (0.221) & (0.149) & (0.240) \\ 
    
    SM No      & -0.098  & -0.262   & 0.407+ & -0.283+ & 0.473+  & -0.389* & 0.240   \\ 
               & (0.154) & (0.164) & (0.237) & (0.169) & (0.269) & (0.175) & (0.289) \\ 
     
    SEG C1     & 0.098   & \multicolumn{2}{c}{0.086}   & 0.055   & 0.221   & 0.023   & 0.223     \\ 
               & (0.179) & \multicolumn{2}{c}{(0.180)} & (0.193) & (0.326) & (0.199) & (0.346)   \\ 
    SEG C2     & 0.282   & \multicolumn{2}{c}{0.276}   & 0.254   & 0.391   & 0.200   & 0.503     \\ 
               & (0.190) & \multicolumn{2}{c}{(0.191)} & (0.207) & (0.332) & (0.214) & (0.357)   \\ 
    SEG DE     & 0.415*  & \multicolumn{2}{c}{0.405*}  & 0.462*  & 0.356   & 0.433*  & 0.618+    \\ 
               & (0.180) & \multicolumn{2}{c}{(0.181)} & (0.199) & (0.312) & (0.205) & (0.337)   \\ 
    
    AGE 35-54  & 0.214    & \multicolumn{2}{c}{0.234}    & 0.204    & 0.337   & 0.165    & 0.435     \\ 
               & (0.169)  & \multicolumn{2}{c}{(0.170)}  & (0.180)  & (0.316) & (0.186)  & (0.334)   \\ 
    AGE 55-64  & 0.924*** & \multicolumn{2}{c}{0.955***} & 1.129*** & 0.607+  & 1.098*** & 1.313***  \\ 
               & (0.206)  & \multicolumn{2}{c}{(0.208)}  & (0.241)  & (0.352) & (0.247)  & (0.390)   \\ 
    AGE 65+    & 0.705*** & \multicolumn{2}{c}{0.728***} & 0.750*** & 0.645+  & 0.695**  & 1.061**   \\ 
               & (0.190)  & \multicolumn{2}{c}{(0.191)}  & (0.209)  & (0.331) & (0.216)  & (0.358)   \\ 
    
    DSA MR     & -0.041   & -0.085   & 0.124  & -0.094  & 0.059   & -0.111  & -0.005    \\ 
               & (0.195)  & (0.209) & (0.295) & (0.212) & (0.297) & (0.222) & (0.325)   \\ 
    DSA TP     & 0.009    & 0.127  & -0.337   & 0.102   & -0.371  & 0.175   & -0.256    \\ 
               & (0.150)  & (0.164) & (0.240) & (0.167) & (0.245) & (0.172) & (0.268)   \\ 
    
    IHD High   & -0.105  & \multicolumn{2}{c}{-0.128}   & -0.129  & -0.118  & -0.124  & -0.189  \\ 
               & (0.175) & \multicolumn{2}{c}{(0.179)}  & (0.190) & (0.360) & (0.194) & (0.377) \\ 
    
    TVT Yes    & -0.035  & \multicolumn{2}{c}{-0.033}   & 0.056   & -0.637  & 0.131   & -0.542  \\ 
               & (0.215) & \multicolumn{2}{c}{(0.216)}  & (0.237) & (0.486) & (0.241) & (0.508) \\ 
    
    WTS NVW    & -0.955**  & \multicolumn{2}{c}{-0.840*}    & -0.712    & -0.965*   & -0.336    & -1.229*     \\ 
               & (0.335)   & \multicolumn{2}{c}{(0.336)}    & (0.512)   & (0.390)   & (0.534)   & (0.578)     \\ 
    WTS IND    & -1.684*** & \multicolumn{2}{c}{-1.568***}  & -1.467*** & -1.792*** & -0.956*   & -2.500***   \\ 
               & (0.277)   & \multicolumn{2}{c}{(0.277)}    & (0.430)   & (0.331)   & (0.450)   & (0.489)     \\ 
    WTS QW     & -2.284*** & \multicolumn{2}{c}{-2.162***}  & -2.199*** & -2.021*** & -1.702*** & -3.149***   \\ 
               & (0.275)   & \multicolumn{2}{c}{(0.276)}    & (0.423)   & (0.329)   & (0.443)   & (0.483)     \\ 
    WTS VW     & -2.479*** & \multicolumn{2}{c}{-2.375***}  & -2.391*** & -2.242*** & -1.875*** & -3.462***   \\ 
               & (0.284)   & \multicolumn{2}{c}{(0.285)}    & (0.428)   & (0.366)   & (0.449)   & (0.510)     \\  
    \cmidrule(lr){1-8}
    \multicolumn{8}{l}{Intercepts}\\
    \cmidrule(lr){1-8}
    ML $\mid$ ND & -1.710*** & -1.602*** &          & 1.576***  &         & 0.978*  &           \\
                 &  (0.337)  & (0.340)   &          & (0.468)   &         & (0.489) &           \\
    ND $\mid$ LL & 0.979**   &           & 1.255*** &           & -1.192* &         & 0.216     \\
                 & (0.329)   &           & (0.378)  &           & (0.481) &         & (0.605)   \\
    \cmidrule(lr){1-8}
    $n_{ind}$/$n_{obs}$           & 965       & \multicolumn{2}{c}{965}       & 965       & 965       & \multicolumn{2}{c}{965}       \\ 
    df          & 18        & \multicolumn{2}{c}{21}        & 17        & 17        & \multicolumn{2}{c}{34}        \\ 
    AIC         & 1762      & \multicolumn{2}{c}{1752}      & 1217      & 649       & \multicolumn{2}{c}{1767}      \\ 
    BIC         & 1849      & \multicolumn{2}{c}{1855}      & 1300      & 732       & \multicolumn{2}{c}{1933}      \\ 
    logLik          & -862.83   & \multicolumn{2}{c}{-855.22}   & -591.6    & -307.62   & \multicolumn{2}{c}{-849.43}   \\ 
    $R_{MF}^2$  & 0.085     & \multicolumn{2}{c}{0.093}     & 0.095     & 0.137     & \multicolumn{2}{c}{0.099}     \\ 
    $R_{CS}^2$  & 0.152     & \multicolumn{2}{c}{0.166}     & 0.121     & 0.096     & \multicolumn{2}{c}{0.176}     \\ 
    $R_{NK}^2$  & 0.178     & \multicolumn{2}{c}{0.193}     & 0.163     & 0.184     & \multicolumn{2}{c}{0.205}     \\ 
    \bottomrule
    \multicolumn{8}{@{}P@{}}{Note: + $p <$ 0.1, * $p <$ 0.05, ** $p <$ 0.01, *** $p <$ 0.001. Standard errors in parentheses. McFadden (MF), Cox \& Snell (CS) and Nagelkerke (NK) pseudo-$R^2$. PO - Proportional Odds model, ML - More Likely, ND - No Difference, LL - Less Likely, NAW - Not at all willing, NVW - Not very willing, IND - Indifferent, QW - Quite willing, VW - Very willing, GEN - Gender, SM - Smart Meter Ownership, DSA - Data Sharing Attitude, MR - Marketing \& Research, TP - Third Party, IHD - Engagement with In-Home Display more than once a week, TVT - Time-Varying Tariff, WTS - Initial Willingness-to-Share. Base categories: GEN=Male, SM=Yes, AGE=18-34, DSA=Basic, IHD=Low, TVT=No, WTS=NAW.}
\end{tabularx}
}
\end{table} 
\begin{landscape}
\begin{table}[H]
\centering
\caption{Logistic Regression Models for Change in WTS with Treatment Effects}
\label{tab:wtsna.ord}
\setlength\targetwidth{1.4\textwidth} 
\resizebox{0.9\textwidth}{!}{
\begin{tabularx}{\targetwidth}{@{} r CCCCCCC CCCCCCC @{}}
    \toprule
    & \multicolumn{2}{c}{\multirow{2}{*}{PO Model}} & \multicolumn{4}{c}{Partial PO Model}
    & \multicolumn{4}{c}{Binary Logistic Models} & \multicolumn{4}{c}{Multinomial Logit Model} \\
     \cmidrule(lr){4-7} \cmidrule(lr){8-11}  \cmidrule(lr){12-15}
    & & & \multicolumn{2}{c}{ND} & \multicolumn{2}{c}{LL} & \multicolumn{2}{c}{$>$ ML} & \multicolumn{2}{c}{$>$ ND} & \multicolumn{2}{c}{ND}  & \multicolumn{2}{c}{LL} \\ 
    \midrule
    \multicolumn{15}{l}{Covariates}\\
    \midrule
    Anon        & -0.706*   & (0.335) & -0.645 & (0.515) & -1.317** & (0.472) & -1.029+ & (0.597) & -2.410*** & (0.703)  & -0.370  & (0.518)   & -1.584**  & (0.581)\\ 
                
    Treat       & 1.423***  & (0.399) & 0.554  & (0.574)   & 1.837*** & (0.473) & 0.712 & (0.768) & 3.289*** & (0.986) & -0.265  & (0.575) & 1.231*   & (0.536) \\ 
               
    GEN Female  & 0.536***  & (0.132) & &\multicolumn{2}{c}{0.522*** (0.135)} & & 0.768***  & (0.216)  & 0.935**  & (0.358)  & 0.472** & (0.156) & 0.716***  & (0.159)   \\ 
                
    SM No       & 0.401*    & (0.182) &  0.273  & (0.225)  &   0.555*  & (0.219) & 0.381  & (0.298)  & 0.909* & (0.437)  & 0.075  & (0.230) & 0.466* & (0.228)  \\ 
                
    SEG C1      & 0.133    & (0.179)  &  0.034  & (0.203)  &  0.264 & (0.237)   & 0.052   & (0.281)   & 0.391  & (0.474)  & -0.020  & (0.208)   & 0.141  & (0.212)    \\ 

    SEG C2      & 0.133     & (0.190) &  0.277  & (0.220)  & -0.074   & (0.257)   & 0.363    & (0.306)   & -0.167  & (0.516)   & 0.253   & (0.223)   & 0.105   & (0.232)   \\ 
    
    SEG DE      & 0.410*    & (0.179) & 0.623** & (0.212)  & 0.149   & (0.236)  & 0.834**   & (0.295) & 0.318  & (0.473)   & 0.573** & (0.214)  & 0.513*   & (0.216)   \\ 
                
    AGE 35-54   & 0.455**   & (0.169) & 0.566**  & (0.189) & 0.249  & (0.235) & 0.686** & (0.262)  & 0.603  & (0.474)  & 0.410* & (0.191)   & 0.530*    & (0.208)   \\ 
                               
    AGE 55-64   & 1.396***  & (0.211) & 1.918*** & (0.262)  &  0.783** & (0.274)  & 2.493***  & (0.378) & 1.498**   & (0.567) & 1.610*** & (0.256)  & 1.732***  & (0.264)  \\ 

    AGE 65+     & 1.446***  & (0.195) &  1.622*** & (0.226) &  1.134***  & (0.252) & 2.128*** & (0.326)  & 2.065*** & (0.534) & 1.259*** & (0.226)  & 1.693***  & (0.231)   \\ 
                                            
    DSA MR      & 0.333     & (0.303) &  0.273   & (0.396)   &  0.449  & (0.352)   & 0.286  & (0.516)   & 0.656   & (0.678)  & 0.130  & (0.403)  & 0.381    & (0.372)   \\ 
                          
    DSA TP      & -0.766**  & (0.239) & -0.571+  & (0.303)  & -0.878** & (0.298)   & -0.883*   & (0.401) & -1.550** & (0.580)  & -0.358  & (0.305)  & -0.970**   & (0.297)   \\                        
    IHD High    & -0.320+   & (0.177) & &\multicolumn{2}{c}{-0.303+ (0.182)} & & -0.382  & (0.278)  & -0.465  & (0.495)   & -0.223  & (0.203) & -0.309 & (0.223)   \\ 
                          
    TVT Yes     & -0.457*   & (0.219) & &\multicolumn{2}{c}{-0.436+ (0.225)} & & -0.408  & (0.341)   & -1.987**  & (0.715)   & -0.100   & (0.247) & -1.056*** & (0.317)   \\  
                                         
    WTS IND     & -0.508    & (0.334) & -0.849+  & (0.513)  & -0.604  & (0.407)   & -0.892   & (0.665)   & -1.096  & (0.791)   & -0.500  & (0.509)   & -0.951+   & (0.496)   \\ 
                           
    WTS WIL     & -0.805**  & (0.299) & -1.775*** & (0.465) & -0.282  & (0.352)  & -2.164*** & (0.606) & -0.544  & (0.681)  & -1.541*** & (0.459)  & -1.383**   & (0.440)   \\ 
                              
    \midrule
    \multicolumn{15}{l}{Interactions}\\
    \midrule
    Anon × DSA MR   & -0.103    & (0.295)  & -0.305  & (0.386)  & 0.155  & (0.416)    & -0.277 & (0.451)   & 0.122  & (0.617)   & -0.299 & (0.386)   & -0.104   & (0.453)   \\ 
    
    Anon × DSA TP   & 0.962***  & (0.227) & 0.890**   & (0.290) & 0.572+ & (0.334)  & 1.262*** & (0.342)  & 1.112*  & (0.493)  & 0.574*  & (0.288)  & 0.873*  & (0.357)   \\ 
    
    Anon × WTS IND  & -0.526+ & (0.316)  & -0.403  & (0.508)   & -0.713+ & (0.429) & -0.593 & (0.586)  & -1.242* & (0.619)  & -0.257 & (0.504)   & -0.693   & (0.553)   \\ 
                              
    Anon × WTS WIL  & -0.952*** & (0.289) & -0.513 & (0.463) & -1.097**  & (0.382) & -0.670 & (0.533) & -2.091*** & (0.568) & -0.168 & (0.463)  & -0.949+  & (0.498)   \\ 
                                   
    Anon × SM No    & -0.533** & (0.202) & -0.765** & (0.253) & 0.200  & (0.313)  & -1.011*** & (0.297) & 0.099 & (0.450) & -0.651** & (0.250) & -0.252  & (0.323)   \\ 
                                    
    Anon × Treat    & 0.224  & (0.194)  & 0.102  & (0.244) & 0.628*  & (0.303) & 0.234   & (0.282) & 1.249** & (0.452)   & 0.047  & (0.241) & 0.715*  & (0.319)   \\ 
    
    Treat × DSA MR  & -0.671+  & (0.386)  & -0.197   & (0.459)  & -1.144* & (0.475)   & -0.314 & (0.632)    & -1.890*  & (0.962) & 0.144  & (0.471)  & -0.829+  & (0.446)   \\  
                                 
    Treat × DSA TP  & -0.493+   & (0.296) & -0.178 & (0.350)   & -0.696+  & (0.383)  & -0.228  & (0.484)   & -1.334+ & (0.775)   & 0.156 & (0.358)    & -0.503   & (0.355)   \\ 
    
    Treat × WTS IND & -0.804+   & (0.419) & -0.234    & (0.590)  & -0.903+  & (0.521)  & -0.461  & (0.800)   & -1.722  & (1.067)    & 0.086   & (0.589)  & -0.640  & (0.557)   \\ 
    
    Treat × WTS WIL & -1.137** & (0.373)  & -0.527  & (0.534)   & -1.338** & (0.451)  & -0.762 & (0.718)   & -2.448** & (0.938)   & -0.013  & (0.531) & -1.031*  & (0.494)   \\ 
                               
    \midrule
    \multicolumn{15}{l}{Intercepts}\\
    \midrule
    ML $\mid$ ND    & -1.227*** & (0.370) & -1.682** & (0.524)  &        &               & 2.157**   & (0.697)   &        &       & 0.977+  & (0.533)    &   \\

    ND $\mid$ LL    & 1.924*** & (0.373) &     &         & 1.853*** & (0.452) &     &       & -3.164*** & (0.925) &   &       & 0.192   & (0.520)   \\
                                                   
    ID              & 1.246   &  & 1.284    &      &&   & 2.164     & & 3.512     & &  1.399 &  & 0.001   \\
    \midrule
    $n_{ind}$           & \multicolumn{2}{c}{965}       & \multicolumn{4}{c}{965}       & \multicolumn{2}{c}{965}       &  \multicolumn{2}{c}{965}      & \multicolumn{4}{c}{965}     \\ 
    $n_{obs}$           & \multicolumn{2}{c}{1930}      & \multicolumn{4}{c}{1930}      & \multicolumn{2}{c}{1930}      & \multicolumn{2}{c}{1930}      & \multicolumn{4}{c}{1930}     \\ 
    df                  & \multicolumn{2}{c}{29}        & \multicolumn{4}{c}{52}        & \multicolumn{2}{c}{28}        & \multicolumn{2}{c}{28}        & \multicolumn{4}{c}{57}       \\ 
    AIC                 & \multicolumn{2}{c}{3571}      & \multicolumn{4}{c}{3526}      & \multicolumn{2}{c}{2046}      & \multicolumn{2}{c}{1500}      & \multicolumn{4}{c}{3373}     \\ 
    BIC                 & \multicolumn{2}{c}{3733}      & \multicolumn{4}{c}{3815}      & \multicolumn{2}{c}{2202}      & \multicolumn{2}{c}{1655}      & \multicolumn{4}{c}{3690}     \\ 
    logLik              & \multicolumn{2}{c}{-1756.73}  & \multicolumn{4}{c}{-1710.77}  & \multicolumn{2}{c}{-995.12}   & \multicolumn{2}{c}{-721.84}   & \multicolumn{4}{c}{-1629.41} \\ 
    $R_{MF}^2$          & \multicolumn{2}{c}{0.129}     & \multicolumn{4}{c}{0.152}     & \multicolumn{2}{c}{0.196}     & \multicolumn{2}{c}{0.249}     & \multicolumn{4}{c}{0.192}    \\ 
    $R_{CS}^2$          & \multicolumn{2}{c}{0.236}     & \multicolumn{4}{c}{0.271}     & \multicolumn{2}{c}{0.222}     & \multicolumn{2}{c}{0.220}      & \multicolumn{4}{c}{0.330}     \\ 
    $R_{NK}^2$          & \multicolumn{2}{c}{0.269}     & \multicolumn{4}{c}{0.310}      & \multicolumn{2}{c}{0.308}     & \multicolumn{2}{c}{0.349}     & \multicolumn{4}{c}{0.377}    \\ 
    Marginal $R_{NK}^2$ & \multicolumn{2}{c}{0.227}     & \multicolumn{4}{c}{0.269}     & \multicolumn{2}{c}{0.223}     & \multicolumn{2}{c}{0.263}     & \multicolumn{4}{c}{0.275}    \\ 
    \bottomrule
    \multicolumn{15}{@{}P@{}}{Note: + $p <$ 0.1, * $p <$ 0.05, ** $p <$ 0.01, *** $p <$ 0.001. Standard errors in parentheses. McFadden (MF), Cox \& Snell (CS) and Nagelkerke (NK) pseudo-$R^2$. PO - Proportional Odds model, ML - More Likely, ND - No Difference, LL - Less Likely, Anon - Anonymised, GP - Group, GEN - Gender, SM - Smart Meter Ownership, DSA - Data Sharing Attitude, MR - Marketing \& Research, TP - Third Party, IHD - Engagement with In-Home Display more than once a week, TVT - Time-Varying Tariff, WTS - Initial Willingness-to-Share, NAW - Not at all willing, NVW - Not very willing, IND - Indifferent, QW - Quite willing, VW - Very willing. Base categories: Anon=No, GP=Control, GEN=Male, SM=Yes, SEG=DE, AGE=18-34, DSA=Basic, IHD=Low, TVT=No, WTS=NAW. Marginal $R_{NK}^2$ is calculated against null model which includes random effects.}
\end{tabularx}
}
\end{table} 
\end{landscape}
\begin{landscape}
    \begin{table}[H]
    \centering
    \caption{z-test $p$-values for Heterogeneity in Willingness-to-Share and Demand for Smart Metering} 
    \label{tab:wts.het.tests}
    \setlength\targetwidth{\textwidth}
    \resizebox{0.9\textwidth}{!}{
    \begin{tabularx}{\targetwidth}{@{}CCCCCCCCCCCC@{}}
    \toprule
    \multicolumn{2}{c}{Group} & \multicolumn{3}{c}{Initial WTS} & \multicolumn{3}{c}{Non-Anon Change in WTS} & \multicolumn{3}{c}{Anon Change in WTS} & \multirow{2}{*}{\makecell{Demand \\for SM}}\\
    \cmidrule(lr){1-2} \cmidrule(lr){3-5} \cmidrule(lr){6-8} \cmidrule{9-11}
    1 & 2 & Unwilling & Indifferent & Willing & More & Indifferent & Less & More & Indifferent & Less & \\
    \midrule
    \multicolumn{12}{l}{\textbf{Gender (GEN)}}\\
    \midrule
    F & M & 0.847 & \textbf{0.096}+ & 0.182 & \textbf{0.000}*** & 0.478 & \textbf{0.010}** & \textbf{0.000}*** & \textbf{0.029}* & \textbf{0.004}** & \textbf{0.002}** \\   
    \midrule
    \multicolumn{12}{l}{\textbf{Age (AGE)}}\\
    \midrule
    35-54 & 18-34 & 0.111 & 0.813 & 0.181 & \textbf{0.006}* & 0.286 & 0.156 & \textbf{0.013}* & \textbf{0.073} & 0.105 & \textbf{0.010}* \\
    55-64 & 18-34 & 0.415 & \textbf{0.008}* & 0.104 & \textbf{0.000}*** & \textbf{0.002}* & \textbf{0.018}+ & \textbf{0.000}*** & \textbf{0.000}*** & \textbf{0.005}* & \textbf{0.003}* \\ 
    65+ & 18-34 &0.887 & \textbf{0.060} & \textbf{0.072} & \textbf{0.000}*** & \textbf{0.066} & \textbf{0.000}*** & \textbf{0.000}*** & \textbf{0.000}*** & \textbf{0.000}*** & \textbf{0.000}*** \\ 
    55-64 & 35-54  &  \textbf{0.021} & \textbf{0.007}* & 0.571 & \textbf{0.000}*** & \textbf{0.018}+ & 0.172 & \textbf{0.000}*** & \textbf{0.000}*** & \textbf{0.072} & 0.263 \\ 
    65+ & 35-54 & 0.131 & \textbf{0.061} & 0.517 & \textbf{0.000}*** & 0.331 & \textbf{0.005}* & \textbf{0.000}*** & \textbf{0.004}* & \textbf{0.002}** & \textbf{0.010}* \\ 
    65+ & 55-64 & 0.322 & 0.362 & 0.993 & 0.288 & 0.159 & 0.293 & 0.220 & 0.144 & 0.341 & 0.259 \\ 
    \midrule
    \multicolumn{12}{l}{\textbf{Socio-Economic Group (SEG)}}\\
    \midrule
    AB & C1 &0.650 & 0.651 & 0.650 & 0.834 & 0.653 & 0.460 & 0.966 & 0.807 & 0.455 & 0.159 \\ 
    AB & C2 & \textbf{0.054} & \textbf{0.050} & \textbf{0.049} & 0.280 & 0.337 & 0.776 & 0.258 & 0.269 & 0.854 & 0.313 \\ 
    AB & DE & \textbf{0.022} & \textbf{0.024} & \textbf{0.021} & \textbf{0.003}* & 0.126 & 0.587 & \textbf{0.004}* & \textbf{0.021} & 0.412 & \textbf{0.026} \\ 
    C1 & C2 & 0.121 & 0.112 & 0.114 & 0.365 & 0.158 & 0.317 & 0.266 & 0.175 & 0.367 & 0.752 \\ 
    C1 & DE& \textbf{0.058} & \textbf{0.059} & \textbf{0.056} & \textbf{0.004}* & \textbf{0.042} & 0.841 & \textbf{0.004}* & \textbf{0.009}+ & 0.934 & 0.436 \\ 
    C2 & DE & 0.834 & 0.835 & 0.835 & \textbf{0.079} & 0.624 & 0.411 & 0.109 & 0.276 & 0.325 & 0.285 \\ 
    \midrule
    \multicolumn{12}{l}{\textbf{Smart Meter Ownership (SM)}}\\
    \midrule
    No & Yes & \textbf{0.000}*** & \textbf{0.004}** & \textbf{0.000}*** & 0.300 & 0.396 & \textbf{0.042}* & \textbf{0.020}* & \textbf{0.002}** & \textbf{0.031}* & \textbf{0.000}*** \\ 
    \midrule
    \multicolumn{12}{l}{\textbf{Time-Varying Tariff (TVT)}}\\
    \midrule
    Yes & No & 0.450 & 0.143 & 0.376 & 0.135 & 0.136 & \textbf{0.000}*** & 0.354 & 0.738 & \textbf{0.000}*** & 0.989 \\ 
    \midrule
    \multicolumn{12}{l}{\textbf{IHD Engagement (IHD)}}\\
    \midrule
    High & Low & \textbf{0.017}* & \textbf{0.088}+ & \textbf{0.002}** & 0.161 & 0.726 & 0.465 & 0.203 & 0.406 & 0.390 & 0.548 \\ 
  \midrule
    \multicolumn{12}{l}{\textbf{Data Sharing Attitudes (DSA)}}\\
    \midrule
    BA & MR & 0.312 & 0.323 & 0.315 & 0.758 & 0.408 & 0.471 & 0.607 & 0.869 & 0.550 & \textbf{0.059} \\ 
    BA & TP & \textbf{0.006}* & \textbf{0.004}* & \textbf{0.004}* & \textbf{0.002}** & \textbf{0.066} & \textbf{0.000}*** & 0.398 & \textbf{0.056} & \textbf{0.023}+ & \textbf{0.004}* \\ 
    MR & TP & 0.183 & 0.161 & 0.169 & \textbf{0.002}** & 0.531 & \textbf{0.001}** & 0.199 & \textbf{0.072} & 0.218 & 0.622 \\ 
    \bottomrule  
    \multicolumn{12}{@{}P@{}}{
      Note:+ $p<$ 0.1, * $p<$ 0.05, ** $p<$ 0.01, *** $p<$ 0.001. Significance labels based on adjusted $p$-values using Holm correction for multiple comparisons within each socio-demographic group. Anon - Anonymised. Initial WTS based on estimated marginal mean probabilities from partial proportional odds model in Table \ref{tab:wtsg.lr}.} 
    \end{tabularx}
    }
    \end{table}
    \end{landscape}

\begin{figure}
    \centering
    \subfloat[\gls{iwts}\label{fig:wtsg_het}]{
    \includegraphics[width=0.3\linewidth,valign=t]{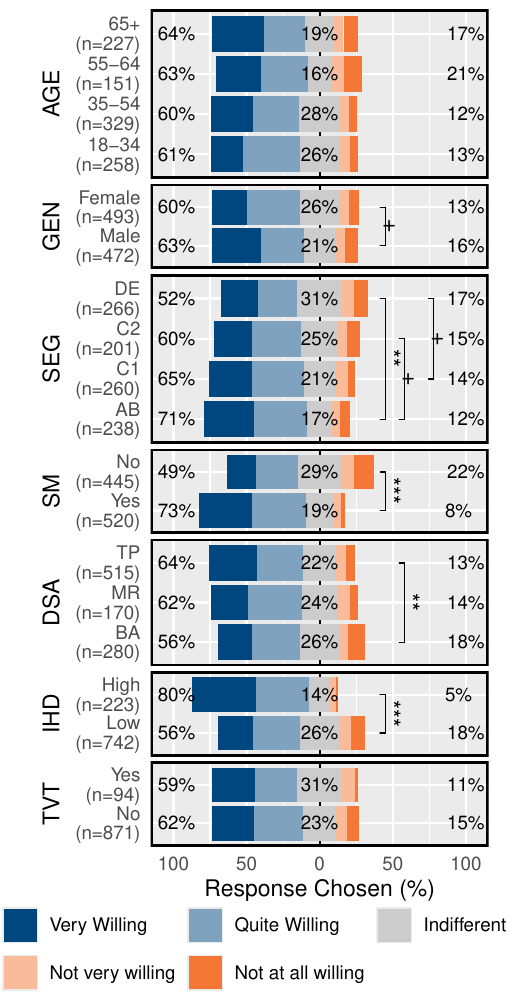}} 
    \subfloat[Change in \gls{wts} for Non-Anonymised Data\label{fig:wtsn_het}]{\includegraphics[width=0.3\linewidth,valign=t]{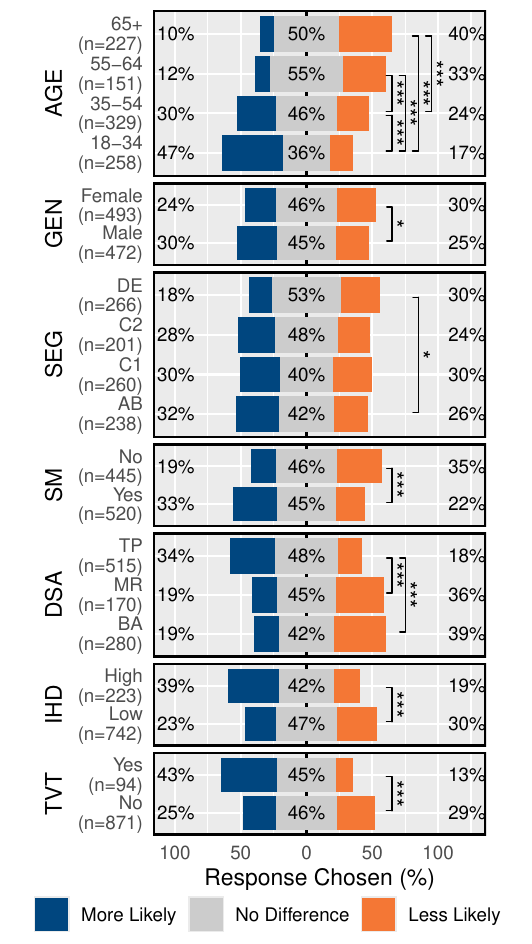}} 
    \subfloat[Change in \gls{wts} for Anonymised Data\label{fig:wtsa_het}]{\includegraphics[width=0.3\linewidth,valign=t]{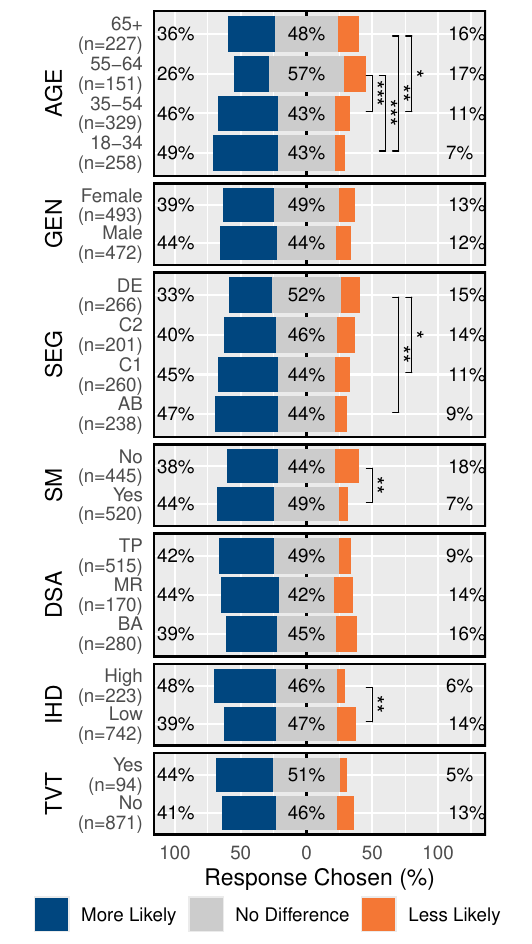}} 
    \caption{\gls{wts} by Socio-Demographic and \gls{dsa}. Significance levels indicate results of non-parametric testing of response distributions. Kruskal-Wallis/Mann-Whitney tests for each grouping, followed by Dunn's test with Holm correction for pairwise comparisons where appropriate. Details of tests can be found in Tables \ref{tab:wtsg.sig}, \ref{tab:wtsn.sig}, and \ref{tab:wtsa.sig}.}
    \label{fig:wtsg.sig}
\end{figure}

\newpage
\section{Supporting Tables for Willingness-to-Pay/Accept Analysis}\label{app:wtp}

\begin{table}[H]
\centering
\caption{Mixed Logit Model Estimates for Discrete Choice Experiment} 
\label{tab:mixl.split}
\setlength\targetwidth{1.3\textwidth}
\resizebox{\textwidth}{!}{
\begin{tabularx}{\targetwidth}{rCCCCCCCCC}
  \toprule
    & \multicolumn{3}{c}{Full Sample} &\multicolumn{3}{c}{Basic + Marketing \& Research} & \multicolumn{3}{c}{Third-Party}\\
    \cmidrule(lr){2-4} \cmidrule(lr){5-7} \cmidrule(lr){8-10}
     & Pooled & Control & Treat & Pooled & Control & Treat & Pooled & Control & Treat \\ 
  \midrule
    \multicolumn{10}{l}{Means}\\
    \midrule
    Fee(\%) &-0.137 *** & -0.178 *** & -0.090 *** & -0.181 *** & -0.217 *** & -0.106 *** & -0.106 *** & -0.148 *** & -0.080 *** \\ 
     & (0.012) & (0.018) & (0.011) & (0.021) & (0.029) & (0.018) & (0.015) & (0.023) & (0.016)  \\ 
    Discount(\%) & 0.079 *** & 0.078 *** & 0.065 *** & 0.107 *** & 0.115 *** & 0.068 *** & 0.060 *** & 0.058 *** & 0.063 *** \\ 
     & (0.007) & (0.009) & (0.008) & (0.013) & (0.016) & (0.012) & (0.008) & (0.012) & (0.010)  \\ 
    Anon & 0.373 * & 0.510 * & 0.189 & 0.920 ** & 0.516 & 0.983 ** & 0.017 & 0.444 & -0.349 \\
\end{tabularx}
}
\end{table}

\begin{table}[H]
\centering
\caption{Mean Willingness-to-Pay/Accept Estimates by Treatment and Data Sharing Attitude} 
\label{tab:wtpa.conf}
\setlength\targetwidth{1.1\textwidth}
\resizebox{\textwidth}{!}{
\begin{tabularx}{\targetwidth}{@{}CCCCCCC@{}}
  \toprule
    & \multicolumn{2}{c}{Full Sample} & \multicolumn{2}{c}{Basic + Marketing \& Research} & \multicolumn{2}{c}{Third-Party}\\
    \cmidrule(lr){2-3} \cmidrule(lr){4-5} \cmidrule(lr){6-7}
    & Control & Treatment& Control & Treatment & Control & Treatment\\ 
  \midrule
  \multicolumn{7}{l}{Willingness-to-Accept (\%/month)}\\
  \midrule
    HH & 3.57 & -0.92 & -4.73 & 4.22 & 13.50 & -5.50 \\ 
    & [-5.06, 12.48] & [-9.85, 7.93] & [-15.62, 5.34] & [-9.7, 19.06] & [-1.31, 31.7] & [-18.01, 6.28] \\ 
   
    Daily & 8.26 & 15.02 & 10.95 & 17.94 & 5.23 & 13.60 \\ 
    & [1.93, 14.24] & [8.49, 21.83] & [4.31, 17.67] & [8.1, 29.12] & [-8.05, 15.83] & [4.47, 22.79] \\  

    Anon RT& 9.16 & 4.16 & 6.32 & 20.55 & 11.16 & -7.88 \\
\end{tabularx}
}
\end{table}

\begin{table}[H]
\centering
\caption{Complete Combinatorial Test $p$-values for Mean Willingness-to-Pay/Accept by Treatment and Data Sharing Attitude} 
\label{tab:poe.freq}
\setlength\targetwidth{1.2\textwidth}
\resizebox{\textwidth}{!}{
\begin{tabularx}{\targetwidth}{@{}CCCCCCCC@{}}
  \toprule
    && \multicolumn{2}{c}{Full Sample} & \multicolumn{2}{c}{Basic + Marketing \& Research} & \multicolumn{2}{c}{Third-Party}\\
    \cmidrule(lr){3-4} \cmidrule(lr){5-6} \cmidrule(lr){7-8}
     && WTP & WTA& WTP & WTA& WTP & WTA\\
    \midrule
    \multicolumn{8}{l}{H1: Treatment $>$ Control$^1$}\\
    \midrule
    \multicolumn{2}{c}{HH}          & 0.724             & 0.762             & 0.166             & 0.154             & \textbf{0.043}+    & \textbf{0.027}+\\
    \multicolumn{2}{c}{Daily}       & \textbf{0.021}+   & \textbf{0.068}+   & 0.115             & 0.126             & 0.946             & 0.873\\
    \multicolumn{2}{c}{Anon RT}     & 0.606             & 0.769             & \textbf{0.041}+    & \textbf{0.073}    & \textbf{0.041}+    & \textbf{0.034}+\\
    \multicolumn{2}{c}{Anon HH}     & \textbf{0.065}+   & 0.512             & \textbf{0.005}**   & \textbf{0.024}*   & 0.299             & 0.111\\
    \multicolumn{2}{c}{Anon Daily}  & 0.641             & 0.775             & \textbf{0.011}*   & \textbf{0.074}    & \textbf{0.036}+    & \textbf{0.066}\\
\end{tabularx}
}
\end{table}
\begin{table}[H]
\centering
\caption{Mixed Logit Model with Socio-Demographic Interactions} 
\label{tab:mixl.het}
\setlength\targetwidth{0.9\textwidth}
\begin{tabularx}{\targetwidth}{@{}rCCCCCCC@{}}
  \toprule
  \textbf{Main Effects} & \multicolumn{2}{c}{Mean} &  \multicolumn{2}{c}{St. Dev.}\\ 
  \midrule
  Fee (\%)&-0.126*** &(0.010) \\
  Discount (\%)& 0.072*** &(0.006)  \\
  Anon&  0.030 &(0.278) &1.538*** &(0.068)\\
  HH & 0.164 &(0.224) & 0.521*** &(0.081)\\
  Daily &   0.543* &(0.250)   & 0.014 &(0.142)\\
  Anon x HH &  0.273 &(0.303) & 0.023 &(0.185)\\
  Anon x Daily &-0.670** &(0.232) & 1.174*** &(0.186)\\
\end{tabularx}
\end{table}

\begin{landscape}
    \begin{table}[H]
    \centering
    \caption{Complete Combinatorial Test $p$-values for Heterogeneity in Willingness-to-Pay/Accept} 
    \label{tab:wtpa.het.tests}
    \setlength\targetwidth{1.3\textwidth}
    \resizebox{\textwidth}{!}{
    \begin{tabularx}{\targetwidth}{@{}CCCCCCCCCCCC@{}}
      \toprule
      \multicolumn{2}{c}{Group} & \multicolumn{5}{c}{WTP} & \multicolumn{5}{c}{WTA}\\
      \cmidrule(lr){1-2} \cmidrule(lr){3-7} \cmidrule(lr){8-12}
        1 & 2 & HH & Anon HH & Daily & Anon Daily & Anon RT & HH & Anon HH & Daily & Anon Daily & Anon RT\\
      \midrule
      \multicolumn{12}{l}{\textbf{Gender (GEN)}}\\
      \midrule
    F & M & 0.143 & 0.385 & \textbf{0.044}* & \textbf{0.000}*** & \textbf{0.009}** & 0.143 & 0.356 & \textbf{0.046}* & \textbf{0.000}*** & \textbf{0.014}* \\ 
    \midrule
    \multicolumn{12}{l}{\textbf{Age (AGE)}}\\
    \midrule
      35-54 & 18-34 & 0.774 & 0.206 & 0.487 & 0.803 & 0.205 & 0.773 & 0.162 & 0.487 & 0.807 & 0.210 \\
      55-64 & 18-34 & 0.797 & 0.463 & 0.232 & 0.492 & 0.211 & 0.797 & 0.456 & 0.234 & 0.492 & 0.216 \\ 
      65+ & 18-34 & 0.663 & 0.372 & \textbf{0.058} & \textbf{0.067} & \textbf{0.026} & 0.662 & 0.351 & \textbf{0.061} & \textbf{0.067} & \textbf{0.032} \\
\end{tabularx}
    }
    \end{table}
    \end{landscape}
\begin{landscape}
\begin{table}[H]
\centering
\caption{Combinatorial Test $p$-values for Segmented Mixed Logit Models} 
\label{tab:seg.het}
\setlength\targetwidth{1.3\textwidth}
\resizebox{\textwidth}{!}{
\begin{tabularx}{\targetwidth}{@{}CCCCCCCCCCCC@{}}
  \toprule
  \multicolumn{2}{c}{Group} & \multicolumn{5}{c}{WTP} & \multicolumn{5}{c}{WTA}\\
  \cmidrule(lr){1-2} \cmidrule(lr){3-7} \cmidrule(lr){8-12}
    1 & 2 & HH & Anon HH & Daily & Anon Daily & Anon RT & HH & Anon HH & Daily & Anon Daily & Anon RT\\
  \midrule
  \multicolumn{12}{l}{\textbf{Gender (GEN)}}\\
  \midrule
F & M & 0.193 & \textbf{0.001}** & 0.522 & \textbf{0.059}+ & 0.125 & 0.185 & \textbf{0.000}*** & 0.341 & \textbf{0.056}+ & \textbf{0.099}+ \\ 
\midrule
\multicolumn{12}{l}{\textbf{Age (AGE)}}\\
\midrule
  35-54 & 18-34 & 0.948 & 0.662 & 0.413 & \textbf{0.040} & 0.834 & 0.944 & 0.763 & 0.505 & \textbf{0.036} & 0.830 \\ 
  55-64 & 18-34 & 0.611 & 0.215 & 0.343 & \textbf{0.007}* & 0.213 & 0.736 & 0.591 & 0.599 & \textbf{0.012}+ & 0.406 \\ 
  65+ & 18-34 & 0.759 & \textbf{0.024} & 0.308 & \textbf{0.000}** & 0.217 & 0.791 & 0.114 & 0.423 & \textbf{0.004}* & 0.327 \\
\end{tabularx}
}
\end{table}
\end{landscape}
\begin{table}[H]
\centering
\caption{Complete Combinatorial Test $p$-values for Mean Willingness-to-Pay/Accept by Treatment and Data Sharing Attitude} 
\label{tab:poe.freq.robust}
\setlength\targetwidth{1.2\textwidth}
\resizebox{\textwidth}{!}{
\begin{tabularx}{\targetwidth}{@{}CCCCCCCC@{}}
  \toprule
    && \multicolumn{2}{c}{Full Sample} & \multicolumn{2}{c}{Basic + Marketing \& Research} & \multicolumn{2}{c}{Third-Party}\\
    \cmidrule(lr){3-4} \cmidrule(lr){5-6} \cmidrule(lr){7-8}
     && WTP & WTA& WTP & WTA& WTP & WTA\\
    \midrule
    \multicolumn{8}{l}{H1: Treatment $>$ Control$^1$}\\
    \midrule
    \multicolumn{2}{c}{HH}          & 0.749 & 0.791 & \textbf{0.034} & \textbf{0.021}+ & 0.155 & 0.152 \\ 
    \multicolumn{2}{c}{Daily}       &  \textbf{0.026}+ & \textbf{0.089} & 0.957 & 0.837 & 0.167 & 0.121 \\
    \multicolumn{2}{c}{Anon RT}     & 0.631 & 0.802 & \textbf{0.041} & \textbf{0.041} & \textbf{0.050} & \textbf{0.067} \\ 
    \multicolumn{2}{c}{Anon HH}     & 0.133 & 0.704 & 0.286 & \textbf{0.085} & \textbf{0.014}* & \textbf{0.028}+ \\ 
    \multicolumn{2}{c}{Anon Daily}  & 0.715 & 0.826 & \textbf{0.038} & 0.123 & \textbf{0.024}+ & \textbf{0.080} \\
\end{tabularx}
}
\end{table}

\newpage
\section{Supporting Tables for Demand for Smart Meters Analysis}\label{app:wsm}

\begin{table}[H]
\centering
\caption{Non-Parametric Testing of Heterogeneity for Demand on Smart Meters} 
\label{tab:wsm.sig}
\setlength\targetwidth{\textwidth}
\resizebox{0.9\textwidth}{!}{
\begin{tabularx}{\targetwidth}{@{}CCCCCCCCCC@{}}
  \toprule
  \multicolumn{2}{c}{\multirow{2}{*}{Grouping}} & \multirow{2}{*}{\makecell{Test\\Stat}} & \multicolumn{3}{c}{$p$} & \multicolumn{4}{c}{Effect} \\ 
\cmidrule(lr){4-6} \cmidrule(lr){7-10}
 &  &  & unadj. & adj. & sig. & Est. & Low & High & Size \\
 \midrule
    \multicolumn{10}{l}{\textbf{Age (AGE)}}\\
  \midrule
  \multicolumn{2}{c}{Omnibus (3)} & 38.40 & 0.000 &  & *** & 0.04 & 0.03 & 1.00 & small \\ 
  18-34 & 35-54 & 3.94 & 0.000 & 0.000 & *** & -0.15 & -0.24 & -0.06 & small \\ 
  18-34 & 55-64 & 4.44 & 0.000 & 0.000 & *** & -0.21 & -0.32 & -0.10 & small \\ 
  18-34 & 65+ & 5.75 & 0.000 & 0.000 & *** & -0.24 & -0.34 & -0.15 & small \\ 
  35-54 & 55-64 & 1.29 & 0.196 & 0.391 & n.s. &  &  &  &  \\
\end{tabularx}
}
\end{table}

\begin{table}[H]
\centering
\caption{Non-Parametric Testing of Treatment Effect on Demand for SM} 
\label{tab:wsm.gp.sig}
\setlength\targetwidth{\textwidth}
\begin{tabularx}{\targetwidth}{@{}CCCCCCCCCCC@{}}
  \toprule
\multicolumn{2}{c}{\multirow{2}{*}{Grouping}} & \multirow{2}{*}{\makecell{Test\\Stat}} & \multicolumn{4}{c}{$p$} & \multicolumn{4}{c}{Effect} \\ 
\cmidrule(lr){4-7} \cmidrule(lr){8-11}
 &  &  & unadj. & adj. & full & sig. & Est. & Low & High & Size \\
  \midrule
   \multicolumn{10}{l}{\textbf{Experimental Group}}\\
  \midrule
    Control & Treatment  & 115793 & 0.865 &  & & n.s. &  &  &  &  \\ 
 \midrule
   \multicolumn{10}{l}{\textbf{Data Sharing Attitudes (DSA)}}\\
  \midrule
  \multicolumn{2}{c}{Omnibus (5)} & 24.88 & 0.000 &  &  & *** & 0.03 & 0.02 & 1.00 & small \\ 
  \multicolumn{2}{c}{Basic} & 0.71 & 0.478 & 1.000 & 1.000 & n.s. &  &  &  &  \\ 
  \multicolumn{2}{c}{Marketing} & 0.68 & 0.498 & 1.000 & 1.000 & n.s. &  &  &  &  \\
\end{tabularx}
\end{table}

\begin{table}[H]
\centering
\caption{Binary Logistic Regression Model for Demand for Smart Meter} 
\label{tab:wsm.lr}
\setlength\targetwidth{0.6\textwidth}
\begin{tabularx}{\targetwidth}{@{}rCCCC@{}}
  \toprule
    & \multicolumn{2}{c}{Base} & \multicolumn{2}{c}{w. Interactions} \\ 
  \midrule
  GEN Female    & 0.565**  & (0.187) & 0.584**  & (0.188) \\  
  SM No         & 2.409*** & (0.231) & 2.431*** & (0.232) \\  
  SEG C1        & 0.373    & (0.263) & 0.370    & (0.265) \\  
  SEG C2        & 0.275    & (0.281) & 0.286    & (0.282) \\
  SEG DE        & 0.545*   & (0.252) & 0.553*   & (0.254) \\  
  AGE 35-54     & 0.630*   & (0.250) & 0.627*   & (0.251) \\  
  AGE 55-64     & 0.928**  & (0.298) & 0.932**  & (0.298) \\
\end{tabularx}
\end{table}

\newpage
\printglossaries

\end{document}